\documentclass[a4paper,titlepage,onecolumn,twoside,12pt,english]{scrbook}

\usepackage[T1]{fontenc}

\usepackage{amsmath, amssymb, mathtools, proof}
\usepackage{nicefrac}
\usepackage{xcolor,graphicx}
\usepackage{scalerel}
\usepackage{xspace}
\usepackage{paralist}
\usepackage{wrapfig}
\usepackage{tikz}
\usetikzlibrary{trees,decorations,decorations.shapes,arrows,automata,positioning,plotmarks,shapes,backgrounds,petri}

\usepackage[english,ngerman]{babel} 
\usepackage[utf8]{inputenc}
\usepackage{amsthm}
\usepackage{tabularx,booktabs,multirow}
\usepackage[margin=0pt]{subcaption}
\usepackage{rotating}
\usepackage[square, authoryear]{natbib}

\usepackage{thesis}
\usepackage{compositional_probabilistic_MCMPST}

\let\tmp\oddsidemargin
\let\oddsidemargin\evensidemargin
\let\evensidemargin\tmp
\reversemarginpar
\allowdisplaybreaks

\begin{document}

\selectlanguage{english}

\maketitle
\cleardoublepage

\section*{Usage of Artificial Intelligence}
The large language model ChatGPT, \url{chatgpt.com}, was sparingly used throughout this thesis for general language help concerning tone, grammar, and vocabulary and as a guide finding literature. All usage is contained in one publicly accessible chat, whose link is found in the bibliography \citep{chatgpt}.

\cleardoublepage

\selectlanguage{english}

\section*{Abstract}

Multiparty session types (MPST) are a robust typing framework that ensures safe and deadlock-free communication within distributed protocols.
As these protocols grow in complexity, compositional modelling becomes increasingly important to scalably verify their behaviour.
Therefore, we propose using a refinement-based subtyping approach to facilitate the modularity needed for compositional verification.
Subtyping in classic MPST systems inherently represents a notion of refinement: A larger type may be \emph{safely substituted} by a smaller, refined type.
The aim of this thesis is to significantly extend this concept and discover just how flexible and expressive subtyping relations can be.\medbreak

We present a probabilistic extension for MPST, the \emph{\calculusname{} $\pi$-calculus}, with a novel, flexible subtyping system which allows one channel (the interface) to be substituted by several channels (the refinement).
Our subtyping is remarkably expressive; any selection of well-typed channels as the refinement has a corresponding interface in a single channel type.
To facilitate this generality, we base our system on a powerful variant of MPST, mixed choice multiparty session types (MCMP), which offers greater flexibility in communication choices. 

We establish soundness of the \emph{\calculusname{} system} through several key results.
In particular, we prove subject reduction, error-freedom and deadlock-freedom, ensuring that well-typed processes are well-behaved.\medbreak

This work demonstrates subtyping to possess great previously untapped potential for stepwise refinement and compositional verification.
The presented framework enables highly expressive, compositional, and verifiable modelling of probabilistic distributed communication.
A promising avenue for further research is \emph{imperfect refinement}, a logical extension of the system which leverages the strengths of the probabilistic setting: We can allow the refinement to be deadlock-free with bounded probability instead of in $100\%$ of cases.

\newpage

\tableofcontents
\listofexamples

\newpage

\chapter{Introduction}
In general, the term \emph{interface} refers to the observable---or externally accessible---aspects of a component, module, or system; the exact representation of an interface inherently depends on the  formalism at hand.
In many cases, interfaces appear at various levels of abstraction and granularity, which may be developed through \emph{stepwise refinement} starting from some high-level description and adding details with each refinement step; reasonable notions of refinement relations should obviously be transitive.
Ideally, refinement can be done in a \emph{compositional} manner, where single components may be refined independently, while yielding a refinement of the full system.
There is a large body of research on notions of refinement and accompanying methods, as it is generally quite specific to the formalism at hand.
Consequently, there is no real consensus on definitions and interpretations, not even on terminology.

State-based representations of systems are likely the class of formalisms with the most diverse contributions.
There, refinement relations typically compare behaviours (in its simplest form traces), often expressed as a mapping from the refined version back to its initial specification~\citep{DBLP:conf/lics/AbadiL88}.
The concepts of (forward and/or backward) \emph{simulation}~\citep{DBLP:journals/iandc/LynchV95}, which roughly correspond to completeness and soundness requirements of an implementation \wrt its specification, are possibly even more wide-spread and come in many flavors.

In \emph{multiparty session types} (MPST)~\citep{yoshida2019very}, where our work is situated, the interface of a process captures the type of the session it participates in, which describes the sequence of input/output actions (messages), along with their types, direction, and order of communication.
Here, often enough, the notion of \emph{channel} comprises a particular \emph{session} of a protocol among several participants, including all of the roles that act within the respective protocol to achieve a common goal \citep{honda98,hondaYoshidaCarbone08, scalas2019less}.
If an interface is represented as a type, then refinement is naturally and best realized as a subtyping relation.

In this thesis, interfaces are (sets of) channels, which are themselves constructed from sessions and the roles used within them.
As usual these channels are given as implementation in a session calculus and as specification in \emph{local contexts}, where a local context $ \Delta $ is a collection of channel names and their respective channel types.
The type system provides rules to check the implementation against its specification and ensures safety and deadlock-freedom (among other properties that could be defined).

As part of our type system, we introduce a novel subtyping relation $ \leq_{\prob} $ that allows to also verify probabilities $ \prob $ and specifies a refinement relation on local contexts.
Starting from $ \Delta $, if $ \Delta' \leq_1 \Delta $ then the \emph{refinment} $ \Delta' $ refines the \emph{interface} $ \Delta $ by distributing some behaviour (on single channels) on sets of interacting channels.
Starting from $ \Delta' $, if $ \Delta' \leq_1 \Delta $ then the \emph{interface} $ \Delta $ provides an abstract specification of the \emph{refinment} $ \Delta' $, where abstract means from an external point of view abstracting from interactions within $ \Delta' $.

\medbreak
To gain some initial understanding, we highlight said dynamic of a refinement distributing the behaviour of a single channel, the interface, to a set of interacting channels with an example.
We will keep revisiting this setting throughout the thesis as we build up the theory of our system.
We use the symbol $\bigcirc$ to indicate the end of an example.
\begin{example}[Running Example]
\addxcontentsline{loe}{example}[\theexample]{Running Example}
	\label{ex:motivatingExample}
	Consider a \participant{p}laintiff bringing a lawsuit to \participant{c}ourt, which, after possibly taking \participant{w}itness testimony, announces a verdict. To initiate the protocol,{\parfillskip=0pt\par}
	\vspace{3pt}\noindent
	\begin{minipage}{\textwidth}
	\begin{wrapfigure}[12]{L}{0.32\textwidth}
		\vspace*{-20pt}
		\scalebox{0.95}{
		  \hspace{0pt}
        \begin{tikzpicture}[auto]
			\node (i) at (2, 1) {\textcolor{blue}{interface}};
			\node[state] (a) at (0, 0) {\participant{p}};
			\node[state] (b) at (2, 0) {\participant{c}};
			\node[state] (c) at (4, 0) {\participant{w}};
			\draw[<->] (a) edge (b);
			\draw[<->] (b) edge (c);
			\draw[-,thick, color=blue] (1, 0.75) -- (3, 0.75) -- (3, -0.75) -- (1, -0.75) -- (1, 0.75);
		\end{tikzpicture}
		}
        \vspace{1em}
	\end{wrapfigure}

	the \participant{p}laintiff sends a message to the \participant{c}ourt, announcing the lawsuit ($\mathtt{lws}$).
	Upon receiving this, the \participant{c}ourt decides with a total $70\%$ chance that it has enough information to announce the verdict to the \participant{p}laintiff. Of those $70\%$, half{\parfillskip=0pt\par}
	\end{minipage}\vspace{2pt}
    
    \noindent of the time, \ie with a probability of $35\%$, the \participant{c}ourt will find the \participant{d}efendant \emph{guilty} ($\mathtt{glt}$) and else \emph{not guilty}.
	Otherwise, with $30\%$, the \participant{c}ourt requests ($\mathtt{rqs}$) a \participant{w}itness to testify.
	Upon receiving this request, the \participant{w}itness sends a message to the \participant{c}ourt representing the statement ($\mathtt{st}$) after which the \participant{c}ourt announces the verdict \emph{not guilty} to the \participant{p}laintiff.
	In case no \participant{w}itness is called, the \participant{c}ourt will send them a message releasing ($\mathtt{rls}$) them and the protocol terminates.
    
	This protocol has an issue: The \participant{d}efendant on whom the judgement is passed is not represented as a participant.
	To obtain its verdict, the \participant{d}efendant could implicitly be assumed part of the \participant{c}ourt.
	But then the \participant{c}ourt has a clear conflict of interest: they hold the power to pass judgement on themselves.
    A solution is splitting the \participant{c}ourt, thereby dividing the power.
	Hence, consider the \participant{c}ourt to be an interface that must be refined into two separate participants: \participant{d}efendant and \participant{j}udge. The \participant{p}laintiff is the same as before but now\parfillskip=0pt\par
	\vspace{3pt}\noindent
	\begin{minipage}{\textwidth}
	\begin{wrapfigure}{R}{0.32\textwidth}
		\vspace*{-20pt}
		\scalebox{0.95}{
		\begin{tikzpicture}[auto]
			\node (i) at (2, 1) {\textcolor{blue}{refinement}};
			\node[state] (a) at (0, 0) {\participant{p}};
			\node[state] (b) at (2, 0.225) {\participant{j}};
			\node[state] (d) at (2, -1.225) {\participant{d}};
			\node[state] (c) at (4, 0) {\participant{w}};
			\draw[<->] (a) edge (b);
			\draw[<->] (b) edge (c);
			\draw[<->] (b) edge (d);
			\draw[-,thick, color=blue] (1, 0.75) -- (3, 0.75) -- (3, -1.75) -- (1, -1.75) -- (1, 0.75);
		\end{tikzpicture}
		}
	\end{wrapfigure}
    
    interacts with a \participant{j}udge instead of the \participant{c}ourt.
	This \participant{j}udge, after receiving the lawsuit, waits for a message from the \participant{d}efendant.
%	Which message is sent is modeled as a probabilistic choice.
	The \participant{d}efendant will send a weak ($\mathtt{wk}$) and strong ($\mathtt{str}$) defence with a $50\%$ and $20\%$ likelihood, respectively.
	Otherwise, it requests to call upon a \participant{w}itness, in $30\%$ of cases. If the defence is strong, the verdict is always{\parfillskip=0pt\par}
	\end{minipage}\vspace{2pt}
    
	\noindent\emph{not guilty}.
	With $70\%$ a weak defence results in \emph{guilty} and with $30\%$ in \emph{not guilty}.
	If no statement is given and a \participant{w}itness is requested, the \participant{j}udge receives the \participant{w}itness testimony, and decides as before.
    \exampleDone
\end{example}
\medbreak

The target audience for this thesis is familiar with transition systems and type systems, including $\pi$-calculus, but we do not assume a close familiarity with (multiparty) session types.
We begin by introducing our calculus in Chapter~\ref{chap:sessionCalculus}, \ie its syntax, operational semantics, and typing system.
The latter includes basic subtyping which does not include any of our novel refinement ideas, to ease the reader into the system as a whole.
Conceptually, this chapter can be considered to be our preliminaries.
Afterwards, in Chapter~\ref{chap:subtyping}, we move on to the core of this work, the refinement-based multi-channel subtyping.
We will first build intuition for the novel subtyping system by providing an intermediary set of subtyping rules which bridge the gap between the functionality of basic subtyping and the complex form of the advanced subtyping.
Only then do we finally present the novel subtyping rules.
The chapter contains plenty of examples to guide the reader through the, at times, convoluted syntax.
Additionally, we show that the intermediary subtyping subsumes the basic one and is in turn subsumed itself by the novel, advanced one. 
In Chapter~\ref{chap:properties}, we then prove several essential properties of the presented \emph{\calculusname{} $\pi$-calculus}, its type system, and multi-channel subtyping.
These include subject reduction (§~\ref{subsec:subjectReduction}) and deadlock-freedom (§~\ref{subsec:errorAndDeadlockFreedom}), both standard in MPST.
Additionally, we prove the flexibility of our subtyping: \emph{Any} selection of well-typed, safe, and deadlock-free channels has a corresponding interface in a single channel type (§~\ref{subsec:interfaceExists}).

Large parts of the work presented in this thesis have also been submitted as a paper to the 22nd International Colloquium on Theoretical Aspects of Computing (ICTAC 2025) \citep{blechschmidt25}.

\section{Related Work}
The history of multiparty session types begins with the inception of the process calculus CCS (calculus of communication systems) in 1980 \citep{DBLP:books/sp/Milner80,DBLP:books/daglib/0067019}.
Extended with name-passing, this framework later became the now well-known $\pi$-calculus \citep{Milner93, DBLP:books/daglib/0098267}.
Based on this, the original binary session type theory was conceived in the 1990s \citep{honda93typesfordyadicinteraction, DBLP:conf/parle/TakeuchiHK94, honda98}.
Inspired also by linear logic \citep{girardLinearLogic87}, session type theory can handle binary communication in which the two communication partners essentially act as duals of each other.
With the goal of allowing more than two communicating partners, session types were further developed into the classic MPST framework in \citep{hondaYoshidaCarbone08} (asynchronous) and \citep{DBLP:journals/entcs/BejleriY09} (synchronous).
The slightly simplified MPST system of \citep{DBLP:conf/concur/BettiniCDLDY08}, introduced shortly afterwards, gained wide acceptance as the canonical version of MPST.
(Multiparty) session types have been extensively researched in many contexts since then, see surveys of the field \citep{huettel16survey, gay17survey}.
The duality-based approach of binary sessions carried over into multiparty sessions, which was overhauled by \citep{scalas2019less}, becoming the new standard.
Hence, we, too, base our framework on theirs.

Our system is additionally heavily influenced by \citep{peters2024separation}, who presented a mixed choice multiparty (MCMP) calculus (based on \citep{DBLP:journals/tcs/CasalMV22}).
We integrate their approach into our system.
As shown in \citep{DBLP:journals/corr/abs-2209-06819, peters2024separation}, MCMP is strictly more expressive than both standard MPST and the system of \citep{DBLP:journals/tcs/CasalMV22}.

There is a significant history of integrating probabilities into transition systems, as early as 1991 with work on probabilistic bisimulation \citep{DBLP:journals/iandc/LarsenS91}.
A number of process calculi have been reimagined with probabilities in literature, including probabilistic CCS \citep{DBLP:books/daglib/0074822} and probabilistic $\pi$-calculus \citep{DBLP:conf/fossacs/HerescuP00}.
Binary sessions have seen probabilistic extension in \citep{inverso_et_al:LIPIcs.CONCUR.2020.14}.
We take inspiration mostly from \citep{aman2019probabilities}, who extended the MPST theory of \citep{hondaYoshidaCarbone08, DBLP:journals/jacm/HondaYC16}.

Subtyping has been a fundamental concept in these systems since the introduction of subtyping for $\pi$-calculus in \citep{DBLP:journals/mscs/PierceS96}.
But even earlier, there were precursors to subtyping for CCS in, for example, simulation \citep{DBLP:books/sp/Milner80, DBLP:journals/iandc/LynchV95} and testing preorders capturing safe replacement \citep{DBLP:journals/tcs/NicolaH84}.
Later, in \citep{DBLP:journals/acta/GayH05}, this concept has been adapted for the subsequent binary session theory. 
Modern MPST subtyping is largely based on the framework of \citep{DBLP:journals/corr/Dezani-Ciancaglini16}.
Our subtyping fundamentally builds on \citep{peters2024separation} (itself based on \citep{chenPrecisenessOfSubtyping17}), in particular for their handling of mixed choices, though the additional complexity of our multi-channel system makes the connection less immediately apparent.

Another approach to integrate refinement in subtyping, though based on a very different motivation, can be found in \citep{DBLP:conf/concur/Horne20}.
In comparison, our subtyping is much more flexible, as we will prove in Section~\ref{subsec:interfaceExists}.
For instance, we do not exclude races, \ie two senders that want to communicate with the same receiver at the same time.

%%%%%%%%%%%%%%%%%%%%%%%%%%%%
%%  the session calculus  %%
%%%%%%%%%%%%%%%%%%%%%%%%%%%%

\chapter[Preliminaries: The Probabilistic Mixed Choice MPST pi-Calculus]{Preliminaries: The Probabilistic Mixed Choice MPST {\scalebox{1.2}{$\pi$}}-Calculus}
\label{chap:sessionCalculus}
After giving a small introduction to multiparty session types (MPST), this chapter introduces the process syntax (§~\ref{sec:processSyntax}), the operational semantics (§~\ref{sec:operationalSemantics}) and the typing system of the \emph{\calculusname{} $\pi$-calculus}.\medbreak

Given the complexity and breadth of the field, we will refrain from explaining MPST in full technicality, neither will we expect detailed background knowledge.
For the curious reader, we recommend \citep{yoshida2019very} as a starting point.

In general terms, MPST is a formal framework for specification and verification of communication protocols involving more than two participants.
Participants are modelled as processes which are formed by a certain grammar, the process syntax.
The core functionality of these processes is communication: Sending and receiving messages to and from other processes.
These messages are sent and received via \emph{channels}.
A channel comprises a common name for a \emph{session} both participants are in, and a \emph{role}, which is an identifier within said session.
In communication, by using the channel and the syntax of the sending action, it is therefore specified which role is sending what message to which other role, where both communication partners must share the same session. 
Interactions happen in synchronous steps, both communication partners ``use up'' the action at the same time: The process makes a \emph{reduction} step (asynchronous MPST exists too, see \cite{hondaYoshidaCarbone08}, we use synchronous MPST). 
To verify or specify interactions between processes, channels are assigned \emph{session types} using typing rules.
Processes built from these types are then said to be \emph{justified} by them.
A key property of these systems, called subject reduction, is that a process justified by some types will be justified by some other types once it reduces.
In other words, once we establish a process to be well-behaved in some sense, we know that no matter which steps it performs, well-behaved-ness remains.
Furthermore by verifying that a collection of types fulfils certain properties, we can infer that processes justified by that collection also fulfil them.
One of the most common properties of interest is deadlock-freedom, the guarantee that whenever no more interaction can occur, all participants are finished.
In line with modern MPST theory, our system \emph{does not} have projection (of global types onto local types) and thus much of the understanding for general $\pi$-calculus is transferable to the work of this thesis.\medbreak

At its core, our system is a standard multiparty session calculus (as in \citep{scalas2019less,scalasYoshida18}) without session initialisation and delegation, \ie without transferring access to certain channels across participants.
We extend this by the mixed choices of \citep{peters2024separation} (based on \citep{Milner93,milnerParrowWalker92}).
Classically, when a participant seeks to communicate, it will be able to do so with only exactly one other participant and only in one mode (sending \emph{or} receiving).
Mixed choices allow the participant to offer sending actions and receiving actions at the same time, to and from arbitrarily different participants.
Mixed choice calculi have increased expressiveness and flexibility.
Finally, our system includes probabilities in the outputs (as in \citep{aman2019probabilities,inverso_et_al:LIPIcs.CONCUR.2020.14}, based on \citep{DBLP:conf/fossacs/HerescuP00, DBLP:journals/entcs/VaraccaY07}).

\section{Process Syntax}
\label{sec:processSyntax}
We begin by defining which shapes our processes may assume.
After explaining the definition in detail, we follow by formalizing the courthouse example from the introduction using this syntax.
\begin{definition}[Process Syntax]
	\label{def:processSyntax}
	The syntax of the \emph{\calculusname\ $ \pi $-calculus} is inductively given by:
	\begin{alignat*}{3}
		v & \defT && x, y, z,\mydots \gbar 1,2,\mydots \gbar \true, \false \quad\kern-2pt && \text{(variables, numbers, booleans)}\\
		c & \defT && \channel{s}{r} && \text{(session with set of roles)}\\
		%\prob_i & \defT && p && \text{(probability $ p \in \mathbb{R} $)}\\
		P, Q, P_i & \defT && \0 \gbar P \inparallel Q \gbar \res{s}P\quad\kern-2pt && \text{(inaction, parallel composition, restriction)}\\
		&& \gbar & \cond{v}{P}{Q} && \text{(conditional)}\\
		&& \gbar & \procdef{D}{P} \gbar \processcall{X}{\vecnew{v}}{\vecnew{c}} && \text{(process definition, process call)}\\
		&& \gbar & c\mc{i}{I} M_i && \text{(mixed choice on $ c $ with finite $ I \neq \emptyset $)}\\
		M_i & \defT && \inp{p}{q}{l}{x}{P} && \text{($ \participant{p} $ receives message $ l{\left( x\right)} $ from $ \participant{q} $)}\\
		&& \gbar & \send{i}{I} \probOut{\prob_i}{N_i \cont P_i} \quad && \text{(probabilistic choice with finite $ I \neq \emptyset $)}\\
		N_i & \defT && \outPrefix{p}{q}{l}{v} \gbar \nonsend && \text{($ \participant{p} $ sends message $ l{\left\langle v \right\rangle} $ to $ \participant{q} $, internal action)}\\
        D & \defT && \procdecl{X}{\vecnew{x}}{\vecnew{c}} \defeq P && \text{(declaration of process constant } X \text{)}
	\end{alignat*}
\end{definition}
We will first explain the syntax.
\emph{Values} $v$ are variables, numbers, or booleans.
Note that the inclusion of more complex values, such as floating point numbers, would be a straightforward, orthogonal extension of the system.
A \emph{channel} $ c = \channel{s}{r} $ specifies a session $s$ being used as a communication endpoint by the roles $\vecparticipant{r} $ to interact with other roles within that session.
The term ``participant'' is overloaded in MPST, referring both to a participant in a communication protocol, and a participant in a session.
In our system, a single process may use several different channels $\channel{s}{r}$, possibly even from different sessions.
Thus, to avoid confusion, we will refer to $\vecparticipant{r}$ only as ``roles''.
In standard session typing, instead of sets, single roles are used.
The reason why we chose these role sets will become apparent when the multi-channel subtyping is introduced later.

\emph{Inaction} $ \0 $ is the representation of a terminated process, we sometimes omit trailing $ \0 $.
\emph{Composition} $ P \inparallel Q $ allows the processes $ P $ and $ Q $ to run in parallel.
We often use the previously mentioned notion of ``participant'' to denote parallel processes.
Parallelly composed processes, \ie different participants may interact.
\emph{Restriction} $ \res{s}P $ encapsulates a \emph{session} $ s $ within the scope of $P$.
The \emph{conditional} $ \cond{v}{P}{Q} $ behaves as $ P $ if $ v $ is true and else as $ Q $.
\emph{Process definition} $\procdef{D}{P}$ models recursion in our calculus through recursive process calls.
Summands of \emph{mixed choices} $c\mc{i}{I} M_i$ are inputs $\inp{p}{q}{l}{x}{P}$ and output sums $\probOut{\prob_i}{N_i \cont P_i}$.
They differ from \citep{peters2024separation} in the fact that our output sums are probabilistic.
The channel $c$ in front of the sum specifies the session $s$ within which the entire mixed choice is taking place.
In \emph{inputs} $\inp{p}{q}{l}{x}{P}$, role $ \participant{p} $ receives a message with label $l$ from role $ \participant{q} $.
The transmitted payload will be stored in variable $x$.
After receiving the input, the participant will continue as the process $P$.
Classically, a sum of inputs is called \emph{branching}, as each different input a participant receives, allows them to continue differently.
In other words, their behaviour branches depending on the communication partner('s choices).
\emph{Probabilistic output choices} $\send{i}{I} \probOut{\prob_i}{N_i.P_i}$, or {probabilistic sums}, function similarly to typical non-probabilistic output choices.
Where classically one output action is chosen nondeterministically from a sum of outputs (called \emph{selection}), in our system one probabilistic output \emph{sum} is chosen nondeterministically from several occurring in a mixed choice.
Clearly differentiating the nondeterministic choices from probabilistic ones is key for probabilistic systems (see \citep{DBLP:conf/concur/SegalaL94}).
Then the output action is selected according to the probabilities $\prob_i \in \mathbb{R}$.
These two selection steps, however, functionally occur simultaneously, not successively.
\emph{Output actions} $N_i$ may either be $\outPrefix{p}{q}{l}{v}$, where role $\participant{p}$ sends a message with label $l$ and payload $v$ to role $\participant{q}$, or $\nonsend$, an \emph{internal action}.
Afterwards, the participant continues as process $P_i$.
\emph{Declarations} $ D $ provide definitions of processes that may be invoked by a \emph{process call} $ \processcall{X}{\vecnew{v}}{\vecnew{c}} $ with some values $ \vecnew{v} $, where $ \vecnew{c} $ lists the channels used by the process that is declared by $ X $.\medbreak

Next, let us specify the binders.
We adopt a form of Barendregt convention: We assume that alpha-conversion is implicitly applied to ensure that all bound variables are pairwise distinct and different from free ones.
Restriction $ \res{s}P $ binds session $ s $ in $ P $.
Declaration $ \procdecl{X}{\vecnew{x}}{\vecnew{c}} \defeq P $ binds process constants $ X $ and variables $ \vecnew{x} $ in $ P $. The vector $\widetilde{c}$ lists the channels used in $ P $.
Message receiving $ \inp{p}{q}{l}{x}{P} $ binds the variable $ x $ in $ P $.
All other occurrences of process constants, variables, and sessions are free.
Let $ \fs{P} $ and $ \fs{D} $ denote the set of free sessions in $ P $ and $ D $, respectively.
Let $ \dpv{D} $ be the set of process variables declared in $ D $ and let $ \fpv{P} $ and $ \fpv{D} $ denote the set of free process variables in $ P $ and $ D $.
Substitution $ P\subst{x_1, \ldots, x_n}{v_1, \ldots, v_n} $ simultaneously replaces all free occurrences of $ x_i $ by $ v_i $ in $ P $, possibly applying alpha-conversion to avoid capture.
Substitution $ P\subst{\vecnew{x}}{\vecnew{v}} $ is undefined if $ \length{\vecnew{x}} \neq \length{\vecnew{v}} $.\medbreak

Finally, for legibility and convenience, we introduce some shorthands and abbreviations.
We abbreviate singleton sums $ c \mc{i}{\{1\}} M $ as $ \insess{c}M $ and $ \send{i}{\{1\}} \probOut{\prob}{N.P} $ as $ \probOut{\prob}{N.P} $.
We sometimes omit the probability $ 1 $, \ie abbreviate outputs $ \probOut{1}{N.P} $ as $ N.P $.
Whenever $ I \cap J = \emptyset $ and $ I \cup J \neq \emptyset $, we allow to split a mixed choice $ c \mc{i}{I \cup J} M_i $ into $ \insess{c}\mc{i}{I} M_i + \mc{j}{J} M_j $ and we allow to split a probabilistic choice $ \send{i}{I \cup J} \probOut{\prob_i}{N_i\cont{P_i}} $ into $ \send{i}{I} \probOut{\prob_i}{N_i\cont{P_i}} \oplus \send{j}{J} \probOut{\prob_j}{N_j\cont{P_j}} $.
In particular, we often split sums into a single summand and the rest of the sum, \ie $ c\mc{i}{I} M_i $ becomes $ \insess{c}M_j + \choice $ with $ \choice = c\mc{i}{I \setminus \{j\}} M_i $ and $ \send{i}{I} \probOut{\prob_i}{N_i\cont{P_i}} $ becomes $\probOut{\prob_j}{N_j}\cont{P_j} \oplus \choiceProb $ with $ \choiceProb = \send{i}{I \setminus \{j\}} \probOut{\prob_i}{N_i\cont{P_i}} $.
To simplify the reduction rules in Definition~\ref{def:reductionSemantics}, we allow $\choice$ and $\choiceProb$ to be \emph{empty} mixed/probabilistic choices.
We allow to unify and split similar summands of probabilistic choices, \ie $ \probOut{\prob_i}{N\cont P} \oplus \probOut{\prob_j}{N\cont P} = \probOut{{\left(\prob_i + \prob_j\right)}}{N\cont P} $.\medbreak

Let us return to the running example to show the protocol as a process in our syntax.
\begin{example}[Running Example---Syntax]
\addxcontentsline{loe}{example}[\theexample]{Running Example---Syntax}
	\label{ex:motivatingExampleSyntax}
	The unrefined interface of the courthouse system in Example~\ref{ex:motivatingExample} can be implemented as a process $P_{\interface} = \res{s}{\left( P_{\participant{p}} \inparallel P_{\participant{c}} \inparallel P_{\participant{w}} \right)} $, whose components represent the participants in the described protocol. The \participant{p}laintiff is $P_{\participant{p}}$, the \participant{c}ourt is $P_{\participant{c}}$, and the \participant{w}itness is $P_{\participant{w}}$. The implementations of these processes are given as
	\begin{gather*}
		P_{\participant{p}} = \insess{\channelsingle{s}{p}}{\out{p}{j}{\mathtt{lws}}{}{}} \insess{\channelsingle{s}{p}}{} \inpPrefix{p}{j}{\mathtt{glt}}{x}\\
		P_{\participant{c}} \kern-1pt=\kern-1pt \insess{\channel{s}{c}}{\inp{j}{p}{\mathtt{lws}}{}{}}\insess{\channel{s}{c}}{}  \begin{cases}
			&\probOut{0.35}{\outPrefix{j}{p}{\mathtt{glt}}{\true}} \cont \insess{\channelsingle{s}{j}}{\outPrefix{j}{w}{\mathtt{rls}}{}}\\
			\!\oplus\!\!\! &\probOut{0.35}{\outPrefix{j}{p}{\mathtt{glt}}{\false}} \cont \insess{\channelsingle{s}{j}}{\outPrefix{j}{w}{\mathtt{rls}}{}}\\
			\!\oplus\!\!\! &\probOut{0.3}{\outPrefix{j}{w}{\mathtt{rqs}}{}}\cont \insess{\channel{s}{c}}{\inpPrefix{j}{w}{\mathtt{st}}{}} \cont \insess{\channel{s}{c}}{\outPrefix{j}{p}{\mathtt{glt}}{\false}}
		\end{cases}\\
		P_{\participant{w}} = \insess{\channelsingle{s}{w}}{\begin{cases}
			& \inp{w}{j}{\mathtt{mtg}}{}{} \insess{\channelsingle{s}{w}}{\outPrefix{w}{j}{\mathtt{st}}{}} \\
			+ & \inpPrefix{w}{j}{\mathtt{rls}}{}
		\end{cases}}
	\end{gather*}
	where the role set of the channel $\channel{s}{c}$ used by the \participant{c}ourt process $P_{\participant{c}}$ is $\vecparticipant{c} = \{\participant{j},\participant{d}\}$, as it embodies both the \participant{j}udge and \participant{d}efendant.
    Avid readers might notice that the \participant{d}efendant is not found within the actions of $P_{\participant{c}}$.
    As, however, this participant gets refined into two, both roles are already found in the role set.
    Syntactically, this does not pose a problem; unused roles are indeed allowed to occur in role sets, as we will see later on.  

	The refined system can be implemented as $ P_{\refinement} = \res{s}{\left( P_{\participant{p}} \inparallel P_{\participant{j}} \inparallel P_{\participant{d}} \inparallel P_{\participant{w}} \right)} $.
	The processes $P_{\participant{p}}$ and $P_{\participant{w}}$ stay the same and we have:
	\begin{gather*}
        P_{\mathtt{rls}} = \insess{\channel{s}{c}}{\outPrefix{j}{w}{\mathtt{rls}}{}}\\
		P_{\participant{d}} = \insess{\channelsingle{s}{d}}
		\begin{cases}
			& \probOut{0.5}{\outPrefix{d}{j}{\mathtt{wk}}{}} \\
			\oplus & \probOut{0.2}{\outPrefix{d}{j}{\mathtt{str}}{}} \\
			\oplus & \probOut{0.3}{\outPrefix{d}{j}{\mathtt{wit}}{}}
		\end{cases}\\
		P_{\participant{j}} = \insess{\channelsingle{s}{j}}{\inp{j}{p}{\mathtt{lws}}{}{}} \insess{\channelsingle{s}{j}}
		\begin{cases}
			& \inpPrefix{j}{d}{\mathtt{wk}}{} \cont \insess{\channelsingle{s}{j}}
			\begin{cases}
				&\probOut{0.7}{\outPrefix{j}{p}{\mathtt{glt}}{\true}} \cont P_{\mathtt{rls}} \\
				\oplus &\probOut{0.3}{\outPrefix{j}{p}{\mathtt{glt}}{\false}} \cont P_{\mathtt{rls}}
			\end{cases}\\
			\!+\!\!\!\! & \inpPrefix{j}{d}{\mathtt{str}}{} \cont \insess{\channelsingle{s}{j}}{\outPrefix{j}{p}{\mathtt{glt}}{\false}} \cont P_{\mathtt{rls}} \\
			\!+\!\!\!\! & \inpPrefix{j}{d}{\mathtt{wit}}{} \cont \insess{\channelsingle{s}{j}}{\outPrefix{j}{w}{\mathtt{rqs}}{}}\cont \\
			& \hspace{4em} \insess{\channelsingle{s}{j}}{\inpPrefix{j}{w}{\mathtt{st}}{}} \cont \insess{\channelsingle{s}{j}}{\outPrefix{j}{p}{\mathtt{glt}}{\false}}
		\end{cases}
	\end{gather*}
	for the \participant{d}efendant and \participant{j}udge, where $P_{\mathtt{rls}}$ is used as a helper process for legibility.
    \exampleDone
\end{example}\medbreak

With the process syntax defined, we will now move on to specifying how processes behave and interact.

\section{Operational Semantics}
\label{sec:operationalSemantics}
In this section, we introduce the operational semantics of the \emph{\calculusname{} $\pi$-calculus}.
First, we define the structural congruence, which we need for the reduction semantics to specify all possible reductions of processes.

\begin{definition}[Structural Congruence $ \equiv $]
	\label{def:structuralCongruence}
	Structural congruence $ \equiv $ is the smallest congruence on processes that includes alpha conversion $ \equiv_{\alpha} $ and:
	\begin{gather*}
		P \inparallel \0 \equiv P \qquad
		P \inparallel Q \equiv Q \inparallel P \qquad
		\left( P \inparallel Q \right) \inparallel R \equiv P \inparallel \left( Q \inparallel R \right) \qquad
		\res{s}\0 \equiv \0
		\vspace*{0.25em}\\
		\res{s}\res{s'}P \equiv \res{s'}\res{s}P \qquad
		P \inparallel \res{s}Q \equiv \res{s}\left( P \inparallel Q \right) \quad \text{if } s \notin \fs{P}
		\vspace*{0.25em}\\
		\procdef{D}{\res{s}P} \equiv \res{s}\procdef{D}{P} \quad \text{if } s \notin \fs{D}
		\vspace*{0.25em}\\
		\left( \procdef{D}{P} \right) \inparallel Q \equiv \procdef{D}{\left( P \inparallel Q \right)} \quad \text{if } \dpv{D} \cap \fpv{Q} = \emptyset
		\vspace*{0.25em}\\
		\begin{array}{c}
			\procdef{D}{\left( \procdef{D'}{P} \right)}
			\equiv \procdef{D \cup D'}{P}
		\end{array}
		\quad \text{if }
		\begin{array}{c}
			\dpv{D} \cap \dpv{D'} = \emptyset \text{ and}\\
			\fpv{D} \cap \dpv{D'} = \emptyset
		\end{array}
    \end{gather*}
\end{definition}

\begin{definition}[Probabilistic Reduction Semantics]
\label{def:reductionSemantics}
The reduction semantics is given as the relation $ \longrightarrow_{\prob} $ inductively defined as follows.
	\begin{gather*}
		\cond{\true}{P}{Q} \longrightarrow_{1} P \; \sidenote{[R-Cond-$ \true $]} \qquad
		\cond{\false}{P}{Q} \longrightarrow_{1} Q \; \sidenote{[R-Cond-$ \false $]}
		\vspace{0.25em}\\
		\procdef{D}{\0} \longrightarrow_{1} \0 \; \sidenote{[R-Def-$ \0 $]} \qquad
		\insess{c}\left( \probOut{\prob}{\nonsend{.}P} \oplus \choiceProb \right) + \choice \longrightarrow_{\prob} P \; \sidenote{[R-$ \nonsend $]}
		\vspace*{0.25em}\\
		\begin{aligned}
			\insess{\channelk{s}{r}{1}}\left( \probOut{\prob}{\out{q}{p}{l}{v}{Q}} \oplus \choiceProb \right) + \choice_Q \inparallel \insess{\channelk{s}{r}{2}} \inp{p}{q}{l}{x}{P} + \choice_P\\
			  \longrightarrow_{\prob} Q \inparallel P\subst{x}{v}
		\end{aligned}
		\; \sidenote{[R-Com]} \vspace{0.25em}\\
		\infer[\sidenote{[R-Def]}]{\procdef{D}{\left( \processcall{X}{\vecnew{v}}{\vecnew{c}} \inparallel Q \right)} \longrightarrow_{1} \procdef{D}{\left( P\subst{\vecnew{x}}{\vecnew{v}} \inparallel Q \right)}}{\procdecl{X}{\vecnew{x}}{\vecnew{c}} \defeq P \in D}
		\vspace{0.25em}\\
		\infer[\sidenote{[R-Par]}]{P \inparallel Q \longrightarrow_{\prob} P' \inparallel Q}{P \longrightarrow_{\prob} P'} \qquad
		\infer[\sidenote{[R-Struct]}]{P \longrightarrow_{\prob} Q}{P \equiv P' \quad P' \longrightarrow_{\prob} Q' \quad Q' \equiv Q} \vspace{0.25em}\\
		\infer[\sidenote{[R-Res]}]{\res{s}P \longrightarrow_{\prob} \res{s}P'}{P \longrightarrow_{\prob} P'}
		\qquad
		\infer[\sidenote{[R-Def-In]}]{\procdef{D}{P} \longrightarrow_{\prob} \procdef{D}{P'}}{P \longrightarrow_{\prob} P'}
	\end{gather*}
\end{definition}

The statement $ P\longrightarrow_{\prob} P' $ is meant to be read as ``process $P$ reduces to the continuation $P'$ with probability $ \prob $''.
Let us now explain each reduction rule.
[R-Cond-$\true$] and [R-Cond-$\false$] allow conditional processes to take a reduction step to their intended continuation depending on the boolean value in the clause.
Rule [R-Def-$ \0 $] allows to garbage collect disused declarations.
With [R-$\nonsend$], an internal action is performed with the probability $ \prob $.
Rule [R-Com] allows communication between two parallel mixed choices, one containing an input and the other an output, with matching roles $ \participant{p}, \participant{q} $ and matching label $ l $, where the probability $ \prob $ of this step is determined by the sender.
By [R-Def], a process call may be executed.
Given the declaration of $X$ as $\procdecl{X}{\vecnew{x}}{\vecnew{c}} \defeq P$ being contained in the declarations $D$, the process call is replaced by the substitution $P\subst{\vecnew{x}}{\vecnew{v}}$.
Often, the reduction semantics is defined ``up-to'' structural congruence, however, we instead choose to include an explicit rule, [R-Struct].
From the remaining rules [R-Par], [R-Res], and [R-Def-$\inpT$], we get that processes can still reduce in different contexts, namely in parallel composition, under session restriction, and within a process definition.

We write $ P \longrightarrow_{\prob} $ if $ P \longrightarrow_{\prob} P' $ for some $ P' $, and $ P \nrightarrow $ if there is no $ \prob $ such that $ P \longrightarrow_{\prob} $.
Let $ \Longrightarrow_{\prob} $ be inductively defined as (a)~$ P \Longrightarrow_{1} P $ and (b)~if $ P \longrightarrow_{\prob_1} P' $ and $ P' \Longrightarrow_{\prob_2} P'' $ then $ P \Longrightarrow_{\prob_1 \prob_2} P'' $.\medbreak

To aid understanding, let us revisit the running example.
\begin{example}[Running Example---Semantics]
\addxcontentsline{loe}{example}[\theexample]{Running Example---Semantics}
	\label{ex:motivatingExampleSemantics}
    Let us examine possible reductions of Example~\ref{ex:motivatingExampleSyntax}.
    Recall that the interface process was $P_{\interface} = \res{s}{\left( P_{\participant{p}} \inparallel P_{\participant{c}} \inparallel P_{\participant{w}} \right)}$ with $\vecparticipant{c} = \{\participant{j},\participant{d}\}$.
    For convenience, we reordered the parallel processes, \ie $P_{\interface} = \res{s}{\left( P_{\participant{p}} \inparallel P_{\participant{c}} \inparallel P_{\participant{w}} \right)} \equiv P_{\interface} = \res{s}{\left( P_{\participant{p}} \inparallel P_{\participant{w}} \inparallel P_{\participant{c}} \right)}$.
    To highlight which components interact in each reduction step, we will alternately highlight them and the corresponding arrow $\longrightarrow_{\prob}$ in \highlight{\highlightname{}} and \highlightdark{\highlightdarkname{}}.\par
    The sequence of reductions of the interface process $P_{\interface}$ we chose begins with the \participant{p}laintiff sending a \emph{lawsuit} to the \participant{c}ourt, specifically the \participant{j}udge.
    The \participant{c}ourt then sends the verdict \emph{guilty} to the \participant{p}laintiff with a $35\%$ probability and afterwards releases the \participant{w}itness.\vspace{-1em}
    \begin{gather*}
        P_{\interface} = \res{s}{\left( P_{\participant{p}} \inparallel P_{\participant{w}} \inparallel P_{\participant{c}} \right)} =\\
		\res{s}{} \Biggl( \insess{\channelsingle{s}{p}}{\highlight{\outPrefix{p}{j}{\mathtt{lws}}{}}} \cont \insess{\channelsingle{s}{p}}{} \inpPrefix{p}{j}{\mathtt{glt}}{x} \;\inparallel\; \insess{\channelsingle{s}{w}}{\begin{cases}
			& \inp{w}{j}{\mathtt{mtg}}{}{} \insess{\channelsingle{s}{w}}{\outPrefix{w}{j}{\mathtt{st}}{}} \\
			+\!\!\!\! & \inpPrefix{w}{j}{\mathtt{rls}}{}
		\end{cases}}\\
		\inparallel\; \insess{\channel{s}{c}}{\highlight{\inpPrefix{j}{p}{\mathtt{lws}}{}}} \cont \insess{\channel{s}{c}}{}  \begin{cases}
			&\probOut{0.35}{\outPrefix{j}{p}{\mathtt{glt}}{\true}} \cont \insess{\channelsingle{s}{j}}{\outPrefix{j}{w}{\mathtt{rls}}{}}\\
			\oplus\!\!\!\! &\probOut{0.35}{\outPrefix{j}{p}{\mathtt{glt}}{\false}} \cont \insess{\channelsingle{s}{j}}{\outPrefix{j}{w}{\mathtt{rls}}{}}\\
			\oplus\!\!\!\! &\probOut{0.3}{\outPrefix{j}{w}{\mathtt{rqs}}{}}\cont \insess{\channel{s}{c}}{\inpPrefix{j}{w}{\mathtt{st}}{}} \cont \insess{\channel{s}{c}}{\outPrefix{j}{p}{\mathtt{glt}}{\false}}
		\end{cases}\; \Biggr)\\[+1em]
        \highlight{\longrightarrow_1} \res{s}{} \Biggl( \insess{\channelsingle{s}{p}}{} \highlightdark{\inpPrefix{p}{j}{\mathtt{glt}}{x}} \;\inparallel\; \insess{\channelsingle{s}{w}}{\begin{cases}
			& \inp{w}{j}{\mathtt{mtg}}{}{} \insess{\channelsingle{s}{w}}{\outPrefix{w}{j}{\mathtt{st}}{}} \\
			+\!\!\!\! & \inpPrefix{w}{j}{\mathtt{rls}}{}
		\end{cases}}\\
		\inparallel\; \insess{\channel{s}{c}}{}  \begin{cases}
			&\highlightdark{\probOut{0.35}{\outPrefix{j}{p}{\mathtt{glt}}{\true}} }\cont \insess{\channel{s}{c}}{\outPrefix{j}{w}{\mathtt{rls}}{}}\\
			\oplus &\probOut{0.35}{\outPrefix{j}{p}{\mathtt{glt}}{\false}} \cont \insess{\channel{s}{c}}{\outPrefix{j}{w}{\mathtt{rls}}{}}\\
			\oplus &\probOut{0.3}{\outPrefix{j}{w}{\mathtt{rqs}}{}}\cont \insess{\channel{s}{c}}{\inpPrefix{j}{w}{\mathtt{st}}{}} \cont \insess{\channel{s}{c}}{\outPrefix{j}{p}{\mathtt{glt}}{\false}}
		\end{cases} \Biggr) \\[+1em]
		\highlightdark{\longrightarrow_{0.35}} \res{s}{} \Biggl( \0 \;\inparallel\; \insess{\channelsingle{s}{w}}{\begin{cases}
			& \inp{w}{j}{\mathtt{mtg}}{}{} \insess{\channelsingle{s}{w}}{\outPrefix{w}{j}{\mathtt{st}}{}} \\
			+\!\!\!\! & \highlight{\inpPrefix{w}{j}{\mathtt{rls}}{}}
		\end{cases}} \; \inparallel \;
		\insess{\channel{s}{c}}{\highlight{\outPrefix{j}{w}{\mathtt{rls}}{}}} \Biggr) \\[+1em]
        \highlight{\longrightarrow_1} \res{s}{} \biggl( \0 \;\inparallel\; \0 \;\inparallel\; \0 \biggr)
	\end{gather*}
	Hence, $ P_{\interface} $ has a sequence $ P_{\interface}\Longrightarrow_{0.35} \0 $, where the \participant{c}ourt finds the \participant{d}efendant~\emph{guilty}.
    
    Let us find a comparable sequence of reductions in the refined system $ P_{\refinement} $.
	The initial \emph{lawsuit} message is sent from \participant{p}laintiff to \participant{j}udge.
	Afterwards, the \participant{d}efendant delivers a \emph{weak defense} to the \participant{j}udge with $50\%$ probability.
    The \participant{j}udge then sends \emph{guilty} to the \participant{p}laintiff with $70\%$ probability and releases the \participant{w}itness.\vspace{-1em}
    \begin{gather*}
        P_{\refinement} = \res{s}{\left( P_{\participant{p}} \inparallel P_{\participant{j}} \inparallel P_{\participant{d}} \inparallel P_{\participant{w}} \right)} = \res{s}{} \Biggl( \insess{\channelsingle{s}{p}}{\highlight{\outPrefix{p}{j}{\mathtt{lws}}{}}} \cont \insess{\channelsingle{s}{p}}{} \inpPrefix{p}{j}{\mathtt{glt}}{x}\\
		\inparallel\; \insess{\channelsingle{s}{j}}{\highlight{\inpPrefix{j}{p}{\mathtt{lws}}{}}} \cont \insess{\channelsingle{s}{j}}
		\begin{cases}
			& \inpPrefix{j}{d}{\mathtt{wk}}{} \cont \insess{\channelsingle{s}{j}}
			\begin{cases}
				&\probOut{0.7}{\outPrefix{j}{p}{\mathtt{glt}}{\true}} \cont P_{\mathtt{rls}} \\
				\oplus &\probOut{0.3}{\outPrefix{j}{p}{\mathtt{glt}}{\false}} \cont P_{\mathtt{rls}}
			\end{cases}\\
			\!+\!\!\!\! & \inpPrefix{j}{d}{\mathtt{str}}{} \cont \insess{\channelsingle{s}{j}}{\outPrefix{j}{p}{\mathtt{glt}}{\false}} \cont P_{\mathtt{rls}} \\
			\!+\!\!\!\! & \inpPrefix{j}{d}{\mathtt{wit}}{} \cont \insess{\channelsingle{s}{j}}{\outPrefix{j}{w}{\mathtt{rqs}}{}}\cont \\
			& \hspace{4em} \insess{\channelsingle{s}{j}}{\inpPrefix{j}{w}{\mathtt{st}}{}} \cont \insess{\channelsingle{s}{j}}{\outPrefix{j}{p}{\mathtt{glt}}{\false}}
		\end{cases}\\
        \inparallel\; \insess{\channelsingle{s}{d}}
        \begin{cases}
            & \probOut{0.5}{\outPrefix{d}{j}{\mathtt{wk}}{}} \\
            \oplus & \probOut{0.2}{\outPrefix{d}{j}{\mathtt{str}}{}} \\
            \oplus & \probOut{0.3}{\outPrefix{d}{j}{\mathtt{wit}}{}}
        \end{cases}%
        \;\inparallel\; \insess{\channelsingle{s}{w}}{
        \begin{cases}
            & \inp{w}{j}{\mathtt{mtg}}{}{} \insess{\channelsingle{s}{w}}{\outPrefix{w}{j}{\mathtt{st}}{}} \\
            +\!\!\!\! & \inpPrefix{w}{j}{\mathtt{rls}}{}
        \end{cases}}\Biggr)\\[+1em]
        \highlight{\longrightarrow_1} \res{s}{} \Biggl( \insess{\channelsingle{s}{p}}{} \inpPrefix{p}{j}{\mathtt{glt}}{x}
		\;\inparallel\; \insess{\channelsingle{s}{j}}
		\begin{cases}
			& \highlightdark{\inpPrefix{j}{d}{\mathtt{wk}}{}} \cont \insess{\channelsingle{s}{j}}
			\begin{cases}
				&\probOut{0.7}{\outPrefix{j}{p}{\mathtt{glt}}{\true}} \cont P_{\mathtt{rls}} \\
				\oplus\!\!\!\! &\probOut{0.3}{\outPrefix{j}{p}{\mathtt{glt}}{\false}} \cont P_{\mathtt{rls}}
			\end{cases}\\
			\!+\!\!\!\! & \inpPrefix{j}{d}{\mathtt{str}}{} \cont \insess{\channelsingle{s}{j}}{\outPrefix{j}{p}{\mathtt{glt}}{\false}} \cont P_{\mathtt{rls}} \\
			\!+\!\!\!\! & \inpPrefix{j}{d}{\mathtt{wit}}{} \cont \insess{\channelsingle{s}{j}}{\outPrefix{j}{w}{\mathtt{rqs}}{}}\cont \\
			& \hspace{4em} \insess{\channelsingle{s}{j}}{\inpPrefix{j}{w}{\mathtt{st}}{}} \cont \insess{\channelsingle{s}{j}}{\outPrefix{j}{p}{\mathtt{glt}}{\false}}
		\end{cases}\\
        \inparallel\; \insess{\channelsingle{s}{d}}
        \begin{cases}
            & \highlightdark{\probOut{0.5}{\outPrefix{d}{j}{\mathtt{wk}}{}}} \\
            \oplus & \probOut{0.2}{\outPrefix{d}{j}{\mathtt{str}}{}} \\
            \oplus & \probOut{0.3}{\outPrefix{d}{j}{\mathtt{wit}}{}}
        \end{cases}%
        \;\inparallel\; \insess{\channelsingle{s}{w}}{
        \begin{cases}
            & \inp{w}{j}{\mathtt{mtg}}{}{} \insess{\channelsingle{s}{w}}{\outPrefix{w}{j}{\mathtt{st}}{}} \\
            +\!\!\!\! & \inpPrefix{w}{j}{\mathtt{rls}}{}
        \end{cases}}\Biggr)\\[+1em]
        \highlightdark{\longrightarrow_{0.5}} \res{s}{} \Biggl( \insess{\channelsingle{s}{p}}{} \highlight{\inpPrefix{p}{j}{\mathtt{glt}}{x}}
		\;\inparallel\; \insess{\channelsingle{s}{j}}
			\begin{cases}
				&\highlight{\probOut{0.7}{\outPrefix{j}{p}{\mathtt{glt}}{\true}}} \cont P_{\mathtt{rls}} \\
				\oplus &\probOut{0.3}{\outPrefix{j}{p}{\mathtt{glt}}{\false}} \cont P_{\mathtt{rls}}
			\end{cases}\\
        \inparallel\; \0%
        \;\inparallel\; \insess{\channelsingle{s}{w}}{}
        \begin{cases}
            & \inp{w}{j}{\mathtt{mtg}}{}{} \insess{\channelsingle{s}{w}}{\outPrefix{w}{j}{\mathtt{st}}{}} \\
            +\!\!\!\! & \inpPrefix{w}{j}{\mathtt{rls}}{}
        \end{cases}\Biggr)\\[+1em]
        \highlight{\longrightarrow_{0.7}} \res{s}{} \Biggl( \0 \;\inparallel\;  P_{\mathtt{rls}} \;\inparallel\; \0%
        \;\inparallel\; \insess{\channelsingle{s}{w}}{}
        \begin{cases}
            & \inp{w}{j}{\mathtt{mtg}}{}{} \insess{\channelsingle{s}{w}}{\outPrefix{w}{j}{\mathtt{st}}{}} \\
            +\!\!\!\! & \inpPrefix{w}{j}{\mathtt{rls}}{}
        \end{cases}\Biggr)\\
        \stackrel{P_{\mathtt{rls}} = \insess{\channelsingle{s}{j}}{\outPrefix{j}{w}{\mathtt{rls}}{}}}{\equiv} \res{s}{} \Biggl( \insess{\channelsingle{s}{j}}{\highlightdark{\outPrefix{j}{w}{\mathtt{rls}}{}}} \;\inparallel\; \insess{\channelsingle{s}{w}}{}
        \begin{cases}
            & \inp{w}{j}{\mathtt{mtg}}{}{} \insess{\channelsingle{s}{w}}{\outPrefix{w}{j}{\mathtt{st}}{}} \\
            +\!\!\!\! & \highlightdark{\inpPrefix{w}{j}{\mathtt{rls}}{}}
        \end{cases}\Biggr)\\[+1em]
        \highlightdark{\longrightarrow_{1}} \res{s}{} \biggl( \0 \;\inparallel\; \0 \biggr)
	\end{gather*}
    Thus, $P_{\refinement}$, too, has a sequence $ P_{\refinement} \Longrightarrow_{0.35} \0 $, in which the \participant{j}udge finds the \participant{d}efendant \emph{guilty}.
    \hfill\exampleDone
\end{example}

Sometimes, \emph{one} sequence of the interface system will correspond to \emph{several} transition sequences of the refined system. 
We observe this in both processes and types, whenever probabilistic branches with the same continuations are summed up in the interface.

\begin{example}[Running Example---Refinement into Separate Reduction Sequences]
\addxcontentsline{loe}{example}[\theexample]{Refinement into Multiple Reduction Sequences}
    For instance, the sequence of $ P_{\interface} $ that leads to \emph{not guilty} without calling the \participant{w}itness with probability $ 0.35 $ is refined into two sequences with probabilities $ 0.15 $ and $ 0.2 $, respectively.
    Let $ P_{\interface}' $ denote the system obtained after one reduction step of $ P_{\interface} $ in the previous Example~\ref{ex:motivatingExampleSemantics}.
    Once again, to highlight which components interact in each reduction step, we will alternately highlight them and the corresponding arrow $\longrightarrow_{\prob}$ in \highlight{\highlightname{}} and \highlightdark{\highlightdarkname{}}.
    Observe the following reduction step of $ P_{\interface}' $.
    \begin{gather*}
        P_{\interface}' = \res{s}{} \Biggl( \insess{\channelsingle{s}{p}}{} \highlightdark{\inpPrefix{p}{j}{\mathtt{glt}}{x}} \;\inparallel\; P_{\participant{w}}\\
		\inparallel\; \insess{\channel{s}{c}}{}  \begin{cases}
			&\probOut{0.35}{\outPrefix{j}{p}{\mathtt{glt}}{\true}} \cont \insess{\channel{s}{c}}{\outPrefix{j}{w}{\mathtt{rls}}{}}\\
			\oplus &\highlightdark{\probOut{0.35}{\outPrefix{j}{p}{\mathtt{glt}}{\false}}} \cont \insess{\channel{s}{c}}{\outPrefix{j}{w}{\mathtt{rls}}{}}\\
			\oplus &\probOut{0.3}{\outPrefix{j}{w}{\mathtt{rqs}}{}}\cont \insess{\channel{s}{c}}{\inpPrefix{j}{w}{\mathtt{st}}{}} \cont \insess{\channel{s}{c}}{\outPrefix{j}{p}{\mathtt{glt}}{\false}}
		\end{cases} \Biggr) \\[+1em]
		\highlightdark{\longrightarrow_{0.35}} \res{s}{} \biggl( \0 \;\inparallel\; P_{\participant{w}} \; \inparallel \;
		\insess{\channel{s}{c}}{{\outPrefix{j}{w}{\mathtt{rls}}{}}} \biggr)
	\end{gather*}
    Compare this now to the following two reduction sequences obtainable from the system obtained after one reduction step of $ P_{\refinement} $:
    \begin{gather*}
        P_{\refinement}' \!= \res{s}{} \Biggl( \insess{\channelsingle{s}{p}}{} \inpPrefix{p}{j}{\mathtt{glt}}{x}
		\inparallel \insess{\channelsingle{s}{j}}
		\begin{cases}
			& \highlightdark{\inpPrefix{j}{d}{\mathtt{wk}}{}} \cont \insess{\channelsingle{s}{j}}
			\begin{cases}
				&\probOut{0.7}{\outPrefix{j}{p}{\mathtt{glt}}{\true}} \cont P_{\mathtt{rls}} \\
				\oplus\!\!\!\! &\probOut{0.3}{\outPrefix{j}{p}{\mathtt{glt}}{\false}} \cont P_{\mathtt{rls}}
			\end{cases}\\
			\!+\!\!\!\! & \inpPrefix{j}{d}{\mathtt{str}}{} \cont \insess{\channelsingle{s}{j}}{\outPrefix{j}{p}{\mathtt{glt}}{\false}} \cont P_{\mathtt{rls}} \\
			\!+\!\!\!\! & \inpPrefix{j}{d}{\mathtt{wit}}{} \cont \insess{\channelsingle{s}{j}}{\outPrefix{j}{w}{\mathtt{rqs}}{}}\cont \\
			& \hspace{4em} \insess{\channelsingle{s}{j}}{\inpPrefix{j}{w}{\mathtt{st}}{}} \cont \insess{\channelsingle{s}{j}}{\outPrefix{j}{p}{\mathtt{glt}}{\false}}
		\end{cases}\\
        \inparallel\; \insess{\channelsingle{s}{d}}
        \begin{cases}
            & \highlightdark{\probOut{0.5}{\outPrefix{d}{j}{\mathtt{wk}}{}}} \\
            \oplus & \probOut{0.2}{\outPrefix{d}{j}{\mathtt{str}}{}} \\
            \oplus & \probOut{0.3}{\outPrefix{d}{j}{\mathtt{wit}}{}}
        \end{cases}%
        \;\inparallel\; P_{\participant{w}}\Biggr)\\[+1em]
        \highlightdark{\longrightarrow_{0.5}} \res{s}{} \Biggl( \insess{\channelsingle{s}{p}}{} \highlight{\inpPrefix{p}{j}{\mathtt{glt}}{x}}
		\;\inparallel\; \insess{\channelsingle{s}{j}}
			\begin{cases}
				&\probOut{0.7}{\outPrefix{j}{p}{\mathtt{glt}}{\true}} \cont P_{\mathtt{rls}} \\
				\oplus &\highlight{\probOut{0.3}{\outPrefix{j}{p}{\mathtt{glt}}{\false}}} \cont P_{\mathtt{rls}}
			\end{cases} \;\inparallel\; \0 \;\inparallel\; P_{\participant{w}}\Biggr)\\[+1em]
        \highlight{\longrightarrow_{0.3}} \res{s}{} \biggl( \0 \;\inparallel\;  P_{\mathtt{rls}} \;\inparallel\; \0 \;\inparallel\; P_{\participant{w}} \biggr)
	\end{gather*}
    For the reduction sequence $P_{\refinement}' \Rightarrow_{0.15} \res{s}{} \bigl( \0 \;\inparallel\;  P_{\mathtt{rls}} \;\inparallel\; \0 \;\inparallel\; P_{\participant{w}} \bigr)$, and
    \begin{gather*}
        P_{\refinement}' \!= \res{s}{} \Biggl( \insess{\channelsingle{s}{p}}{} \inpPrefix{p}{j}{\mathtt{glt}}{x}
		\inparallel \insess{\channelsingle{s}{j}}
		\begin{cases}
			& \inpPrefix{j}{d}{\mathtt{wk}}{} \cont \insess{\channelsingle{s}{j}}
			\begin{cases}
				&\probOut{0.7}{\outPrefix{j}{p}{\mathtt{glt}}{\true}} \cont P_{\mathtt{rls}} \\
				\oplus\!\!\!\! &\probOut{0.3}{\outPrefix{j}{p}{\mathtt{glt}}{\false}} \cont P_{\mathtt{rls}}
			\end{cases}\\
			\!+\!\!\!\! & \highlightdark{\inpPrefix{j}{d}{\mathtt{str}}{}} \cont \insess{\channelsingle{s}{j}}{\outPrefix{j}{p}{\mathtt{glt}}{\false}} \cont P_{\mathtt{rls}} \\
			\!+\!\!\!\! & \inpPrefix{j}{d}{\mathtt{wit}}{} \cont \insess{\channelsingle{s}{j}}{\outPrefix{j}{w}{\mathtt{rqs}}{}}\cont \\
			& \hspace{4em} \insess{\channelsingle{s}{j}}{\inpPrefix{j}{w}{\mathtt{st}}{}} \cont \insess{\channelsingle{s}{j}}{\outPrefix{j}{p}{\mathtt{glt}}{\false}}
		\end{cases}\\
        \inparallel\; \insess{\channelsingle{s}{d}}
        \begin{cases}
            & \probOut{0.5}{\outPrefix{d}{j}{\mathtt{wk}}{}} \\
            \oplus & \highlightdark{\probOut{0.2}{\outPrefix{d}{j}{\mathtt{str}}{}}} \\
            \oplus & \probOut{0.3}{\outPrefix{d}{j}{\mathtt{wit}}{}}
        \end{cases}%
        \;\inparallel\; P_{\participant{w}} \Biggr)\\[+1em]
        \highlightdark{\longrightarrow_{0.2}} \res{s}{} \Biggl( \insess{\channelsingle{s}{p}}{} \highlight{\inpPrefix{p}{j}{\mathtt{glt}}{x}}
		\;\inparallel\; \insess{\channelsingle{s}{j}}{\highlight{\outPrefix{j}{p}{\mathtt{glt}}{\false}}} \cont P_{\mathtt{rls}} \;\inparallel\; \0 \;\inparallel\; P_{\participant{w}} \Biggr)\\[+1em]
        \highlight{\longrightarrow_{1}} \res{s}{} \biggl( \0 \;\inparallel\;  P_{\mathtt{rls}} \;\inparallel\; \0 \;\inparallel\; P_{\participant{w}} \biggr) \text{ ,}
	\end{gather*}
    a different reduction sequence to the same system, $P_{\refinement}' \Rightarrow_{0.2} \res{s}{} \bigl( \0 \;\inparallel\;  P_{\mathtt{rls}} \;\inparallel\; \0 \;\inparallel\; P_{\participant{w}} \bigr)$.
    These two steps together correspond to the reduction step of $ P_{\interface}' $ with $35\%$ probability we have seen above.
    \exampleDone
\end{example}

\medbreak
Having introduced process syntax and operational semantics, we are finished with introducing the \emph{\calculusname{} $\pi$-calculus} itself.
\section{Typing System}
\label{sec:typeSystem}
This section introduces the typing system of the \emph{\calculusname{} $\pi$-calculus}.
Within our system, akin to other works in MPST (see \citep{scalas2019less}), to type processes, \emph{session types} are assigned to the \emph{channels} through which the processes communicate.
There are three components to our typing system.
First is the \emph{type syntax}, a grammar defining the shape of types and typing contexts.
Then comes \emph{subtyping}, a preorder relation on types which enhances the flexibility of the system.
Finally, \emph{typing rules} will specify which processes are justifiable by which collection of types.

\subsection{Type Syntax}
Not unusually in MPST systems, our type syntax looks quite similar to the process syntax.
At first glance there are nonetheless several differences.
Instead of variables and values, we find base types.
Channels are omitted; each channel can have only exactly one type and need therefore not be explicitly stated within the type.
Conditionals and process definitions/-calls do not have explicit types, as they are not needed.
Instead we find classic $\mu$-recursion, as is typically the case for MPST types.
For mixed choice actions, however, any previously acquired intuition will carry over to the corresponding types.

\begin{definition}[Type Syntax]
	\label{def:typeSyntax}
    The syntax of the \emph{\calculusname\ types} are inductively given by:
	%\vspace*{-0.75em}
	\begin{alignat*}{3}
		U & \defT && \natT \gbar \boolT && \text{(base types for numbers and booleans)}\\
		T, T_i & \defT && \tend \gbar t \gbar {\left( \mu t \right)} T \quad && \text{(inaction type, recursion variable, recursion)}\\
		&& \gbar & \mc{i}{I} L_i \gbar L \gbar H \cont T \quad && \text{(mixed choice type with a finite $ I \neq \emptyset $, output)}\\%\gbar \probOut{\prob}{H \cont T}
		L, L_i & \defT && \inpT \gbar \outT \quad && \text{(mixed choice modes)}\\
		\inpT & \defT && \inp{p}{q}{l}{U}{T} \quad && \text{($ \participant{p} $ receives message $ l $ with type $ U $ from $\participant{q} $)}\\
		\outT & \defT && \send{i}{I} \probOut{\prob_i}{H_i \cont T_i} \quad && \text{(probabilistic choice with finite $ I \kern-1pt \neq \emptyset $ and $ \sum_{i \in I} \prob_i \leq 1 $)}\\
		H, H_i & \defT && \outPrefix{p}{q}{l}{U} \gbar \nonsend \quad && \text{($ \participant{p} $ sends message $ l $ with type $ U $ to $ \participant{q} $, internal action)}\\
		\Delta & \defT && \emptyset \gbar \Delta, \type{c}{T} \quad && \text{(local context)}
		%\Lambda & \defT && \Delta \gbar c \cdot \Delta \quad&& \text{(local context with active channel, $c\in \Delta$)}
	\end{alignat*}
\end{definition}
As usual, we begin by explaining the syntax in chronological order.
The \emph{base types} for numbers and booleans are given as $\natT$ and $\boolT$.
The addition of more base types would be a straightforward extension of the system, similar to values in processes.
\emph{Inaction} holds the same meaning as in the process syntax.
The \emph{recursion variable} $t$ and \emph{recursion} ${{\left( \mu t \right)} T}$ are the standard way to type recursive process calls (see \citep{scalas2019less}).
We require recursion to be guarded: For ${{\left( \mu t \right)} T}$, the type $T$ has to contain meaningful action, \ie $T \neq t'$ for all recursion variables $t'$.
The \emph{mixed choice} type is analogous to the process syntax.
For convenience we use the \emph{mixed choice modes} $\inpT$ and $\outT$ in dense syntax blocks.
\emph{Input} type and \emph{probabilistic choice} type, too, follow the process syntax.
$\prob$ are probabilities, $\prob \in \mathbb{R}$ with $0 < \prob \leq 1$.
The sum of probabilities in a probabilistic choice type is at most one, $ \sum_{i \in I} \prob_i \leq 1 $.
The \emph{output} type is analogous to the process syntax.
Finally, $\Delta$ is a \emph{local context}, \ie the aforementioned ``collection of types'' using which processes are justified.
It may either be empty or contain assignments $\type{c}{T}$ of a session type $T$ to a channel $c$.\medbreak

Let $ c \in \Delta $ iff $ \type{c}{T} \in \Delta $ for some $ T $.
If $\channel{s}{r} \in \Delta$ and $\participant{p} \in \vecparticipant{r}$, we write $\participant{p} \in \Delta$.
Types in local contexts are treated equi-recursively, \ie $\Delta, \type{c}{{{\left(\mu t \right)} T}} = \Delta, \type{c}{T\subst{t}{{\left( \mu t \right)} T}}$.
Additionally, we consider a type assignment of a sum with empty index set to be inaction and an inaction type assignment to be the empty set, \ie $\Delta, \type{\channel{s}{r}}{\mc{i}{\emptyset}L_i} = \Delta, \type{\channel{s}{r}}{\tend} = \Delta$.
Composition of local contexts is required to be linear, \ie $\Delta_1, \Delta_2 $ is defined iff $ \vecparticipantk{r}{1} \cap \vecparticipantk{r}{2} = \emptyset $ for all $ \vecparticipantk{r}{1} \in \Delta_1 $ and $ \vecparticipantk{r}{2} \in \Delta_2 $.\medbreak

We inherit for choice types the abbreviations defined for choices in processes, except that we \emph{do not} allow to omit the probability $ 1 $, \ie $ T = \probOut{1}{ H \cont T'} $ is \emph{not} the same as $ H \cont T' $.
This distinction is necessary for the subtyping rules.
We overload the notations $ \choice $ and $ \choiceProb $ to also use them on choices in types.
We again allow to unify and split similar summands of probabilistic choices, \ie $ \probOut{\prob_i}{H\cont T} \oplus \probOut{\prob_j}{H\cont T} = \probOut{{\left(\prob_i + \prob_j\right)}}{H\cont T} $.\medbreak

The typing and subtyping rules decompose types in a stepwise manner.
Hence, we also have types such as probabilistic choices with $ \sum_{i \in I} \prob_i \leq 1 $ or the output type $ H \cont T$.

\begin{definition}[Well-Formed]
    A type $ T $ is \emph{well-formed} if
    \begin{compactenum}[(a)]
        \item all outputs $H$ are part of a probabilistic choice $\send{i}{I} \probOut{\prob_i}{H_i \cont T_i} \oplus \probOut{\prob}{H \cont T}$ and
        \item for all probabilistic choices $ \sum_{i \in I} \prob_i = 1 $ and
	    \item for all $ \inp{p}{q}{l}{U_i}{T_i} + \inp{p}{q}{l}{U_j}{T_j} $ and $ \probOut{\prob_i}{\out{p}{q}{l}{U_i}{T_i}} \oplus \probOut{\prob_j}{\out{p}{q}{l}{U_j}{T_j}} $ have $ U_i = U_j $ and $ T_i = T_j $.
    \end{compactenum}
\end{definition}
Hence, a type $ H \cont T $ is not well-formed.\medbreak 

Similar to \citep{peters2024separation}, we allow for mixed and probabilistic choices to combine the same label with different messages and/or continuations.
Accordingly, we do not require the labels of a mixed or probabilistic choice to be pairwise distinct.
Instead we specify that in well-formed types, summands which share sender, receiver, label and input/output mode must have the same payload- and continuation type.\medbreak

\subsection{Standard Subtyping}
\label{section:subtypingStandard}
To enhance the flexibility of the typing system, we use subtyping.
The intuitive idea behind classic subtyping rules is that for $T' \leq T$, the subtype $T'$ is \emph{smaller} than its supertype $T$ if it is less demanding, accepting more external choices (messages to be received) and offering fewer internal choices (messages to be sent).
Using the subtyping rules to enrich the typing rules formalizes a notion of safe substitution \citep{DBLP:journals/corr/Dezani-Ciancaglini16, DBLP:conf/birthday/Gay16}.
When we consider types to be a protocol specification where the processes are the concrete implementation, safe substitution states that if the specification calls for a certain type $T$, the implementation may instead use any subtype $T'$.
A natural intuition for this concept can be drawn from programming.
Imagine a programmer calling an external function which takes an input value and then gives back a return value.
For example, say that both the input value and output value may be either a natural number or boolean.
If the programmer writes code that sometimes calls the function by inputting a natural number and sometimes a boolean, the code would run as expected.
Removing one of these modes (offering fewer internal choices) and only ever inputting natural numbers would clearly not cause errors.
If instead, however, the programmer were to sometimes also input another value type, say a string, then we would expect an exception.
For handling the return value, they would initialize variables to store whichever return value they get.
If they \emph{additionally} prepared to receive a floating point number (accepting more external choices), the code would still run.
Failing to prepare for either boolean or natural number return values would instead cause errors.
This concept also extends to base types.
When receiving an integer, for example, initializing a variable as a float would be fine.
Instead initializing the variable as a natural number will cause errors if the value received is below zero.
This subtyping for base types is not uncommon in MPST systems (see \citep{scalas2019less}, based on \citep{DBLP:journals/corr/Dezani-Ciancaglini16}).
Our system does not offer this, but an extension would be rather straightforward.\medbreak

To introduce the subtyping rules, we first define the set of unguarded prefixes of a type, as also seen in \citep{peters2024separation}. 

\begin{definition}[Prefix]
	\label{def:pre}
	The set of unguarded prefixes of a type $T$, $\pre(T)$, is given as follows:
	\begin{gather*}
		\pre(\tend) = \pre(t) = \emptyset \qquad
		\pre({{\left( \mu t \right)} T}) = \pre(T) \qquad
		\pre(\probOut{\prob}{\nonsend}\cont T) = \{\probOut{\prob}{\nonsend}\cont T\}\\
		\pre(\inp{p}{q}{l}{U}{T}) = \{\participant{p}{\leftarrow}\participant{q}\mathop{?}\} \qquad
		\pre(\probOut{\prob}{\out{p}{q}{l}{U}{T}}) = \{\participant{p}{\rightarrow}\participant{q}\mathop{!}\}\\
		\pre(\mc{i}{I} L_i) = \{\pre(L_i)\}_{i\in I} \qquad
		\pre(\send{i}{I} \probOut{\prob_i}{H_i}\cont T_i) = \{\pre(\probOut{\prob_i}{H_i}\cont T_i)\}_{i\in I}
	\end{gather*}
	Probabilities are summed up such that:
	\begin{alignat*}{3}
		\pre(\choiceProb &\oplus \probOut{\prob_1}{\out{p}{q}{l_1}{U_1}{T_1}}&&\\
        &\oplus \probOut{\prob_2}{\out{p}{q}{l_2}{U_2}{T_2}}) &\;=\;& \pre(\choiceProb) \cup \{\probOut{\prob_1}{\participant{p}{\rightarrow}\participant{q}\mathop{!}}\} \cup \{\probOut{\prob_2}{\participant{p}{\rightarrow}\participant{q}\mathop{!}}\}\\
		&&\;=\;& \pre(\choiceProb) \cup \{\probOut{\prob_1+\prob_2}{\participant{p}{\rightarrow}\participant{q}\mathop{!}}\}\\
		\pre(\choiceProb &\oplus \probOut{\prob_1}{\nonsend\cont T} \oplus \probOut{\prob_2}{\nonsend\cont T}) &\;=\;& \pre(\choiceProb) \cup \{\probOut{\prob_1}{\nonsend\cont T}\} \cup \{\probOut{\prob_2}{\nonsend\cont T}\}\\
		&&\;=\;& \pre(\choiceProb) \cup \{\probOut{\prob_1+\prob_2}{\nonsend\cont T}\}
	\end{alignat*}
\end{definition}\medbreak

In our subtyping, we will compare prefix sets of types.
Let us therefore examine some types which have and do not have the same set of prefixes.
\begin{example}[Prefixes---Introduction]
\addxcontentsline{loe}{example}[\theexample]{Prefixes: Introduction}
    Consider the following type and its set of unguarded prefixes.
    \begin{gather*}
        T = \begin{cases}
            &\probOut{0.5}{\outPrefix{a}{b}{\mathtt{one}}{}} \\
            \!\oplus\!\!\!\! & \probOut{0.2}{\outPrefix{a}{c}{\mathtt{two}}{}}\\
            \!\oplus\!\!\!\! & \probOut{0.3}{\nonsend}\cont T'
        \end{cases}  \qquad
        \pre(T) =  \{ \probOut{0.5}{\participant{a}{\rightarrow}\participant{b}\mathop{!}}, \probOut{0.2}{\participant{a}{\rightarrow}\participant{c}\mathop{!}}, \probOut{0.3}{\nonsend}\cont T'\}
    \end{gather*}
    The following types have the same prefixes as $T$, \ie $\pre(T) = \pre(T_1) = \pre(T_2)$ despite the outputs having new continuations in $T_1$ and different labels in $T_2$.
    \begin{gather*}
        T_1 = \begin{cases}
            &\probOut{0.5}{\outPrefix{a}{b}{\mathtt{one}}{}}\cont T_{\mathtt{one}} \\
            \!\oplus\!\!\!\! & \probOut{0.2}{\outPrefix{a}{c}{\mathtt{two}}{}} \cont T_{\mathtt{two}}\\
            \!\oplus\!\!\!\! & \probOut{0.3}{\nonsend}\cont T'
        \end{cases}\qquad
        T_2 = \begin{cases}
            &\probOut{0.5}{\outPrefix{a}{b}{\mathtt{three}}{}} \\
            \!\oplus\!\!\!\! & \probOut{0.2}{\outPrefix{a}{c}{\mathtt{four}}{}}\\
            \!\oplus\!\!\!\! & \probOut{0.1}{\nonsend}\cont T'\\
            \!\oplus\!\!\!\! & \probOut{0.2}{\nonsend}\cont T'
        \end{cases}\\
    \end{gather*}
    These types, on the other hand, do not have the same prefixes as $T$, \ie $\pre(T) \neq \pre(T_3)$ and $\pre(T) \neq \pre(T_4)$, as the continuation after the internal action is different in $T_3$ and the sending to role $\participant{c}$ is absent in $T_4$.
    \begin{gather*}
        T_3 = \begin{cases}
            &\probOut{0.5}{\outPrefix{a}{b}{\mathtt{one}}{}} \\
            \!\oplus\!\!\!\! & \probOut{0.2}{\outPrefix{a}{c}{\mathtt{two}}{}}\\
            \!\oplus\!\!\!\! & \probOut{0.3}{\nonsend}\cont T''
        \end{cases}, \text{ where } T'' \neq T \qquad
        T_4 = \begin{cases}
            &\probOut{0.7}{\outPrefix{a}{b}{\mathtt{one}}{}} \\
            \!\oplus\!\!\!\! & \probOut{0.3}{\nonsend}\cont T'
        \end{cases}\\
        \pre(T_3) = \{ \probOut{0.5}{\participant{a}{\rightarrow}\participant{b}\mathop{!}}, \probOut{0.2}{\participant{a}{\rightarrow}\participant{c}\mathop{!}}, \probOut{0.3}{\nonsend}\cont T''\} \qquad \pre(T_4) = \{ \probOut{0.7}{\participant{a}{\rightarrow}\participant{b}\mathop{!}}, \probOut{0.3}{\nonsend}\cont T'\}
    \end{gather*}
    \exampleDone
\end{example}\medbreak

Despite being based on \citep{peters2024separation}, the subtyping rules in Definition~\ref{def:subtypingSimple} do not follow their exact structure.
\citep{peters2024separation} have standard-form input- and output rules; the additional logic required for properly subtyping mixed choice types is entirely contained in their summation rule.
In our version, some of that logic is transferred to the input- and output rules.
\begin{definition}[Standard Subtyping]
\label{def:subtypingSimple}
The relation $\leqsimple{}$ is coinductively defined:
    \begin{gather*}
    %
		%
		% [S-emptyset]
		%
		\infer=[\sidenote{\sendeasy{}}]{\tend \leqsimple \tend}%
		{}\\[+0.25em]
		%
		% [S-Sigma2] splitting into in's and out's
		%
		\infer=[\sidenote{\ssigmaeasy{}}]{{\mc{i}{I'}\inpT'_i + \mc{j}{J'}\outT'_j} \leqsimple {\mc{i}{I}\inpT_i + \mc{j}{J}\outT_j}}%
		{I \cup J \neq \emptyset \qquad%
            {\mc{i}{I'}\inpT'_i} \leqsimple {\mc{i}{I}\inpT_i}\qquad%
            {\mc{j}{J'}\outT'_j} \leqsimple {\mc{j}{J}\outT_j}}\\[+0.25em]
		%
		% [S-Sigma-In]
		%
		\infer=[\sidenote{\ssigmaineasy{}}]{{\mc{i}{I' \cup J'}\inpk{q}{p}{l_i}{U_i}{T'_i}{i}} \leqsimple {\mc{i}{I'} \inpk{q}{p}{l_i}{U_i}{T_i}{i}}}%
        {I' \neq \emptyset \qquad%
        \forall j \in J'.\; \exists i\in I'.\; \pre(\inpT'_j)=\pre(\inpT'_i) \qquad%
        \forall i \in I'.\;{T'_i} \leqsimple {T_i}}\\[+0.25em]
		%
		% [S-Sigma-Out]
		%
		\infer=[\sidenote{\ssigmaouteasy{}}]{{\mc{i}{I}{\send{k}{K_i}}\probOut{\prob_k}{H_k \cont T'_k}} \leqsimple {\mc{i}{I}{\send{k}{K_i} \probOut{\prob_k}{H_k \cont T_k}} + \mc{j}{J}\outT_j}}%
		{\begin{array}{c}
				I \neq \emptyset \qquad \forall i \in I.\; \forall k \in K_i .\; {T'_k}
				\leqsimple {T_k}\\
				\forall j \in J.\;  \exists i \in I.\; \pre({\send{k}{K_i}}\probOut{\prob_k}{H_k \cont T'_k}) = \pre(\send{k}{K_i} \probOut{\prob_k}{H_k \cont T_k})
			\end{array}}
	\end{gather*}
    We have $\type{c_1}{T_1'}, \dots, \type{c_n}{T_n'} \leqsimple{} \type{c_1}{T_1}, \dots, \type{c_n}{T_n}$ if for all $i \in [1..n]$ it holds that ${T_i'} \leqsimple {T_i}$.
\end{definition}
Here and throughout the thesis, by $[1..n]$ we denote the set $\{1,\dots,n\}$.
The explanations of each rule are as follows.
By our summation-splitting rule \ssigmaeasy{}, a mixed choice type ${\mc{i}{I'}\inpT'_i + \mc{j}{J'}\outT'_j}$ is smaller than another, ${\mc{i}{I}\inpT_i + \mc{j}{J}\outT_j}$, if the sum of all inputs and the sum of all outputs are smaller than the respective sum of the larger type, \ie $\mc{i}{I'}\inpT'_i \leqsimple {\mc{i}{I}\inpT_i}$ and ${\mc{j}{J'}\outT'_j} \leqsimple {\mc{j}{J}\outT_j}$.
The input rule \ssigmaineasy{} lets a smaller type have more external choices as usual, with the requirement that no prefixes are found in the sum of the smaller type which are not also found in the larger type, $\forall j \in J'.\; \exists i\in I'.\; \pre(\inpT'_j)=\pre(\inpT'_i)$.
The output rule \ssigmaouteasy{}, while looking more complex given the additional $\oplus$-layer, is essentially analogous to the input rule.
As is usual, smaller types have fewer internal choices (see the additional output sum in the larger type $\mc{j}{J}\outT_j$).
Additionally, all prefixes found in the sum of the larger type, \ie in ${\mc{i}{I}{\send{k}{K_i} \probOut{\prob_k}{H_k \cont T_k}} + \mc{j}{J}\outT_j}$, must be present in the smaller type, $\forall j \in J.\;  \exists i \in I.\; \pre({\send{k}{K_i}}\probOut{\prob_k}{H_k \cont T'_k}) = \pre(\send{k}{K_i} \probOut{\prob_k}{H_k \cont T_k})$.\medbreak

A second example for prefixes is presented here, to demonstrate their influence on subtyping. 
\begin{example}[Prefixes---Subtyping]
\addxcontentsline{loe}{example}[\theexample]{Prefixes: Subtyping}
    To show why the subtyping rules include the prefix conditions, let us consider the following example types.
    \begin{gather*}
        T_{a} \kern-1pt\!= \probOut{1}{\outPrefix{a}{b}{\mathtt{nat}}{\natT}} \quad\!
        T_{b} \!= \inp{b}{a}{\mathtt{nat}}{\natT}{} \inpPrefix{b}{c}{\mathtt{bool}}{\boolT} \quad\!
		T_{c} \!= \outPrefix{c}{b}{\mathtt{bool}}{\boolT}
	\end{gather*}
    Additionally, without formal introduction of the typing rules, take the following process which is typed by $T_a$, $T_b$, and $T_c$ above, $P = $
    \begin{gather*}
        \insess{\channelsingle{s}{a}}{} \outPrefix{a}{b}{\mathtt{nat}}{2} \;
        | \; \insess{\channelsingle{s}{b}}{}\inp{b}{a}{\mathtt{nat}}{x}{} \insess{\channelsingle{s}{b}}{}\inpPrefix{b}{c}{\mathtt{bool}}{y} \;
		  | \; \insess{\channelsingle{s}{c}}{}\outPrefix{c}{b}{\mathtt{bool}}{\top}
	\end{gather*}
    $P$ has two simple reduction steps, does not deadlock and all transmitted values will be stored in properly typed variables.
    Now, consider the following subtype of $T_b$, which we could derive if the prefix condition in Rule \ssigmaineasy{} was not present. 
    \begin{gather*}
        T_b' = \begin{cases}
            &\inp{b}{a}{\mathtt{nat}}{\natT}{} \inpPrefix{b}{c}{\mathtt{bool}}{\boolT}\\
            \!+\!\!\!\! & \inpPrefix{b}{c}{\mathtt{new}}{\natT}
        \end{cases}
    \end{gather*}
    Finally, let us examine the process $P'$, in which the component typed by $T_b$ is substituted by a component typed by $T_b'$, $P' =$
    \begin{gather*}
        \insess{\channelsingle{s}{a}}{} \outPrefix{a}{b}{\mathtt{nat}}{1}
        | \; \insess{\channelsingle{s}{b}}{}\!\begin{cases}
            &\inp{b}{a}{\mathtt{nat}}{x}{} \insess{\channelsingle{s}{b}}{}\inpPrefix{b}{c}{\mathtt{bool}}{y}\\
            \!+\!\!\!\! & \inpPrefix{b}{c}{\mathtt{new}}{z}
        \end{cases}
		  | \; \insess{\channelsingle{s}{c}}{}\outPrefix{c}{b}{\mathtt{bool}}{\top}
	\end{gather*}
    While the formal definition of an error (Definition~\ref{def:errorProcess}) is yet to come, the well-versed reader will recognize that the presence of $\inpPrefix{b}{c}{\mathtt{new}}{}$ causes $P'$ to be an error process.
    In the explanation of why that is, let us call the participants in $P'$ by the channels that they use, for simplicity.
    We see that $\channelsingle{s}{b}$ is prepared to receive a message from $\channelsingle{s}{c}$, who is also ready to send one.
    However, $\channelsingle{s}{b}$ is expecting a natural number with label $\mathtt{new}$ while $\channelsingle{s}{c}$ is transmitting a boolean value with label $\mathtt{bool}$.
    From programming experience, we can intuitively understand why this would cause an error.\medbreak
    
    Similarly, consider an example which might occur if the prefix condition were removed from Rule \ssigmaouteasy{}.
    \begin{gather*}
        T_{a} = \begin{cases}
            &\probOut{1}{\outPrefix{a}{b}{\mathtt{hi}}{}}\\
            +&\probOut{1}{\outPrefix{a}{c}{\mathtt{oops}}{}}
        \end{cases}\qquad
        T_{b} = \inpPrefix{b}{a}{\mathtt{hi}}{}\\
        P = \insess{\channelsingle{s}{a}}{}\begin{cases}
            &\probOut{1}{\outPrefix{a}{b}{\mathtt{hi}}{}}\\
            +&\probOut{1}{\outPrefix{a}{c}{\mathtt{oops}}{}}
        \end{cases} \;\inparallel\; 
        \insess{\channelsingle{s}{b}}{}\inpPrefix{b}{a}{\mathtt{hi}}{}
	\end{gather*}
    As before, consider the following subtype of $T_a$ obtainable without the prefix condition and the process $P'$ with its substitution.
    \begin{gather*}
        T_{a}' = \probOut{1}{\outPrefix{a}{c}{\mathtt{oops}}{}}\\
        P' = \insess{\channelsingle{s}{a}}{}\probOut{1}{\outPrefix{a}{c}{\mathtt{oops}}{}} \;\inparallel\; 
        \insess{\channelsingle{s}{b}}{}\inpPrefix{b}{a}{\mathtt{hi}}{}
    \end{gather*}
    After the substitution, the new system $P'$ is deadlocked.
    \exampleDone
\end{example}\medbreak

We present a small example showcasing a subtyping derivation using Definition~\ref{def:subtypingSimple}.
\begin{example}[Standard Subtyping Derivation]
\addxcontentsline{loe}{example}[\theexample]{Standard Subtyping Derivation}
\label{ex:standardsubtypingderivation}
    Consider the following session type $T$ and its subtype $T'$, \ie $T' \leqsimple T$.
    \begin{gather*}
        T' = \begin{cases}
			& \inpPrefix{a}{b}{\mathtt{in}_1}{} \\
            & \inpPrefix{a}{b}{\mathtt{in}_2}{} \cont T'' \\
            \!+ \!\!\!\! &\probOut{1}{\outPrefix{a}{c}{\mathtt{out}_3}{}}
		\end{cases}\qquad
        T = \begin{cases}
			& \inpPrefix{a}{b}{\mathtt{in}_1}{} \\
			\!+ \!\!\!\! & \begin{cases}
                &\probOut{0.7}{\outPrefix{a}{c}{\mathtt{out}_1}{}}\\
                \!\oplus \!\!\!\! &\probOut{0.3}{\outPrefix{a}{c}{\mathtt{out}_2}{}}
			\end{cases}\\
            \!+ \!\!\!\! &\probOut{1}{\outPrefix{a}{c}{\mathtt{out}_3}{}}
		\end{cases}
    \end{gather*}
    The subtyping derivation tree of $T' \leqsimple T$ is as follows.
    We have abbreviated the rule names to their suffixes, \textcolor{easy}{[$\tend$]} refers to \sendeasy{}, \textcolor{easy}{[$\Sigma$]} to \ssigmaeasy{} and so on.
    \small{
    \begin{gather*}
        \infer=[\textcolor{easy}{[\Sigma]}]{\begin{cases}
			& \inpPrefix{a}{b}{\mathtt{in}_1}{} \\
            & \inpPrefix{a}{b}{\mathtt{in}_2}{} \cont T'' \\
            \!+ \!\!\!\! &\probOut{1}{\outPrefix{a}{c}{\mathtt{out}_3}{}}
		\end{cases} \leqsimple \begin{cases}
			& \inpPrefix{a}{b}{\mathtt{in}_1}{} \\
			\!+ \!\!\!\! & \begin{cases}
                &\probOut{0.7}{\outPrefix{a}{c}{\mathtt{out}_1}{}}\\
                \!\oplus \!\!\!\! &\probOut{0.3}{\outPrefix{a}{c}{\mathtt{out}_2}{}}
			\end{cases}\\
            \!+ \!\!\!\! &\probOut{1}{\outPrefix{a}{c}{\mathtt{out}_3}{}}
		\end{cases}}%
		{\infer=[\textcolor{easy}{[\inpT]}]{\begin{cases}
			& \inpPrefix{a}{b}{\mathtt{in}_1}{} \\
            \!+ \!\!\!\!& \inpPrefix{a}{b}{\mathtt{in}_2}{} \cont T''
		\end{cases} \leqsimple \inpPrefix{a}{b}{\mathtt{in}_1}{}}%
        {\infer=[\textcolor{easy}{[\tend]}]{\tend \leqsimple \tend}%
		{}} \quad\hspace{-0.7pt}%
            \infer=[\textcolor{easy}{[\outT]}]{\probOut{1}{\outPrefix{a}{c}{\mathtt{out}_3}{}} \leqsimple \begin{cases}
                & \kern-3pt\begin{cases}
                &\probOut{0.7}{\outPrefix{a}{c}{\mathtt{out}_1}{}}\\
                \!\oplus \!\!\!\! &\probOut{0.3}{\outPrefix{a}{c}{\mathtt{out}_2}{}}
			\end{cases}\\
            \!+ \!\!\!\! &\probOut{1}{\outPrefix{a}{c}{\mathtt{out}_3}{}}
		      \end{cases}}%
		{\infer=[\textcolor{easy}{[\tend]}]{\tend \leqsimple \tend}%
		{}}}
    \end{gather*}}
    \exampleDone
\end{example}

\subsection{Typing Rules}
\label{section:typingRules}
To finish off the typing system, we will now finally present the \emph{\calculusname{} typing rules}, specifying exactly how and which processes are justified by a given local context. We begin by defining a global environment, another context which stores information about variables and process variables as type derivation trees are traversed.
\begin{definition}[Global Environment]
	\label{def:globalEnvironment}
	$ $\\
	$ \Gamma \; \defT \; \emptyset \gbar \Gamma, \type{x}{U} \gbar \Gamma, \type{X}{\langle \vecnew{U}, \vecnew{T} \rangle} \quad\quad \text{(empty set, variable, process variable)} $
\end{definition}

Global environments $ \Gamma $ contain assignments $ \type{x}{U} $ of variables to base types and assignments $ \type{X}{\langle \vecnew{U}, \vecnew{T} \rangle} $ of process constants to the base types and session types used in the respective declaration.\medbreak

Our type judgements are of the form $\Gamma \vdash P \triangleright \Delta$, meaning ``given the variable- and process types in global environment $\Gamma$, process $P$ is well-typed according to local context $\Delta$''.
We are mainly interested in ``self-contained'' processes without free (process) variables, as their behaviour is predictable.
For these processes, the global environment is empty at the bottom of the type derivation tree.
When $\Gamma = \emptyset$, we sometimes write $\vdash P \triangleright \Delta$.
We will additionally encounter the judgement $\Gamma \vdash \type{v}{U}$, meaning ``global environment $\Gamma$ justifies value $v$ to have type $U$''.
\begin{definition}[Typing Rules]
\label{def:typingRules}
For local context $\Delta$ containing \emph{only well-formed types}, the type judgement $ \Gamma \vdash P \triangleright \Delta $ is defined inductively as follows.
	\begin{gather*}
		\Gamma \vdash \type{1, 2, \mydots}{\natT} \; \sidenote{[Nat]} \quad
		\Gamma \vdash \type{\true, \false}{\boolT} \; \sidenote{[Bool]}
		\quad
		\Gamma, \type{x}{U} \vdash \type{x}{U} \; \sidenote{[Base]} \quad
		\Gamma \vdash \0 \smalltriangleright \emptyset \; \sidenote{[T-$ \0 $]}
		\vspace*{0.25em}\\
        \infer[\sidenote{[T-Res]}]{\Gamma \vdash \res{s}P \smalltriangleright \Delta}{\safe{\left\lbrace \type{\channelk{s}{r}{i}}{T_i} \right\rbrace_{i \in I}} \quad \Gamma \vdash P \smalltriangleright \Delta, \left\lbrace \type{\channelk{s}{r}{i}}{T_i} \right\rbrace_{i \in I}}
		\vspace*{0.25em}\\
        \infer[\sidenote{[T-If]}]{\Gamma \vdash \cond{v}{P}{Q} \smalltriangleright \Delta}{\Gamma \vdash \type{v}{\boolT} \quad \Gamma \vdash P \smalltriangleright \Delta \quad \Gamma \vdash Q \smalltriangleright \Delta}\quad
		\infer[\sidenote{[T-Par]}]{\Gamma \vdash P \inparallel Q \smalltriangleright \Delta_1, \Delta_2}{\Gamma \vdash P \smalltriangleright \Delta_1 \quad \Gamma \vdash Q \smalltriangleright \Delta_2}
		\vspace*{0.25em}\\
		\infer[\sidenote{[T-Def]}]{\Gamma \vdash \procdef{\procdecl{X}{\vecnew{x}}{c_1, \mydots, c_n} \defeq P}{Q} \smalltriangleright \Delta}{\Gamma, \type{X}{\left\langle  \vecnew{U}, T_1, \mydots, T_n  \right\rangle}, \type{\vecnew{x}}{\vecnew{U}} \!\vdash\! P \smalltriangleright \type{c_1}{T_1}, \mydots, \type{c_n}{T_n} \quad \Gamma, \type{X}{\left\langle \vecnew{U}, T_1, \mydots, T_n \right\rangle} \!\vdash\! Q \smalltriangleright \Delta}
		\vspace*{0.25em}\\
		\infer[\sidenote{[T-Var]}]{\Gamma, \type{X}{\left\langle \vecnew{U}, T_1, \mydots, T_n \right\rangle} \vdash \processcall{X}{\vecnew{v}}{c_1, \mydots, c_n} \smalltriangleright \type{c_1}{T_1}, \mydots, \type{c_n}{T_n}}{\Gamma \vdash \type{\vecnew{v}}{\vecnew{U}}}
		\vspace*{0.25em}\\
		\infer[\sidenote{[T-Sum]}]{\Gamma \vdash c\mc{i}{I} M_i \smalltriangleright \Delta, \type{c}{\mc{i}{I} L_i}}{\forall i \in I.\; \Gamma \vdash \insess{c}M_i \smalltriangleright \Delta, \type{c}{L_i}}\qquad
		\infer[\sidenote{[T-Sub]}]{\Gamma \vdash P \smalltriangleright \Delta}{\Gamma \vdash P \smalltriangleright \Delta' \quad \Delta' \leq_1 \Delta}
		\vspace*{0.25em}\\
		\infer[\sidenote{[T-$ \nonsend $]}]{\Gamma \vdash \insess{c}{\probOut{\prob}{\nonsend \cont P}} \smalltriangleright \Delta, \type{c}{\probOut{\prob}{\nonsend \cont T}}}{\Gamma \vdash P \smalltriangleright \Delta, \type{c}{T}}
		\vspace*{0.25em}\\
		\infer[\sidenote{[T-Inp]}]{\Gamma \vdash \insess{c}\inp{p}{q}{l}{x}{P} \smalltriangleright \Delta, \type{c}{\inp{p}{q}{l}{U}{T}}}{\participant{p} \in c \quad \Gamma, \type{x}{U} \vdash P \smalltriangleright \Delta, \type{c}{T}}
		\vspace*{0.25em}\\
		\infer[\sidenote{[T-Out]}]{\Gamma \vdash \insess{c}\probOut{\prob}{\out{p}{q}{l}{v}{P}} \smalltriangleright \Delta, \type{c}{\probOut{\prob}{\out{p}{q}{l}{U}{T}}}}{\participant{p} \in c \quad \Gamma \vdash \type{v}{U} \quad \Gamma \vdash P \smalltriangleright \Delta, \type{c}{T}}
		\vspace*{0.25em}\\
		\infer[\sidenote{[T-Prob]}]{\Gamma \vdash \insess{c}\send{i}{I} \probOut{\prob_i}{N_i.P_i} \smalltriangleright \Delta, \type{c}{\send{i}{I}\probOut{\prob_i}{H_i \cont T_i}}}{\forall i \in I.\; \Gamma \vdash \insess{c}\probOut{\prob_i}{N_i{\cont}P_i} \smalltriangleright \Delta, \type{c}{\probOut{\prob_i}{H_i \cont T_i}}}
	\end{gather*}
\end{definition}

The typing rules are mostly standard.
According to [Nat] and [Bool], the natural numbers $1,2,\mydots$ being of type $\natT$ and the truth values $\top,\bot$ being of type $\boolT$ will be justified by any global environment.
[Base] justifies the variable $x$ to be of type $U$ if $\Gamma$ contains that type assignment.
Rule [T-$\emptyset$] states that given any $\Gamma$, the terminated process $\0$ is justified by an empty local context.
By [T-Res], a restricted process $\res{s} P$ is justified by $\Delta$ if $P$ can be justified by the composition of $\Delta$ and all channel types within the restricted session $\left\lbrace \type{\channelk{s}{r}{i}}{T_i} \right\rbrace_{i \in I}$, given that these types fulfil the predicate $\safe{}$.
Said predicate is introduced with the properties of the type system in Chapter~\ref{chap:properties}.
Intuitively, it verifies that during communication no label mismatch can occur.
[T-If] states that given $\Gamma$, a conditional $\cond{v}{P}{Q}$ is justified according to $\Delta$ if the assignment of the $v$ to $\boolT$ is justified by $\Gamma$, and the continuation processes $P$ and $Q$ each justified by $\Delta$ (given $\Gamma$).
[T-Par] handles parallel composition, in which two participant processes $P$ and $Q$ may be composed if they are each justified by local contexts $\Delta_1$ and $\Delta_2$, whose composition $\Delta_1, \Delta_2$ is defined (\ie respects linearity).
Rule [T-Def] handles process definitions $ \procdef{\procdecl{X}{\vecnew{x}}{c_1, \mydots, c_n} \defeq P}{Q}$.
Firstly, the process $P$ must be justifiable by the local context containing the channels used in the definition, $\type{c_1}{T_1}, \mydots, \type{c_n}{T_n}$ given the process variable and variable assignments $\type{X}{\left\langle \vecnew{U}, T_1, \mydots, T_n \right\rangle}, \type{\vecnew{x}}{\vecnew{U}} $.
Secondly, given the same process variable and variable assignments, the process $Q$ must be justified by $\Delta$.
Rule [T-Var] is a derivation anchor for processes ending in a process call $\processcall{X}{\vecnew{v}}{c_1, \mydots, c_n}$.
The global environment has to contain the process variable assignment for $X$, namely $\type{X}{\left\langle \vecnew{U}, T_1, \mydots, T_n \right\rangle}$, and the local context has to comprise only the channels ${c_1, \mydots, c_n}$, whose types are those from the process variable assignment in the global environment.
Additionally $\Gamma$ must justify the value types, $\Gamma \vdash \type{\vecnew{v}}{\vecnew{U}}$.
With [T-Sum], a process with a mixed choice sum $c\mc{i}{I} M_i$ is justified if the respective channel type is a mixed choice sum type of the same cardinality $\type{c}{\mc{i}{I} L_i}$ such that each process summand is justified by one type summand $\Gamma \vdash \insess{c}M_i \smalltriangleright \Delta, \type{c}{L_i}$.
Rule [T-Sub] allows a process to be justified by a local context $\Delta$ if it can be justified by a subtype-context $\Delta' \leq \Delta$.
This represents the principle of safe substitution mentioned previously.
Internal actions are handled by [T-$\nonsend$], the process can use a channel $c$ to perform an internal action $\nonsend$ with probability $\prob$ if the local context has a session type ${\probOut{\prob}{\nonsend \cont T}}$ assigned to the same channel and the continuation $P$ can be justified by a local context in which $c$ is now assigned the continuation type $T$.
The rules for inputs and outputs, [T-Inp] and [T-Out], function analogously.
Probabilistic choices are typed by \text{[T-Prob]} similarly to how sums are typed with [T-Sum].\medbreak

We will now come back to the running example to showcase the session types of the channels used in our protocol.
\begin{example}[Running Example---Types]
\addxcontentsline{loe}{example}[\theexample]{Running Example---Types}
	\label{ex:motivatingExampleTypes}
	The processes of Example~\ref{ex:motivatingExampleSyntax} can be typed by the types $T_{\participant{p}}, T_{\participant{c}}, T_{\participant{w}}, T_{\participant{j}}$, and $T_{\participant{d}}$, such that we have $\emptyset \vdash P_{\interface} \smalltriangleright \Delta_{\interface}$ with local context $\Delta_{\interface} = \type{\channelsingle{s}{p}}{T_{\participant{p}}}, \type{\channel{s}{c}}{T_{\participant{c}}}, \type{\channelsingle{s}{w}}{T_{\participant{w}}}$ for the interface, and $\emptyset \vdash P_{\refinement} \smalltriangleright \Delta_{\refinement}$ with local context $\Delta_{\refinement} = \type{\channelsingle{s}{p}}{T_{\participant{p}}}, \type{\channelsingle{s}{j}}{T_{\participant{j}}}, \type{\channelsingle{s}{d}}{T_{\participant{d}}}, \type{\channelsingle{s}{w}}{T_{\participant{w}}}$ for the refinement.
	\begin{gather*}
		T_{\participant{p}} = \out{p}{j}{\mathtt{lws}}{}{} \inpPrefix{p}{j}{\mathtt{glt}}{\boolT}
		\vspace*{0.25em}\\
		T_{\participant{c}} = \inp{j}{p}{\mathtt{lws}}{}{} \kern-2pt \begin{cases}
			&\probOut{0.7}{\outPrefix{j}{p}{\mathtt{glt}}{\boolT}} \cont \probOut{1}{\outPrefix{j}{w}{\mathtt{rls}}{}}\\
			\!\oplus\!\!\!\! &\probOut{0.3}{\outPrefix{j}{w}{\mathtt{rqs}}{}}\cont \inpPrefix{j}{w}{\mathtt{st}}{} \cont \probOut{1}{\outPrefix{j}{p}{\mathtt{glt}}{\boolT}}
		\end{cases}
		\vspace*{0.25em}\\
		T_{\participant{w}} = \begin{cases}
			& \inp{w}{j}{\mathtt{mtg}}{}{} \outPrefix{w}{j}{\mathtt{st}}{} \\
			\!+\!\!\!\! & \inpPrefix{w}{j}{\mathtt{rls}}{}
		\end{cases} \qquad
		T_{\participant{d}} = \begin{cases}
			& \probOut{0.5}{\outPrefix{d}{j}{\mathtt{wk}}{}} \\
			\!\oplus\!\!\!\! & \probOut{0.2}{\outPrefix{d}{j}{\mathtt{str}}{}} \\
			\!\oplus\!\!\!\! & \probOut{0.3}{\outPrefix{d}{j}{\mathtt{wit}}{}}
		\end{cases}\vspace*{0.25em}\\
		T_{\participant{j}} = \inp{j}{p}{\mathtt{lws}}{}{} \kern-2pt
		\begin{cases}
			& \inpPrefix{j}{d}{\mathtt{wk}}{} \cont \probOut{1}{\outPrefix{j}{p}{\mathtt{glt}}{\boolT}} \cont \probOut{1}{\outPrefix{j}{w}{\mathtt{rls}}{}} \\
			\!+\!\!\!\! & \inpPrefix{j}{d}{\mathtt{str}}{} \cont \probOut{1}{\outPrefix{j}{p}{\mathtt{glt}}{\boolT}} \cont \probOut{1}{\outPrefix{j}{w}{\mathtt{rls}}{}} \\
			\!+\!\!\!\! & \inpPrefix{j}{d}{\mathtt{wit}}{} \cont \probOut{1}{\outPrefix{j}{w}{\mathtt{rqs}}{}}\cont \\
			& \hspace{6em} \inpPrefix{j}{w}{\mathtt{st}}{} \cont \probOut{1}{\outPrefix{j}{p}{\mathtt{glt}}{\boolT}}
		\end{cases}
	\end{gather*}
    \exampleDone
\end{example}\medbreak

Definition~\ref{def:transitionsLocalContexts} presents rules for transitions of local contexts.
These rules map reductions of processes to their respective types, \ie they are similar to the reduction rules in Definition~\ref{def:reductionSemantics}.
Labelled rules are used on types to highlight the main actors of a step.
\begin{definition}[Labelled Transitions of Local Contexts]
\label{def:transitionsLocalContexts}
Labels $ \alpha $ are of the form $ \type{s}{\inpPrefix{p}{q}{l}{U}} $ for inputs, $ \type{s}{\outPrefix{p}{q}{l}{U}} $ for outputs, $ \type{s}{\tau} $, or $ \type{s}{\comPrefix{p}{q}{l}{U}} $ for communication (from \participant{p} to \participant{q}). The labelled transition $\xlongrightarrow{\alpha}_{\prob}$ is inductively defined by the following rules.
	\begin{gather*}
		\type{\channel{s}{r}}{\inp{p}{q}{l}{U}{T} + \choice} \xlongrightarrow{\type{s}{\inpPrefix{p}{q}{l}{U}}}_{1} \type{\channel{s}{r}}{T} \; \sidenote{[TR-Inp]}
		\vspace*{0.25em}\\
		\type{\channel{s}{r}}{\left( \probOut{\prob}{\out{p}{q}{l}{U}{T}} \oplus \choiceProb \right) + \choice} \xlongrightarrow{\type{s}{\outPrefix{p}{q}{l}{U}}}_{\prob} \type{\channel{s}{r}}{T} \; \sidenote{[TR-Out]}
		\vspace*{0.25em}\\
		\Delta, \type{\channel{s}{r}}{\left( \probOut{\prob}{\nonsend \cont T} \oplus \choiceProb \right) + \choice} \xlongrightarrow{\type{s}{\nonsend}}_{\prob} \Delta, \type{\channel{s}{r}}{T} \; \sidenote{[TR-$ \nonsend $]}
		\vspace*{0.25em}\\
		\infer[\sidenote{[TR-Com]}]{\Delta, \type{c_1}{T_1}, \type{c_2}{T_2} \xlongrightarrow{\type{s}{\comPrefix{p}{q}{l}{U}}}_{\prob} \Delta, \type{c_1}{T_1'}, \type{c_2}{T_2'}}{\type{c_1}{T_1} \xlongrightarrow{\type{s}{\outPrefix{p}{q}{l}{U}}}_{1} \type{c_1}{T_1'} \quad \type{c_2}{T_2} \xlongrightarrow{\type{s}{\inpPrefix{q}{p}{l}{U}}}_{\prob} \type{c_2}{T_2'}}
	\end{gather*}
    We write $ \Delta \xlongrightarrow{\alpha}_{\prob} $ if $ \Delta \xlongrightarrow{\alpha}_{\prob} \Delta' $ for some $ \Delta' $.
    Moreover, we write $ \Delta \mapsto_{\prob} \Delta' $ if $ \Delta \xlongrightarrow{\type{s}{\comPrefix{p}{q}{l}{U}}}_{\prob} \Delta' $ or $ \Delta \xlongrightarrow{\type{s}{\nonsend}}_{\prob} \Delta' $ and lastly, we write $ \Delta \nrightarrow $ if there are no $ \Delta' $ and $ \prob $ such that $ \Delta \mapsto_{\prob} \Delta' $.
    Let $ \mapsto^{*}_{\prob} $ be inductively defined as (a)~$ \Delta \mapsto^{*}_{1} \Delta $ and (b)~if $ \Delta \mapsto_{\prob_1} \Delta' $ and $ \Delta' \mapsto^{*}_{\prob_2} \Delta'' $ then $ \Delta \mapsto^{*}_{\prob_1 \prob_2} \Delta'' $.
\end{definition}

We have now formally introduced all relevant concepts for the \emph{\calculusname{} $\pi$-calculus} and \emph{-types}.
Interspersed with examples, we have presented the process syntax, operational semantics, type syntax, subtyping, and typing rules.
Although we consider the past chapter to merely be preliminaries for the main contribution of this work, it should be stated that the calculus itself is novel in its own right.
While first strides in probabilistic session types have been made (\citep{aman2019probabilities, inverso_et_al:LIPIcs.CONCUR.2020.14}), to the best of our knowledge, our probabilistic mixed choice framework is entirely new.
Even leaving the system as we have defined it so far, and moving on to show central properties of MPST systems for it, we would have created a meaningful scientific contribution.
We will, however, instead introduce another layer: The main idea of this thesis, the multi-channel refinement-based subtyping.
Only after developing our framework further and expanding its capabilities greatly will we prove the major theorems.

%%%%%%%%%%%%%%%
%  subtyping  %
%%%%%%%%%%%%%%%

\chapter{Subtyping with Refinement}
\label{chap:subtyping}
There are two ways to view a typing context and depending on the goal, both are true and useful.
The intuition for subtyping and its intricacies might slightly differ, however.
One might see typing as chronologically ``secondary'' to the processes.
The protocol is given to us as interacting processes and we then assign it types to abstract and verify its behaviour.
Instead, though, we can also view it as chronologically ``primary''.
In this approach, types are an easily verifiable system specification, an \emph{interface}, according to which an implementation is built---the process.
Let us, for now, assume the latter view.
Using classic subtyping, then, one can justify processes that slightly differ from the given specification.
We can \emph{refine} the \emph{interface}, by using a ``smaller'' subtype to justify a part of the protocol.
Our novel system exploits that exact mechanism and completely overhauls what a \emph{refinement} subtype can be.
Behaviour specified by a single typed channel can be distributed among several interacting ones, such that in the abstract, the exact same behaviour is observable from ``the outside''.
With this idea in mind, we have created a subtyping system which allows several channels to be a subtype to a single channel while preserving desirable properties and interaction behaviour.

In the conception of these ideas, we drew inspiration from \citep{DBLP:conf/cav/WatanabeEAH23} and their work on compositional Markov decision processes.
They remarked that given the current trajectory in technological advancements, due to state-space explosion, modern verification targets can be enormous, to the extent that some of these models require more space than the memory size of verification machines.
One alleviation to this, they propose, is compositionality, which offers not only a memory advantage, but, if compositional components get repeated and reused, a performance advantage as well---divide and conquer.
With this in mind, our system offers another great advantage.
Assume for this the other view on subtyping: Processes come first, which are then typed to verify and abstract.
By utilising the subtyping relation in the ``other direction'', \ie from \emph{refinement} to \emph{interface}, we can iteratively and compositionally verify a protocol in parts.
As we will show in Chapter~\ref{chap:properties}, any collection of typed channels can be composed into \emph{one} combined interface supertype.
The subtyping derivation naturally verifies the ``internal'' behaviour (all those interactions which occur only between the typed channels of the respective collection), while preserving the ``external'' behaviour (those interactions whose communication partners are not within the collection).
Applying this principle, we can iteratively create much simpler types for which desired properties are easier to verify.

Note, that usually in MPST, the channels to which the session types are assigned are not considered for the subtyping relation (see \citep{scalas2019less, peters2024separation}).
However for us, as the interactions of the channels are highly relevant to subtyping, we require the subtyping relation to not merely relate types, $T' \leq T$, but typed channels, $\type{c}{T'} \leq \type{c}{T}$ or $\Delta \leq \type{c}{T}$.\medbreak

Given the syntactic density of the multi-channel subtyping system, we have not just opted to dedicate this entire chapter to it, we have also decided to introduce an intermediary subtyping relation which aims to bridge the gap between the standard subtyping (§~\ref{section:subtypingStandard}) and the multi-channel subtyping:
Functionally, it is just as expressive as the standard subtyping, but its syntax and structure is aligned with the advanced subtyping.
As it is intended as a didactic stepping stone, no major properties are shown.
What we \emph{do} show, however, is that the more ``difficult'' subtyping relations subsume the ``easier'' ones.
We first introduce this intermediary subtyping, then adapt the typing system, enhancing it with extra syntax to prepare for the complexities of the new subtyping rules before tackling the multi-channel subtyping.

\section{Single-Channel Subtyping}
\label{sec:singlechannelsubtyping}
As previously stated, the following subtyping is functionally similar to the standard subtyping (Definition~\ref{def:subtypingSimple}).
The form of the rules, however is more akin to those of the upcoming multi-channel subtyping, in fact, these rules are their single-channel variant.
After introducing and explaining the single-channel subtyping, we present an exemplary subtyping derivation, compare this subtyping to the previously encountered standard subtyping, and finally prove that the new subtyping subsumes standard subtyping.
\begin{definition}[Single-Channel Subtyping] 
The relation $\leqmedium{}$ is coinductively defined:
    \label{def:subtypingCompositionalOneChannel}
    \begin{gather*}
        %
		% [S-Sigma2] splitting into in's and out's
		%
		\infer=[\sidenote{\ssigmamedium{}}]{\type{c'}{\mc{i}{I'}\inpT_i + \mc{j}{J'}\outT_j} \leqmedium{\prob} \type{c}{\mc{i}{I}L_i + \mc{j}{J}L_j}}%
		{I \cup J \neq \emptyset \qquad \begin{array}{r c l}
				\type{c'}{\mc{i}{I'}\inpT_i} &\leqmedium{\prob}& \type{c}{\mc{i}{I}L_i}\\
				\type{c'}{\mc{j}{J'}\outT_j} &\leqmedium{\prob}& \type{c}{\mc{j}{J}L_j}
		\end{array}}\\[+0.25em]
		%
		% [S-Sigma-In]
		%
		\infer=[\sidenote{\ssigmainmedium{}}]{\type{c'}{\mc{i}{I' \cup J'}\inpT'_i} \leqmedium{\prob} \type{c}{\mc{i}{I'} \left( \mc{k}{I_i}L_k \right)}}%
		{\begin{array}{c}
				\bigcup_{i\in I'}I_i \neq \emptyset\\
				\forall j \in J'.\; \exists i\in I'.\; \pre(\inpT'_j)=\pre(\inpT'_i)
			\end{array} \qquad \forall i \in I'. \left(%
			\begin{array}{r}
				\type{c'}{\inpT'_i}\\
				\leqmedium{\prob} \type{c}{\mc{k}{I_i}L_k}
			\end{array} \right) }\\[+0.25em]
		%
		% [S-Sigma-Out]
		%
		\infer=[\sidenote{\ssigmaoutmedium{}}]{\type{c'}{\mc{i}{I'}\outT'_i} \leqmedium{\prob} \type{c}{\mc{i}{I'}\left( \mc{k}{I_i}L_k \right) + \mc{j}{J}\outT_j}}%
		{\begin{array}{c}
				\bigcup_{i\in I'}I_i \cup J \neq \emptyset\\
				\forall j \in J.\;  \exists i \in I'.\; \pre(\outT_j) = \pre(\outT'_i)
			\end{array}\qquad \forall i \in I'.\left(%
			\begin{array}{r}
				\type{c'}{\outT'_i}\\
				\leqmedium{\prob} \type{c}{\mc{k}{I_i}L_k}
			\end{array}\right)}\\[+0.25em]
		%
		% [S-oplus] splitting probabilities
		%
		\infer=[\sidenote{\soplusmedium{}}]{\type{c'}{\send{i}{I}}\probOut{\prob'_i}{H'_i \cont T'_i} \leqmedium{\prob^{\Sigma}} \type{c}{\send{j}{J} \probOut{\prob_j}{H_j \cont T_j}}}%
		{\begin{array}{c}
				J = \bigcup_{i \in I}J_i \neq \emptyset \qquad \sum_{j\in J}\prob_j = \prob^{\Sigma} \\
				\forall i\neq j\in I .\; J_i \cap J_j = \emptyset
			\end{array}
			\qquad\kern-10pt \forall i \in I.
			\left(\begin{array}{r}
				\type{c'}{H'_i \cont T'_i}\\
				\leqmedium{\prob'_i\prob^{\Sigma}} \type{c}{\send{j}{J_i} \probOut{\prob_j}{H_j \cont T_j}}
			\end{array}\right)}\\[+0.25em]
		%
		% [S-$\inpT$]
		%
		\infer=[\sidenote{\sinmedium{}}]{\type{\channelk{s}{r}{1}}{\inp{q}{p}{l}{U}{T'}} \leqmedium{1} \type{\channelk{s}{r}{2}}{\inp{q}{p}{l}{U}{T}}}%
		{\vecparticipantk{r}{1} \subseteq \vecparticipantk{r}{2} \qquad \participant{q} \in \vecparticipantk{r}{2} \qquad \type{\channelk{s}{r}{1}}{T'} \leqmedium{1} \type{\channelk{s}{r}{2}}{T}}\\[+0.25em]
        %
		% [S-$\outT$]
		%
		\infer=[\sidenote{\soutmedium{}}]{\type{\channelk{s}{r}{1}}{\out{p}{q}{l}{U}{T'}}\leqmedium{\prob} \type{\channelk{s}{r}{2}}{\probOut{\prob}{\out{p}{q}{l}{U}{T}}}}%
		{\vecparticipantk{r}{1} \subseteq \vecparticipantk{r}{2} \qquad \participant{p} \in \vecparticipantk{r}{2} \qquad \type{\channelk{s}{r}{1}}{T'} \leqmedium{1} \type{\channelk{s}{r}{2}}{T}}\\[+0.25em]
		% %
		% % [Sub-M-e-L] [Sub-M-e-R]
		% %
		% \infer=[\sidenote{\snonsendLmedium{}}]{\type{\channelk{s}{r}{1}}{\nonsend \cont T'} \leqmedium{\prob} \type{\channelk{s}{r}{2}}{T}}%
		% {\type{\channelk{s}{r}{1}}{T'} \leqmedium{\prob} \type{\channelk{s}{r}{2}}{T}}\qquad\!
		% %
		% \infer=[\sidenote{\snonsendRmedium{}}]{\type{c'}{T'} \leqmedium{\prob} \type{\channel{s}{r}}{\probOut{\prob}{\nonsend} \cont T}}%
		% {\type{c'}{T'} \leqmedium{1} \type{\channel{s}{r}}{T}}\\[+0.25em]
        \infer=[\sidenote{\snonsendmedium{}}]{\type{c'}{\nonsend \cont T'} \leqmedium{\prob} \type{c}{\probOut{\prob}{\nonsend} \cont T}}%
        {\type{c'}{T'} \leqmedium{1} \type{c}{T}}\qquad\!
		%
		% [S-emptyset]
		%
		\infer=[\sidenote{\semptysetmedium{}}]{\emptyset \leqmedium{1} \emptyset}%
		{}\qquad\!
		%
		% [S-Split]
		%
		\infer=[\sidenote{\ssplitmedium{}}]{\Delta', \type{c'}{T'} \leqmedium{1} \Delta, \type{c}{T}}%
		{\Delta' \leqmedium{1} \Delta \qquad \type{c'}{T'} \leqmedium{1} \type{c}{T}}
	\end{gather*}
\end{definition}
The first four rules may be understood as one large rule whose purpose is to split mixed choice sums into singular actions.
Not always need all (or any) of these rules be applied, but whenever they do occur in a derivation, they will appear in order---\ssigmamedium{} is applied first (lowermost in the derivation tree) and \soplusmedium{} last (uppermost).
\ssigmamedium{} splits the mixed choice of the subtype into its inputs and probabilistic sums, where each sub-sum must then be subtype of a sub-sum of the mixed choice of the supertype.
As previously, choices on the left of the subtyping may have fewer outputs and more inputs.
This is again implemented by \ssigmainmedium{} and \ssigmaoutmedium{}.
The difference to earlier is, however, that instead of comparing the respective input or output prefixes immediately, the sums are split further.
The rule \ssigmainmedium{} splits an input choice into single summands.
Thereby, it allows for additional inputs on left (via the index set $ J' $) that are not matched on the right.
It is ensured that no additional choice $ \inpT'_j = \inp{p}{q}{l}{U}{T'} $ introduces a prefix $\pre(\inpT'_j) = \participant{p}{\leftarrow}\participant{q}\mathop{?}$ that was not previously justified.
Similarly, \ssigmaoutmedium{} splits a choice into probabilistic choices.
Here, the right hand side may have additional probabilistic choices.
The rule ensures that no prefix of the supertype is lost.
Finally, \soplusmedium{} splits a probabilistic choice into its individual summands and removes their probabilities $ \prob_i' $, by multiplying $ \prob_i' $ with the current probability in the index of the subtype relation $ \leqmedium{\prob} $.
Hence, $\leqmedium{\prob}$ identifies a branch in the subtyping derivation tree in which a subterm of the subtype is considered that occurs in the original, \ie not split, subtype with probability $ \prob $.
In other words, the subterm is subtype of the supertype with probability $\prob$. 

The remaining rules, \sinmedium{} to \ssplitmedium{}, contain the core of the actual subtyping behaviour.
Due to \sinmedium{}, a single input action is the subtype of the same single action if the respective continuations are sub- and supertype.
Both the acting role $\participant{q}$ and the subtype role set $\vecparticipantk{r}{1}$ must be contained in the supertype role set $\vecparticipantk{r}{2}$.
Rule \soutmedium{} is similar.
Additionally it matches the probability on the relation $\leqmedium{\prob}$ in the conclusion to the probability of the sending action of the supertype, $\type{\channelk{s}{r}{2}}{\probOut{\prob}{\out{p}{q}{l}{U}{T}}}$.
For the premise then $ \leq_1 $ is used.
Similarly, \snonsendmedium{} handles internal actions, matching the probability of the relation.
\semptysetmedium{} ensures that $\emptyset \leqmedium{1} \emptyset$.
As $\Delta, \type{c}{\tend} = \Delta$, this rule relates $\tend$-typed channels.
Finally, \ssplitmedium{} splits contexts such that each channel in the supertype context is related to exactly one channel in the subtype context.\medbreak

For better understanding, we will perform a subtyping derivation using these rules.
\begin{example}[Single-Channel Subtyping Derivation]
\addxcontentsline{loe}{example}[\theexample]{Single-Channel Subtyping Derivation}
\label{ex:singlechannelsubtypingderivation}
    Consider the session type $T$ and its subtype $T'$ from Example~\ref{ex:standardsubtypingderivation}, \ie $T' \leqsimple T$.
    \begin{gather*}
        T' = \begin{cases}
			& \inpPrefix{a}{b}{\mathtt{in}_1}{} \\
            & \inpPrefix{a}{b}{\mathtt{in}_2}{} \cont T'' \\
            \!+ \!\!\!\! &\probOut{1}{\outPrefix{a}{c}{\mathtt{out}_3}{}}
		\end{cases}\qquad
        T = \begin{cases}
			& \inpPrefix{a}{b}{\mathtt{in}_1}{} \\
			\!+ \!\!\!\! & \begin{cases}
                &\probOut{0.7}{\outPrefix{a}{c}{\mathtt{out}_1}{}}\\
                \!\oplus \!\!\!\! &\probOut{0.3}{\outPrefix{a}{c}{\mathtt{out}_2}{}}
			\end{cases}\\
            \!+ \!\!\!\! &\probOut{1}{\outPrefix{a}{c}{\mathtt{out}_3}{}}
		\end{cases}
    \end{gather*}
    With the following subtyping derivation, we show that also $\type{c}{T'} \leqmedium{1} \type{c}{T}$.
    We have abbreviated the rule names to their suffixes, \textcolor{medium}{[$\emptyset$]} refers to \semptysetmedium{}, \textcolor{medium}{[$\Sigma$]} to \ssigmamedium{} and so on.
    As the derivation tree is rather wide, we present the two branches separately, beginning on the left-hand side with the inputs.
    \begin{gather*}
        \infer=[\textcolor{medium}{[\Sigma]}]{\type{c}{\begin{cases}
			& \inpPrefix{a}{b}{\mathtt{in}_1}{} \\
            & \inpPrefix{a}{b}{\mathtt{in}_2}{} \cont T'' \\
            \!+ \!\!\!\! &\probOut{1}{\outPrefix{a}{c}{\mathtt{out}_3}{}}
		\end{cases}} \leqmedium{1} \type{c}{\begin{cases}
			& \inpPrefix{a}{b}{\mathtt{in}_1}{} \\
			\!+ \!\!\!\! & \begin{cases}
                &\probOut{0.7}{\outPrefix{a}{c}{\mathtt{out}_1}{}}\\
                \!\oplus \!\!\!\! &\probOut{0.3}{\outPrefix{a}{c}{\mathtt{out}_2}{}}
			\end{cases}\\
            \!+ \!\!\!\! &\probOut{1}{\outPrefix{a}{c}{\mathtt{out}_3}{}}
		\end{cases}}}%
    		{\infer=[\textcolor{medium}{[\Sigma\text{-}\inpT]}]{\type{c}{\begin{cases}
    			& \inpPrefix{a}{b}{\mathtt{in}_1}{} \\
                \!+ \!\!\!\!& \inpPrefix{a}{b}{\mathtt{in}_2}{} \cont T''
    		\end{cases}} \leqmedium{1} \type{c}{\inpPrefix{a}{b}{\mathtt{in}_1}{}} }%
                {\infer=[\textcolor{medium}{[\inpT]}]{ \type{c}{\inpPrefix{a}{b}{\mathtt{in}_1}{}} \leqmedium{1} \type{c}{\inpPrefix{a}{b}{\mathtt{in}_1}{}} }%
        		{ \infer=[\textcolor{medium}{[\emptyset]}]{ \type{c}{\tend} \leqmedium{1} \type{c}{\tend} }%
        		{} } } \hspace{3em} \dots \hspace{5em}}
    \end{gather*}
    \normalsize{Next is the branch on the right, the outputs.}
    \begin{gather*}
        \infer=[\textcolor{medium}{[\Sigma]}]{\type{c}{\begin{cases}
			& \inpPrefix{a}{b}{\mathtt{in}_1}{} \\
            & \inpPrefix{a}{b}{\mathtt{in}_2}{} \cont T'' \\
            \!+ \!\!\!\! &\probOut{1}{\outPrefix{a}{c}{\mathtt{out}_3}{}}
		\end{cases}} \leqmedium{1} \type{c}{\begin{cases}
			& \inpPrefix{a}{b}{\mathtt{in}_1}{} \\
			\!+ \!\!\!\! & \begin{cases}
                &\probOut{0.7}{\outPrefix{a}{c}{\mathtt{out}_1}{}}\\
                \!\oplus \!\!\!\! &\probOut{0.3}{\outPrefix{a}{c}{\mathtt{out}_2}{}}
			\end{cases}\\
            \!+ \!\!\!\! &\probOut{1}{\outPrefix{a}{c}{\mathtt{out}_3}{}}
		\end{cases}}}
            {\hspace{2em} \dots \hspace{4em} \infer=[\textcolor{medium}{[\Sigma\text{-}\outT]}]{ \type{c}{\probOut{1}{\outPrefix{a}{c}{\mathtt{out}_3}{}}} \leqmedium{1} \type{c}{\begin{cases}
                & \begin{cases}
                &\probOut{0.7}{\outPrefix{a}{c}{\mathtt{out}_1}{}}\\
                \!\oplus \!\!\!\! &\probOut{0.3}{\outPrefix{a}{c}{\mathtt{out}_2}{}}
			\end{cases}\\
            \!+ \!\!\!\! &\probOut{1}{\outPrefix{a}{c}{\mathtt{out}_3}{}}
		      \end{cases}} }%
        		{\infer=[\textcolor{medium}{[\oplus]}]{ \type{c}{\probOut{1}{\outPrefix{a}{c}{\mathtt{out}_3}{}}} \leqmedium{1} \type{c}{\probOut{1}{\outPrefix{a}{c}{\mathtt{out}_3}{}}} }%
        		{ \infer=[\textcolor{medium}{[\outT]}]{ \type{c}{\outPrefix{a}{c}{\mathtt{out}_3}{}} \leqmedium{1} \type{c}{\probOut{1}{\outPrefix{a}{c}{\mathtt{out}_3}{}}} }%
        		{ \infer=[\textcolor{medium}{[\emptyset]}]{ \type{c}{\tend} \leqmedium{1} \type{c}{\tend} }%
        		{} } }}}
    \end{gather*}
    \normalsize{We notice that despite there being more steps, the overall tree is similar to that of Example~\ref{ex:standardsubtypingderivation}.}
    \exampleDone
\end{example}
As we have just seen, the functionality of the previous \ssigmaineasy{} and \ssigmaouteasy{} is now distributed over several rules, namely \ssigmainmedium{} and \sinmedium{} for \ssigmaineasy{} and \ssigmaoutmedium{}, \soplusmedium{}, \soutmedium{}, and \snonsendmedium{} for \ssigmaouteasy{}.
Additionally, the previous \sendeasy{} rule became \semptysetmedium{}.
\ssplitmedium{} is a new rule, which in essence accomplishes what the side condition in Definition~\ref{def:subtypingSimple}, ``We have $\type{c_1}{T_1'}, \dots, \type{c_n}{T_n'} \leqsimple{} \type{c_1}{T_1}, \dots, \type{c_n}{T_n}$ if for all $i \in [1..n]$ it holds that ${T_i'} \leqsimple {T_i}$'', did.
Notice that with the single-channel rules, we may subtype different channels, as long as the role set of the subtype channel is a subset of the roles of the supertype channel, \ie $\type{\channelk{s}{r}{1}}{T'} \leqmedium{\prob} \type{\channelk{s}{r}{2}}{T}$ if $\vecparticipantk{r}{1} \subseteq \vecparticipantk{r}{2}$ (by \sinmedium{}, \soutmedium{}). 

We now show that, as demonstrated in the previous example, single-channel subtyping indeed subsumes the standard subtyping of Definition~\ref{def:subtypingSimple}.
\begin{theorem}
\label{thm:mediumSubtypingIncludesSimple}
    If $\Delta' \leqsimple \Delta$, then $\Delta' \leqmedium{1} \Delta$.
\end{theorem}
\begin{proof}
    Assume that $\Delta' \leqsimple \Delta$.
    Note that according to Definition~\ref{def:subtypingSimple}, for contexts $\Delta' = \type{c_1}{T_1'}, \dots, \type{c_n}{T_n'}$ and $\Delta = \type{c_1}{T_1}, \dots, \type{c_n}{T_n}$ we have $\Delta' \leqsimple \Delta$ if for all $k \in [1..n]$ we have ${T_k'} \leqsimple {T_k}$.
    Additionally, from Definition~\ref{def:subtypingCompositionalOneChannel}, specifically rule \ssplitmedium{}, we have that if all $k \in [1..n]$ we have $\type{c_k}{T_k'} \leqmedium{1} \type{c_k}{T_k}$, then also $\Delta' \leqmedium{1} \Delta$.
    It thus suffices to prove that if ${T_k'} \leqsimple {T_k}$ then $\type{c_k}{T_k'} \leqmedium{1} \type{c_k}{T_k}$ which we show by structural induction on the derivation of ${T_k'} \leqsimple {T_k}$.
    For legibility we drop the subscript and continue with $c_k = c$, $T_k' = T'$ and $T_k = T$.
    We proceed by case distinction on the last applied rule.
    \begin{description}
        \item[Case {\sendeasy{}}:]%
        We have $T' = {{\tend }}$ and $T = {\tend }$ with ${\tend \leqsimple \tend}$.
        By the definition of local contexts $\type{c}{{\tend }} = \tend$ thus by \semptysetmedium{} of Definition~\ref{def:subtypingCompositionalOneChannel} we have $\type{c}{\tend} \leqmedium{1} \type{c}{\tend}$ as required.
        \item[Case {\ssigmaeasy{}}:]%
        We have $T' = {{\mc{i}{I'}\inpT'_i + \mc{j}{J'}\outT'_j}}$ and $T = {{\mc{i}{I}\inpT_i + \mc{j}{J}\outT_j}}$ for which ${\mc{i}{I'}\inpT'_i + \mc{j}{J'}\outT'_j} \leqsimple {\mc{i}{I}\inpT_i + \mc{j}{J}\outT_j}$, ${\mc{i}{I'}\inpT'_i} \leqsimple {\mc{i}{I}\inpT_i}$, ${\mc{j}{J'}\outT'_j} \leqsimple {\mc{j}{J}\outT_j}$, and $I \cup J \neq \emptyset$.
        Therefore by induction hypothesis have $\type{c}{\mc{i}{I'}\inpT'_i} \leqmedium{1} \type{c}{\mc{i}{I}\inpT_i}$ and $\type{c}{\mc{j}{J'}\outT'_j} \leqsimple \type{c}{\mc{j}{J}\outT_j}$.
        By \ssigmamedium{} of Definition~\ref{def:subtypingCompositionalOneChannel}, we thus have $\type{c}{\mc{i}{I'}\inpT_i + \mc{j}{J'}\outT_j} \leqmedium{1} \type{c}{\mc{i}{I}L_i + \mc{j}{J}L_j}$ as required.
        \item[Case {\ssigmaineasy{}}:]%
        Have $T' = {\mc{i}{I' \cup J'}\inpk{q}{p}{l_i}{U_i}{T'_i}{i}}$ and $T = {\mc{i}{I'} \inpk{q}{p}{l_i}{U_i}{T_i}{i}}$ with ${\mc{i}{I' \cup J'}\inpk{q}{p}{l_i}{U_i}{T'_i}{i}} \leqsimple {\mc{i}{I'} \inpk{q}{p}{l_i}{U_i}{T_i}{i}}$, $I' \neq \emptyset$, $\forall j \in J'.\; \exists i\in I'.\; \pre(\inpT'_j)=\pre(\inpT'_i)$, and $\forall i \in I'.\; {T'_i} \leqsimple {T_i}$.
        Apply the induction hypothesis on $\forall i \in I'.\; {T'_i} \leqsimple {T_i}$ to obtain $\forall i \in I'.\; \type{c}{T'_i} \leqsimple \type{c}{T_i}$.
        Since both sub- and supertype have the same channel, which thus have the same role vector, we may apply \sinmedium{} of Definition~\ref{def:subtypingCompositionalOneChannel} to obtain $\type{c}{\inpk{q}{p}{l_i}{U_i}{T'_i}{i}} \leqmedium{1} \type{c}{\inpk{q}{p}{l_i}{U_i}{T_i}{i}}$ for $\forall i \in I'$.
        Let then $\forall i \in I'.\; I_i = i$, where as $I' \neq \emptyset$ have $\bigcup_{i\in I'}I_i \neq \emptyset$.
        With this,  $\forall i \in I'.\; \type{c}{\inpk{q}{p}{l_i}{U_i}{T'_i}{i}} \leqmedium{1} \type{c}{\inpk{q}{p}{l_i}{U_i}{T_i}{i}}$, and $\forall j \in J'.\; \exists i\in I'.\; \pre(\inpT'_j)=\pre(\inpT'_i)$, we can apply \ssigmainmedium{} of Definition~\ref{def:subtypingCompositionalOneChannel} to obtain the required fact of $\type{c}{\mc{i}{I' \cup J'}\inpk{q}{p}{l_i}{U_i}{T'_i}{i}} \leqmedium{1} \type{c}{\mc{i}{I'} \inpk{q}{p}{l_i}{U_i}{T_i}{i}}$.
        \item[Case {\ssigmaouteasy{}}:]%
        In this case we have the two types $T' = {\mc{i}{I}{\send{k}{K_i}}\probOut{\prob_k}{H_k \cont T'_k}}$ and $T = {\mc{i}{I}{\send{k}{K_i} \probOut{\prob_k}{H_k \cont T_k}} + \mc{j}{J}\outT_j}$, where ${\mc{i}{I}{\send{k}{K_i}}\probOut{\prob_k}{H_k \cont T'_k}} \leqsimple {\mc{i}{I}{\send{k}{K_i} \probOut{\prob_k}{H_k \cont T_k}} + \mc{j}{J}\outT_j}$, $\forall j \in J.\;  \exists i \in I.\; \pre({\send{k}{K_i}}\probOut{\prob_k}{H_k \cont T'_k}) = \pre(\send{k}{K_i} \probOut{\prob_k}{H_k \cont T_k})$, and $I \neq \emptyset$, $\forall i \in I.\; \forall k \in K_i .\; {T'_k}	\leqsimple {T_k}$.
        From the induction hypothesis we obtain $\forall i \in I.\;\forall k \in K_i .\;\type{c}{T'_k} \leqmedium{1} \type{c}{T_k}$.
        Now $\forall i \in I.\; \forall k \in K_i$ depending on the shape of $H_k$ we have one of two cases.
        If $H_k = \nonsend$, we apply \snonsendLmedium{} and \snonsendRmedium{} of Definition~\ref{def:subtypingCompositionalOneChannel} to obtain $\type{c}{\nonsend \cont T'_k} \leqmedium{\prob_k} \type{c}{\nonsend \cont T_k}$.
        In the case in which $H_k \neq \nonsend$, since both sub- and supertype have the same channel, which thus have the same role vector, we may apply \soutmedium{} of Definition~\ref{def:subtypingCompositionalOneChannel} to obtain $\type{c}{\probOut{\prob_k}{H_k \cont T'_k}} \leqmedium{\prob_k} \type{c}{\probOut{\prob_k}{H_k \cont T_k}}$.
        Thus we have $\forall i \in I.\; \forall k \in K_i.\; \type{c}{H_k \cont T_k'} \leqmedium{\prob_k} \type{c}{\probOut{\prob_k}{H_k \cont T_k'}}$.
        By the definition of local contexts, all types in $\Delta'$ and $\Delta$ are well-formed.
        For $T' = {\mc{i}{I}{\send{k}{K_i}}\probOut{\prob_k}{H_k \cont T'_k}}$ specifically this implies that $\forall i \in I.\; \sum_{k \in K_i}\prob_k = 1$.
        Let then for each $i$ in $I$ $J_i = i$, hence $J = I$.
        From this, $I \neq \emptyset$, $\forall i \in I.\; \sum_{k \in K_i}\prob_k = 1$, and $\forall i \in I.\; \forall k \in K_i.\; \type{c}{H_k \cont T_k'} \leqmedium{\prob_k} \type{c}{\probOut{\prob_k}{H_k \cont T_k'}}$ we may apply \soplusmedium{} of Definition~\ref{def:subtypingCompositionalOneChannel} to obtain $\forall i \in I.\; \type{c}{\send{k}{K_i}\probOut{\prob_k}{H_k \cont T_k'}} \leqmedium{1} \type{c}{\send{k}{K_i}\probOut{\prob_k}{H_k \cont T_k'}}$.
        Finally, from this, $I \neq \emptyset$, and $\forall j \in J.\;  \exists i \in I.\; \pre({\send{k}{K_i}}\probOut{\prob_k}{H_k \cont T'_k}) = \pre(\send{k}{K_i} \probOut{\prob_k}{H_k \cont T_k})$, we can apply \ssigmaoutmedium{} of Definition~\ref{def:subtypingCompositionalOneChannel} to obtain $\type{c}{\mc{i}{I}\send{k}{K_i}\probOut{\prob_k}{H_k \cont T_k'}} \leqmedium{1} \type{c}{\mc{i}{I}\send{k}{K_i}\probOut{\prob_k}{H_k \cont T_k'} + \mc{j}{J}\outT_j}$ as required.\qedhere
    \end{description}
\end{proof}\medbreak

With this, we have finished presenting the single-channel subtyping.
Before making use of the acquired intuition to introduce the multi-channel subtyping, we will first have a small intermission to slightly adapt the typing system.

\section{Adapted Typing System}
\label{sec:adaptedTypingSystem}
Our aim was to keep the subtyping rules as slim as possible.
When the subtype, however, is a collection of several typed channels, a lot of terms must be matched and thus there inevitably exists a large number of branches within the subtyping derivation tree.
To keep track of those, we opted to introduce the concept of the \emph{active channel}.
Only found within subtyping derivation trees and not in type judgements, the active channel essentially marks that channel to which a branch in the derivation is dedicated.
For this, we introduce a new typing context, the local context with an active channel, $\Lambda$.

\begin{definition}[Local Context with an Active Channel]
	\label{def:localContext}
	%\vspace*{-0.75em}
	\begin{alignat*}{3}
		\Lambda & \defT &&\; \Delta \gbar c \cdot \left( \Delta, \type{c}{T} \right) &&
	\end{alignat*}
\end{definition}
A $ \Lambda $ is either a local context $\Delta$ (Definition~\ref{def:typeSyntax}) or a local context with an \emph{active channel}.
In accordance with this new context, we redefine labelled transitions of local contexts, based on Definition~\ref{def:transitionsLocalContexts}.
The labelled transition on local contexts with an active channel $\xlongrightarrow{\alpha}_{\prob}$ is given inductively by the following rules. 
\begin{definition}[Labelled Transitions of Local Contexts with an Active Channel]
	\label{def:transitionsLocalContextsWithActiveChannel}
    Labels $ \alpha $ in these rules are of the form $ \type{s}{\inpPrefix{p}{q}{l}{U}} $ for inputs, $ \type{s}{\outPrefix{p}{q}{l}{U}} $ for outputs, $ \type{s}{\tau} $, or $ \type{s}{\comPrefix{p}{q}{l}{U}} $ for communication (from \participant{p} to \participant{q}).
    \begin{gather*}
		\type{\channel{s}{r}}{\inp{p}{q}{l}{U}{T} + \choice} \xlongrightarrow{\type{s}{\inpPrefix{p}{q}{l}{U}}}_{1} \type{\channel{s}{r}}{T} \; \sidenote{[TR-Inp]}
		\vspace*{0.25em}\\
		\type{\channel{s}{r}}{\left( \probOut{\prob}{\out{p}{q}{l}{U}{T}} \oplus \choiceProb \right) + \choice} \xlongrightarrow{\type{s}{\outPrefix{p}{q}{l}{U}}}_{\prob} \type{\channel{s}{r}}{T} \; \sidenote{[TR-Out]}
		\vspace*{0.25em}\\
		\PBnoName{\type{\channel{s}{r}}{\out{p}{q}{l}{U}{T}} \xlongrightarrow{\type{s}{\outPrefix{p}{q}{l}{U}}}_{1} \type{\channel{s}{r}}{T} \; \sidenote{[TR$\Lambda$-Out]}}
		\vspace*{0.25em}\\
		\Delta, \type{\channel{s}{r}}{\left( \probOut{\prob}{\nonsend \cont T} \oplus \choiceProb \right) + \choice} \xlongrightarrow{\type{s}{\nonsend}}_{\prob} \Delta, \type{\channel{s}{r}}{T} \; \sidenote{[TR-$ \nonsend $]}
		\vspace*{0.25em}\\
		\PBnoName{\channel{s}{r}\cdot \left(\type{\Delta, \channel{s}{r}}{\left( \probOut{\prob}{\nonsend \cont T} \oplus \choiceProb \right) + \choice}\right) \xlongrightarrow{\type{s}{\nonsend}}_{\prob} \Delta, \type{\channel{s}{r}}{T} \; \sidenote{[TR$\Lambda$-$\nonsend$]}
		}\vspace*{0.25em}\\
		\PBnoName{\channel{s}{r}\cdot \left(\Delta, \type{\channel{s}{r}}{\nonsend \cont T }\right) \xlongrightarrow{\type{s}{\nonsend}}_{1} \Delta, \type{\channel{s}{r}}{T} \; \sidenote{[TR$\Lambda$-$\nonsend$-2]}}
		\vspace*{0.25em}\\
		\infer[\sidenote{[TR-Com]}]{\Delta, \type{c_1}{T_1}, \type{c_2}{T_2} \xlongrightarrow{\type{s}{\comPrefix{p}{q}{l}{U}}}_{\prob} \Delta, \type{c_1}{T_1'}, \type{c_2}{T_2'}}{\type{c_1}{T_1} \xlongrightarrow{\type{s}{\outPrefix{p}{q}{l}{U}}}_{1} \type{c_1}{T_1'} \quad \type{c_2}{T_2} \xlongrightarrow{\type{s}{\inpPrefix{q}{p}{l}{U}}}_{\prob} \type{c_2}{T_2'}}
		\vspace*{0.25em}\\
		\PBnoName{\infer[\sidenote{[TR$\Lambda$-Com]}]{c_2 \cdot \left(\Delta, \type{c_1}{T_1}, \type{c_2}{T_2} \right)\xlongrightarrow{\type{s}{\comPrefix{p}{q}{l}{U}}}_{\prob} \Delta, \type{c_1}{T_1'}, \type{c_2}{T_2'}}{\type{c_1}{T_1} \xlongrightarrow{\type{s}{\outPrefix{p}{q}{l}{U}}}_{1} \type{c_1}{T_1'} \quad \type{c_2}{T_2} \xlongrightarrow{\type{s}{\inpPrefix{q}{p}{l}{U}}}_{\prob} \type{c_2}{T_2'}}}
	\end{gather*}
    We write $ \Lambda \xlongrightarrow{\alpha}_{\prob} $ if $ \Lambda \xlongrightarrow{\alpha}_{\prob} \Lambda' $ for some $ \Lambda' $.
    Moreover, we write $ \Lambda \mapsto_{\prob} \Lambda' $ if $ \Lambda \xlongrightarrow{\type{s}{\comPrefix{p}{q}{l}{U}}}_{\prob} \Lambda' $ or $ \Lambda \xlongrightarrow{\type{s}{\nonsend}}_{\prob} \Lambda' $ and $ \Lambda \nrightarrow $ if there are no $ \Lambda' $ and $ \prob $ such that $ \Lambda \mapsto_{\prob} \Lambda' $.
    Let $ \mapsto^{*}_{\prob} $ be inductively defined as (a)~$ \Lambda \mapsto^{*}_{1} \Lambda $ and (b)~if $ \Lambda \mapsto_{\prob_1} \Lambda' $ and $ \Lambda' \mapsto^{*}_{\prob_2} \Lambda'' $ then $ \Lambda \mapsto^{*}_{\prob_1 \prob_2} \Lambda'' $.
\end{definition}
All rules in black, named [TR-*], are taken directly from the previous definition of labelled transitions.
Those highlighted in blue, named \PBnoName{[TR$\Lambda$-*]}, are additions needed to accommodate active channels.
As before, these rules map reductions of processes on their respective types, \ie they are similar to the reduction rules in Definition~\ref{def:reductionSemantics}.
Labelled rules are used on types to highlight the main actors of a step.

\section{Multi-Channel Subtyping}
\label{section:multiChannelSubtypingRules}
With all preliminaries behind us, we now finally present the multi-channel subtyping of the \emph{\calculusname{} $\pi$-calculus}.
In this section, after introducing and explaining the subtyping rules in detail, we will show that multi-channel subtyping subsumes single-channel subtyping, and finish the chapter by revisiting the courthouse protocol with two examples.\medbreak

Structurally, the multi-channel subtyping rules are very similar to those of single-channel subtyping (§~\ref{sec:singlechannelsubtyping}).
Apart from the addition of certain rules, as explained in the following, for most rules the difference lies mostly in the amount of typed channels on the left-hand side of the relation: One for single-channel subtyping, several for multi-channel, as the names imply.
\begin{definition}[Subtyping]
    \label{def:subtypingCompositional}
    The relation $ \leq_{\prob} $ is coinductively defined: 
	\begin{gather*}
    %
		%
		% [S-Sigma1] choosing active actor
		%
		\infer=[\sidenote{[S-$\Sigma$-1]}]{\Delta = \type{\channelk{s}{r}{1}}{T_1}, \ldots,  \type{\channelk{s}{r}{n}}{T_n} \leq_{\prob} \type{\channel{s}{r}}{\mc{j}{J_1} L_j + \ldots + \mc{j}{J_n} L_j}}%
		{\bigcup_{k \in [1..n]}J_k \neq \emptyset \quad \forall k \in \left[ 1 .. n \right].\; \channelk{s}{r}{k} \cdot \Delta \leq_{\prob} \type{\channel{s}{r}}{\mc{j}{J_k} L_j}}\\[+0.25em]
		%
		% [S-Sigma2] splitting into in's and out's
		%
		\infer=[\sidenote{[S-$\Sigma$-2]}]{\channelk{s}{r}{1} \cdot \left( \Delta, \type{\channelk{s}{r}{1}}{\mc{i}{I'}\inpT_i + \mc{j}{J'}\outT_j} \right) \leq_{\prob} \type{\channelk{s}{r}{2}}{\mc{i}{I}L_i + \mc{j}{J}L_j}}%
		{I \cup J \neq \emptyset \quad \begin{array}{r c l}
				\channelk{s}{r}{1} \cdot \left( \Delta, \type{\channelk{s}{r}{1}}{\mc{i}{I'}\inpT_i} \right) &\leq_{\prob}& \type{\channelk{s}{r}{2}}{\mc{i}{I}L_i}\\
				\channelk{s}{r}{1} \cdot \left( \Delta, \type{\channelk{s}{r}{1}}{\mc{j}{J'}\outT_j} \right) &\leq_{\prob}& \type{\channelk{s}{r}{2}}{\mc{j}{J}L_j}
		\end{array}}\\[+0.25em]
		%
		% [S-Sigma-In]
		%
		\infer=[\sidenote{[S-$\Sigma$-$\inpT$]}]{\channelk{s}{r}{1} \cdot \left( \Delta, \type{\channelk{s}{r}{1}}{\mc{i}{I' \cup J'}\inpT'_i} \right) \leq_{\prob} \type{\channelk{s}{r}{2}}{\mc{i}{I'} \left( \mc{k}{I_i}L_k \right)}}%
		{\begin{array}{c}
				\bigcup_{i\in I'}I_i \neq \emptyset\\
				\forall j \in J'.\; \exists i\in I'.\; \pre(\inpT'_j)=\pre(\inpT'_i)
			\end{array} \quad \forall i \in I'. \left(%
			\begin{array}{r}
				\channelk{s}{r}{1} \cdot \left( \Delta, \type{\channelk{s}{r}{1}}{\inpT'_i} \right)\\
				\leq_{\prob} \type{\channelk{s}{r}{2}}{\mc{k}{I_i}L_k}
			\end{array} \right) }\\[+0.25em]
		%
		% [S-Sigma-Out]
		%
		\infer=[\sidenote{[S-$\Sigma$-$\outT$]}]{\channelk{s}{r}{1} \cdot \left( \Delta, \type{\channelk{s}{r}{1}}{\mc{i}{I'}\outT'_i} \right) \leq_{\prob} \type{\channelk{s}{r}{2}}{\mc{i}{I'}\left( \mc{k}{I_i}L_k \right) + \mc{j}{J}\outT_j}}%
		{\begin{array}{c}
				\bigcup_{i\in I'}I_i \cup J \neq \emptyset\\
				\forall j \in J.\;  \exists i \in I'.\; \pre(\outT_j) = \pre(\outT'_i)
			\end{array}\quad \forall i \in I'.\left(%
			\begin{array}{r}
				\channelk{s}{r}{1} \cdot \left( \Delta, \type{\channelk{s}{r}{1}}{\outT'_i} \right)\\
				\leq_{\prob} \type{\channelk{s}{r}{2}}{\mc{k}{I_i}L_k}
			\end{array}\right)}\\[+0.25em]
		%
		% [S-oplus] splitting probabilities
		%
		\infer=[\sidenote{[S-$\oplus$]}]{\channelk{s}{r}{1} \cdot \left( \Delta, \type{\channelk{s}{r}{1}}{\send{i}{I}}\probOut{\prob'_i}{H'_i \cont T'_i} \right) \leq_{\prob^{\Sigma}} \type{\channelk{s}{r}{2}}{\send{j}{J} \probOut{\prob_j}{H_j \cont T_j}}}%
		{\begin{array}{c}
				J = \bigcup_{i \in I}J_i \neq \emptyset \quad \sum_{j\in J}\prob_j = \prob^{\Sigma} \\
				\forall i\neq j\in I .\; J_i \cap J_j = \emptyset
			\end{array}
			\quad \forall i \in I.
			\left(\begin{array}{r}
				\channelk{s}{r}{1} \cdot \left( \Delta, \type{\channelk{s}{r}{1}}{H'_i \cont T'_i} \right)\\
				\leq_{\prob'_i\prob^{\Sigma}} \type{\channelk{s}{r}{2}}{\send{j}{J_i} \probOut{\prob_j}{H_j \cont T_j}}
			\end{array}\right)}\\[+0.25em]
		%
		% [S-$\inpT$]
		%
		\infer=[\sidenote{[S-$\inpT$]}]{\channelk{s}{r}{1} \cdot \left( \Delta, \type{\channelk{s}{r}{1}}{\inp{q}{p}{l}{U}{T'}} \right) \leq_1 \type{\channelk{s}{r}{2}}{\inp{q}{p}{l}{U}{T}}}%
		{\vecparticipantk{r}{1} \subseteq \vecparticipantk{r}{2} \quad \participant{q} \in \vecparticipantk{r}{2} \quad \participant{p} \notin \Delta \quad \Delta, \type{\channelk{s}{r}{1}}{T'} \leq_1 \type{\channelk{s}{r}{2}}{T}}\\[+0.25em]
		%
		% [S-$\outT$]
		%
		\infer=[\sidenote{[S-$\outT$]}]{\channelk{s}{r}{1} \cdot \left( \Delta, \type{\channelk{s}{r}{1}}{\out{p}{q}{l}{U}{T'}} \right) \leq_{\prob} \type{\channelk{s}{r}{2}}{\probOut{\prob}{\out{p}{q}{l}{U}{T}}}}%
		{\vecparticipantk{r}{1} \subseteq \vecparticipantk{r}{2} \quad \participant{p} \in \vecparticipantk{r}{2} \quad \participant{q} \notin \Delta \quad \Delta, \type{\channelk{s}{r}{1}}{T'} \leq_1 \type{\channelk{s}{r}{2}}{T}}\\[+0.25em]
		%
		% [S-Link]
		%
		\infer=[\sidenote{[S-Link]}]{\channelk{s}{r}{1}\cdot \left( \Delta, \type{\channelk{s}{r}{1}}{\out{p}{q}{l}{U}{T'_1}}, \type{\channelk{s}{r}{2}}{\inp{q}{p}{l}{U}{T'_2} + \mc{i}{I} L_i}\right) \leq_{\prob} \type{\channel{s}{r}}{T}}%
		{\participant{p} \in \vecparticipantk{r}{1} \quad \participant{q} \in \vecparticipantk{r}{2} \quad \Delta, \type{\channelk{s}{r}{1}}{T'_1}, \type{\channelk{s}{r}{2}}{T'_2}\leq_{\prob} \type{\channel{s}{r}}{T}}\\[+0.25em]
		%
		% [Sub-M-e-L] [Sub-M-e-R]
		%
		\infer=[\sidenote{[S-$\nonsend$-L]}]{\channelk{s}{r}{1} \cdot \left( \Delta, \type{\channelk{s}{r}{1}}{\nonsend \cont T'} \right) \leq_{\prob} \type{\channelk{s}{r}{2}}{T}}%
		{\Delta, \type{\channelk{s}{r}{1}}{T'} \leq_{\prob} \type{\channelk{s}{r}{2}}{T}}\qquad\!
		\infer=[\sidenote{[S-$\nonsend$-R]}]{\Lambda \leq_{\prob} \type{\channel{s}{r}}{\probOut{\prob}{\nonsend} \cont T}}%
		{\Lambda \leq_{1} \type{\channel{s}{r}}{T}}\\[+0.25em]
		%
		% recursion
		%
%		\infer=[\sidenote{[S-$ \mu $-L]}]{\Delta, \type{c}{{\left( \mu t \right)} T'} \leq_1 \type{c}{T}}%
%		{\Delta, \type{c}{T'\subst{t}{{\left( \mu t \right)} T'}} \leq_1 \type{c}{T}} \qquad\kern-5pt%
%		%
%		\infer=[\sidenote{[S-$ \mu $-R]}]{\Delta, \type{c}{T'} \leq_1 \type{c}{{\left( \mu t \right)} T}}%
%		{\Delta, \type{c}{T'} \leq_1 \type{c}{T\subst{t}{{\left( \mu t \right)} T}}}\\[+0.25em]
		%
		% [S-emptyset]
		%
		\infer=[\sidenote{[S-$\emptyset$-1]}]{\emptyset \leq_1 \emptyset}%
		{}\qquad\!
		\infer=[\sidenote{[S-$\emptyset$]}]{c\cdot \Delta \leq_\prob \emptyset}%
		{\stuck(c\cdot \Delta)}\qquad\!
		%
		% [S-Split]
		%
		\infer=[\sidenote{[S-Split]}]{\Delta'_1, \Delta'_2 \leq_1 \Delta, \type{c}{T}}%
		{\Delta'_1 \leq_1 \Delta \quad \Delta'_2 \leq_1 \type{c}{T}}
	\end{gather*}
\end{definition}
Akin to single-channel subtyping, the first five rules may be understood as one large rule, as the order of these rules within a derivation is fixed.
Among these five rules, [S-$\Sigma$-1] is applied first (lowermost in the derivation tree) and [S-$\oplus$] last (uppermost).
As we are now handling several channels in the refinement, the new branching rule [S-$\Sigma$-1] introduces the aforementioned \emph{active channel} to produce branches for each channel in the refinement.
No other rule sets an active channel.
We have $ \Delta = \type{\channelk{s}{r}{1}}{T_1}, \ldots,  \type{\channelk{s}{r}{n}}{T_n} \leq_{\prob} \type{\channel{s}{r}}{\mc{j}{J_1} L_j + \ldots + \mc{j}{J_n} L_j}$ if there is a way to split the choice of the supertype along the $n$ channels of the refinement.
Therefore, [S-$ \Sigma $-1] creates $ n $ branches, where in each branch exactly one of the $ n $ channels is set active $ \channelk{s}{r}{1} \cdot \Delta $.
Intuitively, $ \channelk{s}{r}{k} \cdot \Delta $ is used to identify the next possible action of $ \channelk{s}{r}{k} $ if it exists.
Else $ \channelk{s}{r}{k} $ is pending (see below) and an empty choice $ \type{\channel{s}{r}}{\sum_{j \in J_k} L_j} $ with $ J_k $ is created.
The side condition $ \bigcup_{k \in \left[1..n\right]} J_k \neq \emptyset $ ensures that not all $ J_k $ are empty.
For each (non-pending) channel in $\Delta$, the derivation tree in the respective branch above [S-$\Sigma$-1] creates the corresponding component of the mixed choice of $\channel{s}{r}$.
Using an active channel and branching in this way, allows us to separate the splitting of a choice onto several rules.

After an active channel is set, the remaining four splitting rules function essentially like those of the single-channel subtyping (Definition~\ref{def:subtypingCompositionalOneChannel}), while keeping the active channel on the left in the precondition.
Rule [S-$\Sigma$-2], corresponds to \ssigmamedium{}, [S-$\Sigma$-$\inpT$] and [S-$\Sigma$-$\outT$] to \ssigmainmedium{} and \ssigmaoutmedium{}, respectively, and [S-$\oplus$] to \soplusmedium{}.
For the latter, the probabilistic choice rules, the purpose of ``storing'' the probability $\prob_i'$ in the relation was not apparent thus far.
Essentially, when several steps of purely internal interaction between channels of the refinement occur, these several steps can be aggregated into one step, whose probability is the product of the individual steps.\medbreak

Again, the remaining rules, [S-$\inpT$] to [S-Split], contain the core of the actual subtyping behaviour; in most of these rules subtype contexts have an active channel whose type is a singular input/output action.
The input rule, [S-$\inpT$], is largely the same as \sinmedium{}: A single input action is the supertype of a context if its active channel offers the same single action and the continuations are subtypes within the local context \emph{without} an active channel.
As in \sinmedium{}, both the acting role $\participant{q}$ and the subtype role set $\vecparticipantk{r}{1}$ must be contained in the supertype role set $\vecparticipantk{r}{2}$.
Note that the communication partner $\participant{p}$ of this input must not be found within $\Delta$.
Intuitively, this means that only ``external'' inputs can occur in the interface type; input actions whose communication partner is within the refinement are handled by [S-Link] or [S-$\emptyset$].
Rule [S-$\outT$] is similar; its parallel is \soutmedium{}.
Here again, it matches the probability on the relation $\leq_{\prob}$ in the conclusion to the probability of the sending action of the supertype, $\type{\channelk{s}{r}{2}}{\probOut{\prob}{\out{p}{q}{l}{U}{T}}}$.
For the premise then $ \leq_1 $ is used.

[S-Link] is an entirely new rule, which unfolds internal communication within the context on the left, if the active channel offers the sending action.
The respective communication is not visible in the supertype.
Similar to [S-$\inpT$] and [S-$\outT$], there is no active channel in the premise.

Similarly, [S-$\nonsend$-L] allows internal actions of the active channel of the subtype context to unfold.
[S-$\nonsend$-R] allows internal actions of the supertype to unfold.
They expand on the \snonsendmedium{} rule: Making two rules out of one allows the supertype to contain internal actions independently from those of the subtype context.
Indeed, sometimes a supertype is required to contain an internal action that does not correspond to an internal action in the subtype.
Consider the case $\Delta \leq_{1} \Delta'$, where $\Delta'$ has a chain of internal communications $\Delta \mapsto^{*}_{\prob} \Delta'$ leading to a context in which an external input $\Delta' \xrightarrow{s:\inpPrefix{p}{q}{l}{U}}$ with $ \participant{q} \notin \Delta' $ is available.
Then the supertype $\Delta$ needs a transition $\Delta \xrightarrow{s:\nonsend}_{\prob}$.
[S-$\nonsend$-R] matches the probabilities of the relation and the action similar to [S-$\outT$].
[S-$ \emptyset $-1] is equivalent to \semptysetmedium{}.

By the new rule [S-$ \emptyset $], the $ \tend $ type is also a supertype of a pending context.
In [S-$\Sigma$-1], [S-$\Sigma$-2], [S-$\Sigma$-$\inpT$], and [S-$\Sigma$-$\outT$], some of the constructed parts of the choice in the supertype may be empty.
This happens if the subtype context with its active channel is pending, $\Stuck{c \cdot \Delta}$.
An action of $c_1 \cdot \Delta$ may be pending, because its communication partner on $ c_2 $ might need to first \emph{link} with some $ c_3 $.
Note that internal actions, treated by [S-Link] and [S-$\nonsend$-L], are resolved only in branches, where the sender is the active channel.
Hence, all internal inputs are pending, because they are handled in the branch of the sender.
Sending actions may also be pending, if the corresponding receiving action is not available yet.
\begin{definition}[Pending]
	\label{def:stuck}
	\label{def:pending}
	A local typing context with an active channel $\Lambda = \channel{s}{p} \cdot \Delta$ is pending, $\Stuck{\Lambda}$, if $\type{\channel{s}{p}}{\mc{i}{I}\inpT_i + \mc{j}{J} \outT_j} \in \Delta$ and
	\begin{enumerate}
		\item for all $ i \in I$ have $\inpT_i = \inpk{p}{q}{l_i}{U_i}{T_i}{i}$ for which exists $\type{\channel{s}{q}}{T} \in \Delta$ with $\participant{q}_i \in \vecparticipant{q}$, and
			\item for all $j \in J$ have $\outT_j = \outk{p}{q}{l_k}{U_k}{T_k}{k} \oplus \choiceProb_j$ for which exists $\type{\channel{s}{q}}{T} \in \Delta$ with $\participant{q}_k \in \vecparticipant{q}$ and
			$T \neq \inpk{q}{p}{l_i}{U_i}{T'_i}{i} + \choice$.
	\end{enumerate}
\end{definition}
In other words, $\Stuck{c\cdot \Delta}$ if the type of $c$ is a mixed choice, where all inputs seek to communicate with each some partner from $\Delta$ which is not available, and all probabilistic choices contain an output seeking to communicate with a partner from $\Delta$ which is not available.

Finally, [S-Split] splits contexts such that there is a single channel on the right-hand side of the subtyping relation.
In contrast to \ssplitmedium{}, the left-hand side may now be a collection of typed channels, a subtype context.
This rule facilitates reflexivity and transitivity of $ \leq_1 $.\medbreak

The subtyping rules ensure that all roles on which actions are performed in the refinement appear in the role set used in the interface---or role sets for more than one channel in the interface.
This may appear as a limitation of our approach, since the interface needs to know about these roles even if it has no further knowledge about the inner structure of the refinement.
However, this design choice was made to simplify the presentation of subtyping and is not crucial.
Instead we may use a new role (for each channel) in the interface, let [S-Split] create a mapping between refinement roles and their corresponding interface role for the bookkeeping in the remaining subtyping rules, and replace in rules such as [S-$ \inpT $] each refinement role in actions on the left by the corresponding interface role in the actions on the right.\medbreak

We will now show that single-channel subtyping subsumes multi-channel subtyping, similar to Theorem~\ref{thm:mediumSubtypingIncludesSimple}.
\begin{theorem}
\label{thm:hardSubtypingIncludesMedium}
    If $\Delta' \leqmedium{\prob} \Delta$ then $\Delta' \leq_{\prob} \Delta$.
\end{theorem}

For the proof, we will utilise the following lemma for dealing with active channels.
\begin{lemma}
\label{lem:activechanneliffnoactivechannel}
    % For non-empty mixed choice type $T = \mc{i}{I} L_i$, it holds that $c' \cdot \left( \type{c'}{T'} \right) \leq_{\prob} \type{c}{T}$ if and only if $\type{c'}{T'} \leq_{\prob} \type{c}{T}$.
    $\type{c'}{T'} \leq_{\prob} \type{c}{T}$ implies $c' \cdot \left( \type{c'}{T'} \right) \leq_{\prob} \type{c}{T}$.
\end{lemma}
\begin{proof}
    % We show the forward and backward direction separately.\todo{fix proof, T always mixed choice type}
    % \begin{description}
    %     \item[``$\Rightarrow$'':]%
    %     By [S-$\Sigma$-1] of Definition~\ref{def:subtypingCompositional}.
    %     \item[``$\Leftarrow$'':]%
        By induction on the derivation of $\type{c'}{T'} \leq_{\prob} \type{c}{T}$.
        Proceed by case distinction on the last applied rule, we may consider only those cases in which the conclusion has no active channel.
        \begin{description}
            \item[Case {[S-$\Sigma$-1]}:]%
            Clear.
            \item[Case {[S-$\nonsend$-R]}:]%
            Have $\type{c'}{T'} \leq_{\prob} \type{c}{\probOut{{\prob}}{\nonsend}\cont T}$ and $\type{c'}{T'} \leq_{1} \type{c}{T}$.
            By induction hypothesis $c' \cdot \left( \type{c'}{T'} \right) \leq_{1} \type{c}{T}$.
            By [S-$\nonsend$-R] thus $c' \cdot \left( \type{c'}{T'} \right) \leq_{\prob} \type{c}{\probOut{{\prob}}{\nonsend}\cont T}$ as required.
            \item[Case {[S-$\emptyset$-1]}:]%
            Have $\emptyset \leq_{\prob} \emptyset$ with ${\prob} = 1$, thus $\type{c'}{\tend} = \emptyset$.
            Since $\Stuck{c' \cdot \left(\type{c'}{\tend} \right)}$ holds vacuously, from [S-$\emptyset$] we obtain ${c' \cdot \left( \emptyset \right)} \leq_{\prob} \emptyset$ as required.
            \item[Case {[S-Split]}:]%
            By induction hypothesis.\qedhere
        \end{description}
    % \end{description}
\end{proof}

\begin{proof}[Proof of Theorem~\ref{thm:hardSubtypingIncludesMedium}]
    Assume that $\Delta' \leqmedium{\prob} \Delta$.
    We show $\Delta' \leq_{\prob} \Delta$ by structural induction on the derivation of $\Delta' \leqmedium{\prob} \Delta$.
    We proceed by case distinction on the last applied rule.
    \begin{description}
        \item[Case {\ssigmamedium{}}:]
        Have $\Delta' \kern-1pt= \type{\channelk{s}{r}{1}}{\mc{i}{I'}\inpT_i + \kern-1pt\mc{j}{J'}\outT_j}$ and $\Delta = \type{\channelk{s}{r}{2}}{\mc{i}{I}L_i + \kern-1pt\mc{j}{J}L_j}$ with $\type{\channelk{s}{r}{1}}{\mc{i}{I'}\inpT_i +\kern-1pt \mc{j}{J'}\outT_j} \leqmedium{{\prob}} \type{\channelk{s}{r}{2}}{\mc{i}{I}L_i +\kern-1pt \mc{j}{J}L_j}$, $\type{\channelk{s}{r}{1}}{\mc{i}{I'}\inpT_i} \leqmedium{{\prob}} \type{\channelk{s}{r}{2}}{\mc{i}{I}L_i}$, $\type{\channelk{s}{r}{1}}{\mc{j}{J'}\outT_j} \leqmedium{{\prob}} \type{\channelk{s}{r}{2}}{\mc{j}{J}L_j}$, and $I \cup J \neq \emptyset$.
        From the induction hypothesis get $\type{\channelk{s}{r}{1}}{\mc{i}{I'}\inpT_i} \leq_{{\prob}} \type{\channelk{s}{r}{2}}{\mc{i}{I}L_i}$ and $\type{\channelk{s}{r}{1}}{\mc{j}{J'}\outT_j} \leq_{{\prob}} \type{\channelk{s}{r}{2}}{\mc{j}{J}L_j}$.
        By Lemma~\ref{lem:activechanneliffnoactivechannel}, we obtain $\channelk{s}{r}{1} \cdot \left( \type{\channelk{s}{r}{1}}{\mc{i}{I'}\inpT_i} \right) \leq_{{\prob}} \type{\channelk{s}{r}{2}}{\mc{i}{I}L_i}$ and $\channelk{s}{r}{1} \cdot \left( \type{\channelk{s}{r}{1}}{\mc{j}{J'}\outT_j} \right) \leq_{{\prob}} \type{\channelk{s}{r}{2}}{\mc{j}{J}L_j}$.
        Since also $I \cup J \neq \emptyset$ we can thus apply [S-$\Sigma$-2] of Definition~\ref{def:subtypingCompositional} to obtain $\channelk{s}{r}{1} \cdot \left( \type{\channelk{s}{r}{1}}{\mc{i}{I'}\inpT_i + \mc{j}{J'}\outT_j} \right) \leq_{{\prob}} \type{\channelk{s}{r}{2}}{\mc{i}{I}L_i + \mc{j}{J}L_j}$.
        With this and $I \cup J \neq \emptyset$ we can use [S-$\Sigma$-1] of Definition~\ref{def:subtypingCompositional} to obtain $\type{\channelk{s}{r}{1}}{\mc{i}{I'}\inpT_i + \mc{j}{J'}\outT_j} \leq_{{\prob}} \type{\channelk{s}{r}{2}}{\mc{i}{I}L_i + \mc{j}{J}L_j}$ as required.
        \item[Case {\ssigmainmedium{}}:]%
        Have $\Delta' = \type{\channelk{s}{r}{1}}{\mc{i}{I' \cup J'}\inpT'_i}$ and $\Delta = \type{\channelk{s}{r}{2}}{\mc{i}{I'} \left( \mc{k}{I_i}L_k \right)}$ for which $\type{\channelk{s}{r}{1}}{\mc{i}{I' \cup J'}\inpT'_i} \leqmedium{\prob} \type{\channelk{s}{r}{2}}{\mc{i}{I'} \left( \mc{k}{I_i}L_k \right)}$, $\forall i \in I'.\; \type{\channelk{s}{r}{1}}{\inpT'_i} \leqmedium{\prob} \type{\channelk{s}{r}{2}}{\mc{k}{I_i}L_k}$, with $\bigcup_{i\in I'}I_i \neq \emptyset$, and $\forall j \in J'.\; \exists i\in I'.\; pre(\inpT'_j)=\pre(\inpT'_i)$.
        Apply induction hypothesis to obtain $\forall i \in I'.\; \type{\channelk{s}{r}{1}}{\inpT'_i} \leq_{\prob} \type{\channelk{s}{r}{2}}{\mc{k}{I_i}L_k}$.
        Thus $\forall i \in I'.\; \channelk{s}{r}{1} \cdot \left( \type{\channelk{s}{r}{1}}{\inpT'_i} \right) \leq_{\prob} \type{\channelk{s}{r}{2}}{\mc{k}{I_i}L_k}$ by Lemma~\ref{lem:activechanneliffnoactivechannel}.
        With this, $\bigcup_{i\in I'}I_i \neq \emptyset$, and $\forall j \in J'.\; \exists i\in I'.\; pre(\inpT'_j)=\pre(\inpT'_i)$, we can apply [S-$\Sigma$-$\inpT$] and [S-$\Sigma$-1] of Definition~\ref{def:subtypingCompositional} to conclude similarly to Case \ssigmamedium{}.
        \item[Case {\ssigmaoutmedium{}:}]%
        In this case we have the two local contexts $\Delta' = \type{\channelk{s}{r}{1}}{\mc{i}{I'}\outT'_i}$ and $\Delta = \type{\channelk{s}{r}{2}}{\mc{i}{I'}\left( \mc{k}{I_i}L_k \right) + \mc{j}{J}\outT_j}$, for which also $\type{\channelk{s}{r}{1}}{\mc{i}{I'}\outT'_i} \leqmedium{\prob} \type{\channelk{s}{r}{2}}{\mc{i}{I'}\left( \mc{k}{I_i}L_k \right) + \mc{j}{J}\outT_j}$, and $\forall i \in I'.\; \type{\channelk{s}{r}{1}}{\outT'_i} \leqmedium{\prob} \type{\channelk{s}{r}{2}}{\mc{k}{I_i}L_k}$, where $\bigcup_{i\in I'}I_i \cup J \neq \emptyset$, and $\forall j \in J.\;  \exists i \in I'.\; \pre(\outT_j) = \pre(\outT'_i)$.
        Similar to previous cases, we apply the induction hypothesis and Lemma~\ref{lem:activechanneliffnoactivechannel} to $\forall i \in I'.\; \type{\channelk{s}{r}{1}}{\outT'_i} \leqmedium{\prob} \type{\channelk{s}{r}{2}}{\mc{k}{I_i}L_k}$ to obtain $\forall i \in I'.\; \channelk{s}{r}{1} \cdot \left( \type{\channelk{s}{r}{1}}{\outT'_i} \right) \leq_{\prob} \type{\channelk{s}{r}{2}}{\mc{k}{I_i}L_k}$.
        Analogous to Case \ssigmainmedium{}, we conclude by [S-$\Sigma$-$\outT$] and [S-$\Sigma$-1] of Definition~\ref{def:subtypingCompositional}.
        \item[Case {\soplusmedium{}}:]%
        Have $\Delta' = \type{\channelk{s}{r}{1}}{\send{i}{I}}\probOut{\prob'_i}{H'_i \cont T'_i}$ and $\Delta = \type{\channelk{s}{r}{2}}{\send{j}{J} \probOut{\prob_j}{H_j \cont T_j}}$ where $\type{\channelk{s}{r}{1}}{\send{i}{I}}\probOut{\prob'_i}{H'_i \cont T'_i} \leqmedium{\prob^{\Sigma}} \type{\channelk{s}{r}{2}}{\send{j}{J} \probOut{\prob_j}{H_j \cont T_j}}$, and $\forall i \in I.\; \type{\channelk{s}{r}{1}}{H'_i \cont T'_i} \leqmedium{\prob'_i\prob^{\Sigma}} \type{\channelk{s}{r}{2}}{\send{j}{J_i} \probOut{\prob_j}{H_j \cont T_j}}$, and $J = \bigcup_{i \in I}J_i \neq \emptyset$, $\sum_{j\in J}\prob_j = \prob^{\Sigma}$, and $\forall i\neq j\in I .\; J_i \cap J_j = \emptyset$.
        Similar to previous cases, we apply the induction hypothesis and Lemma~\ref{lem:activechanneliffnoactivechannel} to $\forall i \in I.\; \type{\channelk{s}{r}{1}}{H'_i \cont T'_i} \leqmedium{\prob'_i\prob^{\Sigma}} \type{\channelk{s}{r}{2}}{\send{j}{J_i} \probOut{\prob_j}{H_j \cont T_j}}$ to obtain $\forall i \in I.\; \channelk{s}{r}{1} \cdot \left( \type{\channelk{s}{r}{1}}{H'_i \cont T'_i} \right) \leq_{\prob'_i\prob^{\Sigma}} \type{\channelk{s}{r}{2}}{\send{j}{J_i} \probOut{\prob_j}{H_j \cont T_j}}$.
        Analogous to Case \ssigmainmedium{}, we conclude by [S-$\oplus$] and [S-$\Sigma$-1] of Definition~\ref{def:subtypingCompositional}.
        \item[Case {\sinmedium{}}:]%
        We have $\Delta' = \type{\channelk{s}{r}{1}}{\inp{q}{p}{l}{U}{T'}}$ and $\Delta = \type{\channelk{s}{r}{2}}{\inp{q}{p}{l}{U}{T}}$, for which $\type{\channelk{s}{r}{1}}{\inp{q}{p}{l}{U}{T'}} \leqmedium{1} \type{\channelk{s}{r}{2}}{\inp{q}{p}{l}{U}{T}}$, and $\type{\channelk{s}{r}{1}}{T'} \leqmedium{1} \type{\channelk{s}{r}{2}}{T}$, where $\participant{q} \in \vecparticipantk{r}{2}$ and $\vecparticipantk{r}{1} \subseteq \vecparticipantk{r}{2}$.
        We can apply the induction hypothesis to obtain $\type{\channelk{s}{r}{1}}{T'} \leq_1 \type{\channelk{s}{r}{2}}{T}$.
        With this and $\participant{q} \in \vecparticipantk{r}{2}$ and $\vecparticipantk{r}{1} \subseteq \vecparticipantk{r}{2}$ we can apply [S-$\inpT$] of Definition~\ref{def:subtypingCompositional} to obtain $\channelk{s}{r}{1} \cdot \left( \type{\channelk{s}{r}{1}}{\inp{q}{p}{l}{U}{T'}} \right) \leq_1 \type{\channelk{s}{r}{2}}{\inp{q}{p}{l}{U}{T}}$.
        Then by [S-$\Sigma$-1] of Definition~\ref{def:subtypingCompositional} we have $\type{\channelk{s}{r}{1}}{\inp{q}{p}{l}{U}{T'}} \leq_1 \type{\channelk{s}{r}{2}}{\inp{q}{p}{l}{U}{T}}$ as required.
        \item[Case {\soutmedium{}}:]%
        We have $\Delta' = \type{\channelk{s}{r}{1}}{\out{p}{q}{l}{U}{T'}}$ and $\Delta = \type{\channelk{s}{r}{2}}{\probOut{\prob}{\out{p}{q}{l}{U}{T}}}$ for which $\type{\channelk{s}{r}{1}}{\out{p}{q}{l}{U}{T'}}\leqmedium{\prob} \type{\channelk{s}{r}{2}}{\probOut{\prob}{\out{p}{q}{l}{U}{T}}}$, $\vecparticipantk{r}{1} \subseteq \vecparticipantk{r}{2}$, $\participant{p} \in \vecparticipantk{r}{2}$, and $\type{\channelk{s}{r}{1}}{T'} \leqmedium{1} \type{\channelk{s}{r}{2}}{T}$.
        We can apply the induction hypothesis to obtain $\type{\channelk{s}{r}{1}}{T'} \leq_1 \type{\channelk{s}{r}{2}}{T}$.
        With this, $\vecparticipantk{r}{1} \subseteq \vecparticipantk{r}{2}$, and $\participant{p} \in \vecparticipantk{r}{2}$ we can apply [S-$\outT$] of Definition~\ref{def:subtypingCompositional} to obtain $\channelk{s}{r}{1} \cdot 
        \left( \type{\channelk{s}{r}{1}}{\out{p}{q}{l}{U}{T'}} \right) \leq_{\prob} \type{\channelk{s}{r}{2}}{\probOut{\prob}{\out{p}{q}{l}{U}{T}}}$.
        Then apply [S-$\Sigma$-1] of Definition~\ref{def:subtypingCompositional} to obtain $\type{\channelk{s}{r}{1}}{\out{p}{q}{l}{U}{T'}} \leq_{\prob} \type{\channelk{s}{r}{2}}{\probOut{\prob}{\out{p}{q}{l}{U}{T}}}$ as required.
        \item[Case {\snonsendmedium{}}:]%
        We have $\Delta' = \type{c'}{\nonsend \cont T'}$ and $\Delta = \type{c}{\probOut{{\prob}}{\nonsend} \cont T}$ with $\type{c'}{\nonsend \cont T'} \leqmedium{{\prob}} \type{c}{\probOut{{\prob}}{\nonsend} \cont T}$ and $\type{c'}{T'} \leqmedium{1} \type{c}{T}$.
        From the induction hypothesis we get $\type{c'}{T'} \leq_{1} \type{c}{T}$.
        We can then apply [S-$\nonsend$-R] and [S-$\nonsend$-L] of Definition~\ref{def:subtypingCompositional} to obtain $c' \cdot \left( \type{c'}{\nonsend \cont T'} \right) \leq_{{\prob}} \type{c}{\probOut{{\prob}}{\nonsend} \cont T}$.
        By [S-$\Sigma$-1] of Definition~\ref{def:subtypingCompositional} then $ \type{c'}{\nonsend \cont T'} \leq_{{\prob}} \type{c}{\probOut{{\prob}}{\nonsend} \cont T}$ as required.
        \item[Case {\semptysetmedium{}}:]%
        Straightforward by [S-$\emptyset$-1] of Definition~\ref{def:subtypingCompositional}.
        \item[Case {\ssplitmedium{}}:]%
        Straightforward by application of the induction hypothesis and [S-Split] of Definition~\ref{def:subtypingCompositional}.\qedhere
    \end{description}
\end{proof}

Let us revisit the courthouse protocol from our running example to help build more intuition for refinement based multi-channel subtyping. 
\begin{example}[Running Example---Subtyping]
\addxcontentsline{loe}{example}[\theexample]{Running Example---Subtyping}
	\label{ex:motivatingExampleSubtyping}
    Looking at the types from Example~\ref{ex:motivatingExampleTypes}, clearly the interactions between \emph{roles} $\participant{j}$ and $\participant{d}$ are concealed to the others.
    Indeed, with the rules of Definition~\ref{def:subtypingCompositional}, we can assert that $ \type{\channelsingle{s}{j}}{T_{\participant{j}}}, \type{\channelsingle{s}{d}}{T_{\participant{d}} \leq_1 \type{\channel{s}{c}}{T_{\participant{c}}} }$, where $\vecparticipant{c} = \{\participant{j},\participant{d}\}$.

	To further illustrate subtyping, consider a more complex version of the courthouse protocol. Here, the \participant{d}efendant, after notifying the \participant{j}udge about wanting a witness statement, seeks out a meeting with the \participant{w}itness themselves. With $T_{\participant{p}}$ and $T_{\participant{w}}$ untouched, we redefine only $T_{\participant{c}}, T_{\participant{d}}$, and $T_{\participant{j}}$ as follows.
	\begin{gather*}
		T^{*}_{\participant{c}} \hspace{-1pt}\mathop{\scalebox{0.9}[1]{$=$}} \inp{j}{p}{\mathtt{lws}}{}{} \kern-2pt \begin{cases}\hspace{-1pt}
			&\probOut{0.7}{\outPrefix{j}{p}{\mathtt{glt}}{\boolT}} \cont \probOut{1}{\outPrefix{j}{d}{\mathtt{rls}}{}}\\
			\!\oplus\!\!\!\! &\probOut{0.3}{\nonsend}\cont \kern-2pt
			\begin{cases}\!\hspace{-1pt}
				&\probOut{1}{\outPrefix{d}{w}{\mathtt{mtg}}{}}\cont \inp{j}{w}{\mathtt{st}}{} \probOut{1}{\outPrefix{j}{p}{\mathtt{glt}}{\boolT}}\\
				\!+\!\!\!\! &\inp{j}{w}{\mathtt{st}}{}\probOut{1}{\outPrefix{d}{w}{\mathtt{mtg}}{}}\cont\probOut{1}{\outPrefix{j}{p}{\mathtt{glt}}{\boolT}}
			\end{cases}
		\end{cases}
		\vspace*{0.25em}\\
		T^{*}_{\participant{d}} \hspace{-1pt}\mathop{\scalebox{0.9}[1]{$=$}} \begin{cases}
			& \probOut{0.5}{\outPrefix{d}{j}{\mathtt{wk}}{}} \\
			\!\oplus\!\!\!\! & \probOut{0.2}{\outPrefix{d}{j}{\mathtt{str}}{}} \\
			\!\oplus\!\!\!\! & \probOut{0.3}{\outPrefix{d}{j}{\mathtt{wit}}{}}  \cont \probOut{1}{\outPrefix{d}{w}{\mathtt{mtg}}{}}
		\end{cases}\vspace*{0.25em}\\
		T^{*}_{\participant{j}} \hspace{-1pt}\mathop{\scalebox{0.9}[1]{$=$}} \inp{j}{p}{\mathtt{lws}}{}{} \kern-2pt
		\begin{cases}
			& \inpPrefix{j}{d}{\mathtt{wk}}{} \cont \probOut{1}{\outPrefix{j}{p}{\mathtt{glt}}{\boolT}} \cont \probOut{1}{\outPrefix{j}{d}{\mathtt{rls}}{}} \\
			\!+\!\!\!\! & \inpPrefix{j}{d}{\mathtt{str}}{} \cont \probOut{1}{\outPrefix{j}{p}{\mathtt{glt}}{\boolT}} \cont \probOut{1}{\outPrefix{j}{d}{\mathtt{rls}}{}} \\
			\!+\!\!\!\! & \inpPrefix{j}{d}{\mathtt{wit}}{} \cont \inpPrefix{j}{w}{\mathtt{st}}{} \cont \probOut{1}{\outPrefix{j}{p}{\mathtt{glt}}{\boolT}}
		\end{cases}
	\end{gather*}
	Where, as before, for $\vecparticipant{c} = \{\participant{j},\participant{d}\}$ it holds that $ \type{\channelsingle{s}{j}}{T^{*}_{\participant{j}}}, \type{\channelsingle{s}{d}}{T^{*}_{\participant{d}} \leq_1 \type{\channel{s}{c}}{T^{*}_{\participant{c}}} }$.\medbreak

	Notice how the interface now contains an \emph{internal action} $\nonsend$.
	The interface is created without knowledge of the specifics of the \participant{w}itness and therefore needs to accommodate for both the \participant{d}efendant-\participant{w}itness \emph{meeting} coming before the \participant{w}itness-\participant{j}udge \emph{statement} and vice versa.
	This leads in $ T^{*}_{\participant{c}} $ to a probabilistic choice being followed in the case with probability $ 0.3 $ by a mixed choice containing receiving actions.
	A $\nonsend$ in between is necessary to connect these choices.
	Also note, that only the interface type $ T^{*}_{\participant{c}} $ has mixed choice but not the refinement.
	\exampleDone
\end{example}

We will now present a subtyping derivation using these rules.
Instead of the types used in Example~\ref{ex:standardsubtypingderivation} and Example~\ref{ex:singlechannelsubtypingderivation}, we will here present the more complex setting from above.
\begin{example}[Running Example---Subtyping Derivation]
\addxcontentsline{loe}{example}[\theexample]{Running Example---Subtyping Derivation}
	\label{ex:subtypingRulesApplied}
	We highlight the following snippets from the subtyping derivation of $ \type{\channelsingle{s}{j}}{T^*_{\participant{j}}}, \type{\channelsingle{s}{d}}{T^*_{\participant{d}} \leq_1 \type{\channel{s}{c}}{T^*_{\participant{c}}} }$ from Example~\ref{ex:motivatingExampleSubtyping}, to illustrate our subtyping rules.
    During the derivation, we will reuse the identifiers $T_{\participant{j}}, T_{\participant{d}}, T_{\participant{c}}$ for legibility; they do not refer to the types from our previous examples.
    Moving bottom-up, we begin at the original statement to demonstrate how [S-$\Sigma$-1] handles cases in which channels are pending.
	\begin{gather*}
		\infer=[\sidenote{[S-$\Sigma$-1]}]{ \Delta = \type{\channelsingle{s}{j}}{\inp{j}{p}{\mathtt{lws}}{}{} T_{\participant{j}}}, \type{\channelsingle{s}{d}}{T^*_{\participant{d}}} \leq_1 \type{\channel{s}{c}}{\inp{j}{p}{\mathtt{lws}}{}{} T_{\participant{c}}} }%
		{\infer=[\sidenote{[S-$\inpT$]}]{\channelsingle{s}{j} \cdot \Delta \leq_1 \type{\channel{s}{c}}{\inp{j}{p}{\mathtt{lws}}{}{} T_{\participant{c}}}}{ \type{\channelsingle{s}{j}}{T_{\participant{j}}}, \type{\channelsingle{s}{d}}{T^*_{\participant{d}}} \leq_1 \type{\channel{s}{c}}{T_{\participant{c}}}} \quad \infer=[\sidenote{[S-$\emptyset$]}]{\channelsingle{s}{d} \cdot \Delta \leq_1 \emptyset}{\Stuck{\channelsingle{s}{d} \cdot \Delta }}}
	\end{gather*}
	Continuing upwards from the left-hand side of this tree, we have the following derivation.
	Note that for the probabilistic sum in the type of $\channel{s}{c}$, the case ``guilty'' with $70\%$ probability is split into a case with $50\%$ and a case with $20\%$.
	[S-$\oplus$] splits the tree into three branches.
	We give the branch that leads to [S-$\nonsend$].
	\begin{gather*}
		\infer=[\sidenote{[S-$\oplus$]}]{\channelsingle{s}{j} \cdot \left( \type{\channelsingle{s}{j}}{T_{\participant{j}}}, \type{\channelsingle{s}{d}}{\begin{cases}
					& \probOut{0.5}{\outPrefix{d}{j}{\mathtt{wk}}{}} \\
					\!\oplus\!\!\!\! & \probOut{0.2}{\outPrefix{d}{j}{\mathtt{str}}{}} \\
					\!\oplus\!\!\!\! & \probOut{0.3}{\outPrefix{d}{j}{\mathtt{wit}}{}}  \cont T'_{\participant{d}}
			\end{cases}} \right)  \leq_{1}  \type{\channel{s}{c}}{\begin{cases}
			&\probOut{0.5}{H_{\mathtt{glt}}} \cont T_{\mathtt{rls}}\\
			\!\oplus\!\!\!\! &\probOut{0.2}{H_{\mathtt{glt}}} \cont T_{\mathtt{rls}}\\
			\!\oplus\!\!\!\! &\probOut{0.3}{\nonsend}\cont T'_{\participant{c}} \end{cases} }}%
			{\ldots \quad \infer=[\sidenote{[S-Link]}]{\channelsingle{s}{j} \cdot \left( \type{\channelsingle{s}{j}}{T_{\participant{j}}}, \type{\channelsingle{s}{d}}{\outPrefix{d}{j}{\mathtt{wit}}{}  \cont T'_{\participant{d}}} \right)  \leq_{0.3}  \type{\channel{s}{c}}{\probOut{0.3}{\nonsend} \cont T'_{\participant{c}} } }%
			{\infer=[\sidenote{[S-$\nonsend$]}]{\type{\channelsingle{s}{j}}{\inpPrefix{j}{w}{\mathtt{st}}{} \cont T'_{\participant{j}}},  \type{\channelsingle{s}{d}}{T'_{\participant{d}}} \leq_{0.3}  \type{\channel{s}{c}}{\probOut{0.3}{\nonsend} \cont T'_{\participant{c}} }}%
			{\type{\channelsingle{s}{j}}{\inpPrefix{j}{w}{\mathtt{st}}{} \cont T'_{\participant{j}}},  \type{\channelsingle{s}{d}}{T'_{\participant{d}}} \leq_{1}  \type{\channel{s}{c}}{ T'_{\participant{c}} } } }}
	\end{gather*}
	Finally, continuing on the previous branch, we see how [Sub-$\Sigma$-1] assembles a mixed choice in the supertype of a subtype context without any mixed choices.
	\begin{gather*}
		\infer=[\sidenote{[S-$\Sigma$-1]}]{\Delta = \begin{array}{c}
				\type{\channelsingle{s}{j}}{\inpPrefix{j}{w}{\mathtt{st}}{} \cont T'_{\participant{j}}}, \\ \type{\channelsingle{s}{d}}{\probOut{1}{\outPrefix{d}{w}{\mathtt{mtg}}{}}}
			\end{array} \leq_{1}  \type{\channel{s}{c}}{ \begin{cases}\!
					&\probOut{1}{\outPrefix{d}{w}{\mathtt{mtg}}{}}\cont T_{1}\\
					\!+\!\!\!\! &\inp{j}{w}{\mathtt{st}}{} T_{2}
			\end{cases} } }%
		{\infer=[\sidenote{[S-$\oplus$]}]{\channelsingle{s}{d}  \cdot  \Delta \leq_{1}  \type{\channel{s}{c}}{\probOut{1}{\outPrefix{d}{w}{\mathtt{mtg}}{}}\cont T_{1}} }{\dots} \quad \infer=[\sidenote{[S-$\inpT$]}]{\channelsingle{s}{j}  \cdot  \Delta \leq_{1}  \type{\channel{s}{c}}{\inp{j}{w}{\mathtt{st}}{} T_{2} } }{\dots}}
	\end{gather*}
    \exampleDone
\end{example}

In this chapter, we have introduced the novel refinement-based multi-channel subtyping of the \emph{\calculusname{} $\pi$-calculus}.
We first introduced the idea and concept, then gave an intermediary single-channel subtyping (§~\ref{sec:singlechannelsubtyping}) to simplify the transition from standard subtyping to our approach.
Afterwards, we presented the concept of active channels and introduced the corresponding new typing context $\Lambda$, needed for keeping track of branches within subtyping derivations (§~\ref{sec:adaptedTypingSystem}).
Finally, we defined the main subtyping.
To affirm the readers intuition, we proved two theorems on the size of the different subtypings in relation to each other.
Firstly, Theorem~\ref{thm:mediumSubtypingIncludesSimple}, stating that types which are related according to standard subtyping $\leqsimple$ are also related according to single-channel subtyping (with probability one) $\leqmedium{1}$.
Secondly, Theorem~\ref{thm:hardSubtypingIncludesMedium} stating the same for single-channel subtyping $\leqmedium{\prob}$ and multi-channel subtyping $\leq_{\prob}$.
Along the way, we gave several examples to aid understanding and highlight important concepts.

Conceptually, the presented theory is already quite interesting; in the upcoming chapter, however, we will prove just how powerful and functional the system actually is. 

\chapter{Properties}
\label{chap:properties}
MPST systems can enforce many desirable properties on the run-time behaviour of processes through the typing contexts which justify them.
This chapter is dedicated to proving such properties, fulfilled by the multi-channel subtyping system and the processes of the \emph{\calculusname\ $\pi$-calculus}.
We show both crucial standard properties, such as subject reduction (§~\ref{subsec:subjectReduction}) and deadlock-freedom (§~\ref{subsec:errorAndDeadlockFreedom}), as well as properties unique to our system.
The latter includes a strong result on flexibility which separates our work from the contributions of \citep{DBLP:conf/concur/Horne20}: \emph{Any} safe and deadlock-free local context has a interface which is a \emph{single} channel type (§~\ref{subsec:interfaceExists}).
To show which good behaviour we can enforce on our processes through their typing, we first need to focus on the types themselves.

\section{Properties of Types and Subtyping}
\label{sec:propertiesOfSubtyping}
We begin with properties of \emph{\calculusname{} types} and their multi-channel subtyping.
Three main results are presented in this section: \begin{inparaenum}[(1)]
    \item The subtyping relation (with probability one) $\leq_{1}$ is a preorder.
    \item Preservation of safety and deadlock-freedom under subtyping and labelled transitions for local contexts with an active channel $\Lambda$.
    \item Any local context $\Delta$ has an interface which is a single channel.
\end{inparaenum}

First, we state three preliminary properties which are used not only for the larger proofs of this section, but also subject reduction in Section~\ref{sec:propertiesOfTypedProcesses}.
From here forward, instead of regular commas `` , '', we sometimes use `` ; '' for punctuation to avoid confusion with composition of local contexts $\Delta_1 , \Delta$.\medbreak

The following result states that an empty context can only be the supertype of an empty context, where the probability of the relation is 1. 
\begin{lemma}
	\label{lem:absorbingEndInSubtypingR}
	$ \Lambda \leq_{\prob} \emptyset $ implies $ \Lambda = \emptyset $ and $ \prob = 1 $.
\end{lemma}

\begin{proof}
	Note that the side condition $ \bigcup_{k \in \left[1..n\right]} J_k \neq \emptyset $ forbids for applications of [S-$ \Sigma $-1] on empty local contexts at the right of $ \leq_{\prob} $.
	Because of that, it is impossible to set an active channel on the left of $ \leq_{\prob} $ if the right is empty.
	By the subtyping rules, then the only rule that can be applied to derive $ \Lambda \leq_{\prob} \emptyset $ is [S-$ \emptyset $-1].
	By [S-$ \emptyset $-1], then $ \Lambda = \emptyset $ and $ \prob = 1 $.
	%\hfill $ \square $
\end{proof}
Note that the opposite direction of the above lemma is not true, \ie $ \emptyset \leq_{\prob} \Delta $ does not imply $ \Delta = \emptyset $ (not even for $ \prob = 1 $).
The subtyping rule [S-$ \nonsend $-R] allows for $ \nonsend $ in $ \Delta $.
However, $ \emptyset \leq_1 \Delta $ implies that all types in $ \Delta $ are constructed from $ \bigoplus $, $ \nonsend $, and $ \tend $ only.\medbreak

Lemma~\ref{lem:splitSubtyping} and Lemma~\ref{lem:composeSubtyping} state that the subtyping relation is preserved if the involved local contexts are split or composed if the probability of the relation is one.
\begin{lemma}
	\label{lem:splitSubtyping}
	$ \Delta \leq_1 \Delta_1, \Delta_2 $ implies that there are $ \Delta_1' $ and $ \Delta_2' $ such that $ \Delta = \Delta_1', \Delta_2' $; $ \Delta_1' \leq_1 \Delta_1 $, and $ \Delta_2' \leq_1 \Delta_2 $.
\end{lemma}

\begin{proof}
	Assume $ \Delta \leq_1 \Delta_1, \Delta_2 $.
	We proceed by induction on the number of assignments in $ \Delta_2 $.
	\begin{description}
		\item[Base Case $ \Delta_2 = \emptyset $:] Then we can choose $ \Delta_1' = \Delta $ and $ \Delta_2' = \emptyset $.
			Clearly, then $ \Delta = \Delta_1', \Delta_2' $.
			Then $ \Delta \leq_1 \Delta_1, \Delta_2 $ implies $ \Delta_1' \leq_1 \Delta_1 $.
			Finally, by [S-$ \emptyset $-1], then $ \Delta_2' \leq_1 \Delta_2 $.
		\item[Induction Step $ \Delta_2 = \Delta_3, \type{c}{T} $:] By the subtyping rules, then the only rule that can be applied in the bottom of the derivation tree for $ \Delta \leq_1 \Delta_1, \Delta_2 $ is then [S-Split].
			By [S-Split], then $ \Delta = \Delta_{1, 1}, \Delta_{1, 2} $ such that $ \Delta_{1, 1} \leq_1 \Delta_1, \Delta_3 $ and $ \Delta_{1, 2} \leq_1 \type{c}{T} $.
			By the induction hypothesis, then $ \Delta_{1, 1} \leq_1 \Delta_1, \Delta_3 $ implies that there are $ \Delta_1' $ and $ \Delta_3' $ such that $ \Delta_{1, 1} = \Delta_1', \Delta_3' $; $ \Delta_1' \leq_1 \Delta_1 $, and $ \Delta_3' \leq_1 \Delta_3 $.
			By [S-Split], then $ \Delta_3' \leq_1 \Delta_3 $ and $ \Delta_{1, 2} \leq_1 \type{c}{T} $ imply $ \Delta_3', \Delta_{1, 2} \leq_1 \Delta_2 $.
			By choosing $ \Delta_2' = \Delta_3', \Delta_{1, 2} $, then  $ \Delta = \Delta_{1, 1}, \Delta_{1, 2} $ and $ \Delta_{1, 1} = \Delta_1', \Delta_3' $ imply $ \Delta = \Delta_1', \Delta_2' $.
			We already have $ \Delta_1' \leq_1 \Delta_1 $.
			Finally, $ \Delta_2' = \Delta_3', \Delta_{1, 2} $ and $ \Delta_3', \Delta_{1, 2} \leq_1 \Delta_2 $ imply $ \Delta_2' \leq_1 \Delta_2 $.\qedhere
	\end{description}
\end{proof}

\begin{lemma}
	\label{lem:composeSubtyping}
	If $ \Delta_1' \leq_1 \Delta_1 $ and $ \Delta_2' \leq_1 \Delta_2 $ such that the compositions $ \Delta_1', \Delta_2' $ and $ \Delta_1, \Delta_2 $ are defined then $ \Delta_1', \Delta_2' \leq_1 \Delta_1, \Delta_2 $.
\end{lemma}

\begin{proof}
	Assume $ \Delta_1' \leq_1 \Delta_1 $ and $ \Delta_2' \leq_1 \Delta_2 $ and that the compositions $ \Delta_1', \Delta_2' $ and $ \Delta_1, \Delta_2 $ are defined.
	We show $ \Delta_1', \Delta_2' \leq_1 \Delta_1, \Delta_2 $ by induction on the number of assignments in $ \Delta_2 $.
	\begin{description}
		\item[Base Case $ \Delta_2 = \emptyset $:] By Lemma~\ref{lem:absorbingEndInSubtypingR}, then $ \Delta_2' \leq_1 \Delta_2 $ implies $ \Delta_2' = \emptyset $.
			Then $ \Delta_1' \leq_1 \Delta_1 $ implies $ \Delta_1', \Delta_2' \leq_1 \Delta_1, \Delta_2 $.
		\item[Induction Step $ \Delta_2 = \Delta_3, \type{c}{T} $:] By the subtyping rules, then the only rule that can be applied in the bottom of the derivation tree for $ \Delta_2' \leq_1 \Delta_2 $ is then [S-Split].
			By [S-Split], then $ \Delta_2' = \Delta_{2, 1}', \Delta_{2, 2}' $ such that $ \Delta_{2, 1}' \leq_1 \Delta_3 $ and $ \Delta_{2, 2}' \leq_1 \type{c}{T} $.
			Since the compositions $ \Delta_1', \Delta_2' $ and $ \Delta_1, \Delta_2 $ are defined, so are the compositions $ \Delta_1', \Delta_{2, 1}' $ and $ \Delta_1, \Delta_3 $.
			By the induction hypothesis, then $ \Delta_1' \leq_1 \Delta_1 $ and $ \Delta_{2, 1}' \leq_1 \Delta_3 $ imply $ \Delta_1', \Delta_{2, 1}' \leq_1 \Delta_1, \Delta_3 $.
			By [S-Split], then $ \Delta_1', \Delta_{2, 1}' \leq_1 \Delta_1, \Delta_3 $ and $ \Delta_{2, 2}' \leq_1 \type{c}{T} $ imply $ \Delta_1', \Delta_2' \leq_1 \Delta_1, \Delta_2 $.\qedhere
	\end{description}
\end{proof}

\subsection{Subtyping is a Preorder}
As is classic for subtyping relations (see \citep{DBLP:books/daglib/0005958}), our $\leq_1$-relation, too, is a preorder and towards this, we show reflexivity and transitivity separately.
We restrict this statement to the relation with probability one, not for general probabilities.
This is not to say that considerations of especially transitivity with non-one probabilities would not be interesting.
For $\Delta_1 \leq_{\prob_1} \Delta_2 \leq_{\prob_2} \Delta_3$ with $\prob_1 \neq 1$ and $\prob_2 \neq 1$ the probability $\prob_3$ for $\Delta_1 \leq_{\prob_3} \Delta_3$ is likely the product $\prob_3 = \prob_1\prob_2$.
However, the usefulness of such a statement is limited in the current setting; any ``classic'' usage of subtyping will be with $\leq_1$, such as in the typing rules.
Clearly, we have to restrict the usage of the relation to local contexts \emph{without} an active channel, as supertype contexts never have an active channel.
We argue that this is hardly a restriction, as active channels are merely a syntactic tool to help with the legibility of our subtyping rules.\medbreak

The proof of reflexivity is rather simple thanks to the [S-Split] rule ensuring that the right-hand side will always be a single channel for the remainder of the rules.  
\begin{proposition}[Reflexivity]
	\label{prop:subtypeRefl}
	The relation $ \leq_1 $ restricted to local contexts (without an active channel) is reflexive.
\end{proposition}

\begin{proof}
	Due to [S-Split], it suffices to show reflexivity on typed channels, not local contexts.
	\begin{gather*}
			\infer=[\sidenote{[S-Split]}]{\Delta, \type{c}{T} \leq_1 \Delta, \type{c}{T}}%
			{\Delta \leq_1 \Delta \quad \type{c}{T} \leq_1 \type{c}{T}}
	\end{gather*}
	We show the reflexivity $\type{c}{T} \leq_1 \type{c}{T}$ by induction on the type $T$.
	For $T=\tend$ it follows immediately from [S-$\tend$].
	For $T = \mc{i}{I}\inpT_i + \mc{j}{J}\outT_j$ we have
	\begin{gather*}
		\infer=[\sidenote{[S-$\Sigma$-1]}]{\type{c}{\mc{i}{I}\inpT_i + \mc{j}{J}\outT_j} \leq_1 \type{c}{\mc{i}{I}\inpT_i + \mc{j}{J}\outT_j}}%
		{\infer=[\sidenote{[S-$\Sigma$-2]}]{c\cdot \left( \type{c}{\mc{i}{I}\inpT_i + \mc{j}{J}\outT_j} \right) \leq_1 \type{c}{\mc{i}{I}\inpT_i + \mc{j}{J}\outT_j}}%
		{\infer=[\sidenote{[S-$\Sigma$-$\inpT$]}]{c\cdot \left( \type{c}{\mc{i}{I}\inpT_i} \right) \leq_1 \type{c}{\mc{i}{I}\inpT_i}}{\forall i \in I.\; c\cdot \left( \type{c}{\inpT_i} \right) \leq_1 \type{c}{\inpT_i}} \quad \infer=[\sidenote{[S-$\Sigma$-$\outT$]}]{c\cdot \left( \type{c}{\mc{j}{J}\outT_j} \right) \leq_1 \type{c}{\mc{j}{J}\outT_j}}{\forall j \in J.\; c\cdot \left( \type{c}{\outT_j} \right) \leq_1 \type{c}{\outT_j}}}}
	\end{gather*}
	For all $\inpT_i = \inpk{p}{q}{l_i}{U_i}{T_i}{i}$, the left-hand side immediately follows from [S-$\inpT$].
	For all $\outT_j = \send{i}{I_j} \probOut{\prob_i}{H_i}\cont T_i$, we have
	\begin{gather*}
		\infer=[\sidenote{[S-$\oplus$]}]{c\cdot \left( \type{c}{\send{i}{I_j} \probOut{\prob_i}{H_i}\cont T_i} \right) \leq_1 \type{c}{\send{i}{I_j} \probOut{\prob_i}{H_i}\cont T_i}}{\forall i \in I_j.\; c\cdot \left( \type{c}{H_i \cont T_i} \right) \leq_{\prob_i} \type{c}{\probOut{\prob_i}{H_i}\cont T_i}}
	\end{gather*}
	where all branches follow straightforwardly from [S-$\outT$] or [S-$\nonsend$].
	%\hfill $ \square $
\end{proof}

For transitivity, too, the rule [S-Split] causes the proof to not be too complex.
\begin{proposition}[Transitivity]
	\label{prop:subtypeTrans}
	The relation $ \leq_1 $ restricted to local contexts (without an active channel) is transitive.
\end{proposition}

\begin{proof}
	Assume $ \Delta_1 \leq_1 \Delta_2 $ and $ \Delta_2 \leq_1 \Delta_3 $.
	We prove $ \Delta_1 \leq_1 \Delta_3 $ by induction on the number of assignments in $ \Delta_3 $.
	\begin{description}
		\item[Base Case $ \Delta_3 = \emptyset $:] By Lemma~\ref{lem:absorbingEndInSubtypingR}, then $ \Delta_2 \leq_1 \Delta_3 $ implies that $ \Delta_2 = \emptyset $.
			By Lemma~\ref{lem:absorbingEndInSubtypingR}, then $ \Delta_1 \leq_1 \Delta_2 $ implies that $ \Delta_1 = \emptyset $.
			By [S-$ \emptyset $-1], then $ \Delta_1 \leq_1 \Delta_3 $.
		\item[Induction Step $ \Delta_3 = \Delta_3', \type{c}{T} $:] By the subtyping rules, then the only rule that can be applied in the bottom of the derivation tree for $ \Delta_2 \leq_1 \Delta_3 $ is then [S-Split].
			By [S-Split], then $ \Delta_2 = \Delta_{2, 1}, \Delta_{2, 2} $ such that $ \Delta_{2, 1} \leq_1 \Delta_3' $ and $ \Delta_{2, 2} \leq_1 \type{c}{T} $.
			By Lemma~\ref{lem:splitSubtyping}, then $ \Delta_1 = \Delta_{1, 1}, \Delta_{1, 2} $; $ \Delta_{1, 1} \leq_1 \Delta_{2, 1} $, and $ \Delta_{1, 2} \leq_1 \Delta_{2, 2} $.
			By the induction hypothesis, then $ \Delta_{1, 1} \leq_1 \Delta_{2, 1} $ and $ \Delta_{2, 1} \leq_1 \Delta_3' $ imply $ \Delta_{1, 1} \leq_1 \Delta_3' $ and, similarly, $ \Delta_{1, 2} \leq_1 \Delta_{2, 2} $ and $ \Delta_{2, 2} \leq_1 \type{c}{T} $ imply $ \Delta_{1, 2} \leq_1 \type{c}{T} $.
			By [S-Split], then $ \Delta_{1, 1} \leq_1 \Delta_3' $ and $ \Delta_{1, 2} \leq_1 \type{c}{T} $ imply $ \Delta_1 \leq_1 \Delta_3 $.\qedhere
	\end{description}
\end{proof}

By Propositions~\ref{prop:subtypeRefl} and~\ref{prop:subtypeTrans}, our subtyping relation fulfils all properties of a preorder.
\begin{corollary}
\label{cor:subtypePreorder}
    The relation $\leq_1$ restricted to local contexts (without an active channel) is a preorder.
\end{corollary}
The relation is not a partial order, however. A simple example for why $\leq_1$ is not antisymmetric involves stacked internal actions $\nonsend$.
\begin{example}[Subtyping is not Antisymmetric]
    \addxcontentsline{loe}{example}[\theexample]{Subtyping is not Antisymmetric}
    Consider the following two types of the channel $\channel{s}{r}$, $\type{\channel{s}{r}}{\probOut{1}{\nonsend}}$ and $\type{\channel{s}{r}}{\probOut{1}{\nonsend}\cont \probOut{1}{\nonsend}}$.
    Then clearly they are not the same, $\type{\channel{s}{r}}{\probOut{1}{\nonsend}} \neq \type{\channel{s}{r}}{\probOut{1}{\nonsend}\cont \probOut{1}{\nonsend}}$, but we have:
    \begin{gather*}
        \infer=[\sidenote{[S-$\nonsend$-R]}]{\type{\channel{s}{r}}{\probOut{1}{\nonsend}} \leq_1 \type{\channel{s}{r}}{\probOut{1}{\nonsend}\cont \probOut{1}{\nonsend}}}%
        {\infer=[\sidenote{Prop.~\ref{prop:subtypeRefl}}]{\type{\channel{s}{r}}{\probOut{1}{\nonsend}} \leq_1 \type{\channel{s}{r}}{\probOut{1}{\nonsend}}}{}} \qquad
        \infer=[\sidenote{[S-$\Sigma$-1]}]{\type{\channel{s}{r}}{\probOut{1}{\nonsend}\cont \probOut{1}{\nonsend}} \leq_1 \type{\channel{s}{r}}{\probOut{1}{\nonsend}}}%
            {\infer=[\sidenote{[S-$\oplus$]}]{\channel{s}{r} \cdot \left( \type{\channel{s}{r}}{\probOut{1}{\nonsend}\cont \probOut{1}{\nonsend}} \right) \leq_1 \type{\channel{s}{r}}{\probOut{1}{\nonsend}}}%
                {\infer=[\sidenote{[S-$\nonsend$-L]}]{\channel{s}{r} \cdot \left( \type{\channel{s}{r}}{{\nonsend}\cont \probOut{1}{\nonsend}} \right) \leq_1 \type{\channel{s}{r}}{\probOut{1}{\nonsend}}}%
                    {\infer=[\sidenote{Prop.~\ref{prop:subtypeRefl}}]{\type{\channel{s}{r}}{\probOut{1}{\nonsend}} \leq_1 \type{\channel{s}{r}}{\probOut{1}{\nonsend}}}%
                        {}}}}
    \end{gather*}
    Therefore they are both subtypes of each other.\exampleDone
\end{example}

\subsection{Safety and Deadlock-Freedom}
\label{subsec:safetyAndDeadlockFreedom}
Being arguably the two most important and commonly shown properties of MPST systems, we, too, show that our system fulfils safety and deadlock-freedom.
Recall that our goal is to infer that typed processes satisfy them from the fact that their types do.
Thus we need to first define these properties for local contexts (with an active channel).
After doing so, we show a key result on their preservation under subtyping and labelled transitions.
\medbreak

We inherit the predicates for \emph{safety} and \emph{deadlock-freedom} from \citep{scalas2019less,peters2024separation}, extended to local contexts with an active channel.
A context is safe if it contains for every unguarded output for which there is an unguarded input on the same prefix also an unguarded matching input.
\begin{definition}[Safety Property]
	\label{def:safetyProperty}
    The co-inductive property $ \varphi $ is a safety property of local contexts $ \Lambda $ if and only if for all $ \varphi{\left( \Lambda \right)} $
    \begin{compactenum}
    	\item if $ \Lambda = \Delta $ (without active channel), then
    		\begin{compactenum}
    			\item transitions $ \Delta \xlongrightarrow{\type{s}{\outPrefix{p}{q}{l}{U}}}_{\prob} $ and $ \Delta \xlongrightarrow{\type{s}{\inpPrefix{q}{p}{l'}{U'}}}_{1} $ imply $ \Delta \xlongrightarrow{\type{s}{\comPrefix{p}{q}{l}{U}}}_{\prob} \Delta' $ and $ \varphi{\left( \Delta' \right)} $, and
    			\item the transition $ \Delta \xlongrightarrow{\type{s}{\nonsend}}_{\prob} \Delta' $ implies $ \varphi{\left( \Delta' \right)} $.
    		\end{compactenum}
    	\item if $ \Lambda = \channel{s}{r} \cdot \Delta $ (with active channel), then
		\begin{compactenum}
			\item transitions $ \Delta \xlongrightarrow{\type{s}{\outPrefix{p}{q}{l}{U}}}_{\prob} $ with $ \participant{p} \in \vecparticipant{r} $ and $ \Delta \xlongrightarrow{\type{s}{\inpPrefix{q}{p}{l'}{U'}}}_{1} $ imply $ \Lambda \xlongrightarrow{\type{s}{\comPrefix{p}{q}{l}{U}}}_{\prob} \Delta' $ and $ \varphi{\left( \Delta' \right)} $, and
			\item the transition $ \Lambda \xlongrightarrow{\type{s}{\nonsend}}_{\prob} \Delta' $ implies $ \varphi{\left( \Delta' \right)} $.
		\end{compactenum}
    \end{compactenum}
	We say that $ \Lambda $ is safe, $ \safe{\Lambda} $, if $ \varphi{\left( \Lambda \right)} $ for some safety property $ \varphi $.
\end{definition}

Following almost immediately from this definition, we show that a safe context $\Delta$ can only transition into a safe context.
\begin{lemma}
	\label{lem:typeReductionSafe}
	If $ \Delta \mapsto^{*}_{\prob} \Delta' $ and $ \safe{\Delta} $ then $ \safe{\Delta'} $.
\end{lemma}

\begin{proof}
	Assume $ \Delta \mapsto_{\prob} \Delta' $ and $ \safe{\Delta} $.
	By Definition~\ref{def:transitionsLocalContexts}, then either $ \Delta \xlongrightarrow{\type{s}{\outPrefix{p}{q}{l}{U}}}_{\prob} $, and $ \Delta \xlongrightarrow{\type{s}{\inpPrefix{p}{q}{l'}{U'}}}_{1} $, and $ \Delta \xlongrightarrow{\type{s}{\comPrefix{p}{q}{l}{U}}}_{\prob} \Delta' $, or $ \Delta \xlongrightarrow{\type{s}{\nonsend}}_{\prob} \Delta' $.
	By Definition~\ref{def:safetyProperty}, then $ \safe{\Delta'} $.
	The proof for $ \Delta \mapsto^{*}_{\prob} \Delta' $ and $ \safe{\Delta} $ follows then by an induction on the number of steps in $ \Delta \mapsto^{*}_{\prob} \Delta' $.
	%\hfill $ \square $
\end{proof}

Next is deadlock-freedom.
A context is deadlock-free if it can terminate only with $ \emptyset $ and similarly a process is deadlock-free if it can terminate modulo $ \equiv $ only in $ \0 $.
Also, we define pending contexts to be deadlock-free, as their actions are treated by another branch of the subtyping derivation tree.
This is merely a technicality with no major impact, however, given that pending contexts appear only in subtyping derivation trees and not type judgements.

\begin{definition}[Deadlock-Freedom]
	\label{def:deadlockFreedom}
	The local context (with an active channel) $ \Lambda $ is \emph{deadlock-free}, denoted as $ \dfree{\Lambda} $, if $ \Lambda \mapsto^{*} \Lambda' \nrightarrow $ implies $ \Lambda' = \emptyset $ for all $ \Lambda' $ or if $\stuck(\Lambda)$.
	A process $ P $ is \emph{deadlock-free} if and only if for all $ P' $ such that $ P \Longrightarrow_{\prob} P' $ either
	\begin{inparaenum}[(a)]
		\item $ P' \nrightarrow $ and $ P \equiv \0 $, or
		\item there are $ P'' $ and $ \prob' $ such that $ P' \longrightarrow_{\prob'} P'' $.
	\end{inparaenum}
\end{definition}

Similar to \citep{scalas2019less,peters2024separation}, we show that our type system satisfies the standard properties of subtyping, subject reduction, and ensures deadlock-freedom.

\begin{theorem}[Subtyping and Properties]
	\label{thm:subtypingProperties}
	\begin{compactenum}
        \item[]
		\item	\label{thm:subtypingProperties:safe}
				If $ \Lambda_1 \leq_{\prob} \Lambda_2 $ and $ \safe{\Lambda_2} $, then
				\begin{enumerate}
					\item $\safe{\Lambda_1}$.
					\item If $\Lambda_1 \mapsto_{\prob_1} \Lambda'_1$ then there exist $\Lambda_2', \prob_2, \prob_3$ such that $\Lambda_2 \mapsto^{*}_{\prob_1\prob_2} \Lambda_2'$ and $\Lambda_1' \leq_{\prob_3} \Lambda_2'$ with $\prob = \prob_2\prob_3$ and $\safe{\Lambda_2'}$.
				\end{enumerate}
		\item \label{thm:subtypingProperties:df}
				If $ \Lambda_1 \leq_{\prob} \Lambda_2 $, $ \safe{\Lambda_2} $, and $ \dfree{\Lambda_2} $, then
				\begin{enumerate}
					\item $ \dfree{\Lambda_1} $.
					\item If $ \Lambda_1 \mapsto_{\prob_1} \Lambda'_1 $ then there exist $ \Lambda_2', \prob_2, \prob_3 $ such that $\Lambda_2 \mapsto^{*}_{\prob_1\prob_2} \Lambda_2'$ and $\Lambda_1' \leq_{\prob_3} \Lambda_2'$ with $\prob = \prob_2\prob_3$ and $\dfree{\Lambda_2'}$.
				\end{enumerate}
		\item Checking $ \safe{\Delta} $ and $ \dfree{\Delta} $ is decidable.
	\end{compactenum}
\end{theorem}

\begin{proof}
	\begin{description}
		\item[1.(a)] Assume $ \Lambda_1 \leq_{\prob} \Lambda_2 $ and $ \safe{\Lambda_2} $.
			We show $ \safe{\Lambda_1} $ by structural induction on the derivation of $ \Lambda_1 \leq_{\prob} \Lambda_2 $.
			\begin{description}
				\item[Case {[S-$ \Sigma $-1]}:]
					We have the two contexts $\Lambda_1 = \Delta = \type{\channelk{s}{r}{1}}{T_1}, \ldots,  \type{\channelk{s}{r}{n}}{T_n}$, and $ \Lambda_2 = \type{\channel{s}{r}}{\mc{j}{J_1} L_j + \ldots + \mc{j}{J_n} L_j} $, where $ \channelk{s}{r}{k} \cdot \Delta \leq_{\prob} \type{\channel{s}{r}}{\mc{j}{J_k} L_j}$ for all $ k \in \left[ 1 .. n \right] $.
					Since $\safe{\Lambda_2}$, we also have $\safe{\type{\channel{s}{r}}{\mc{j}{J_k} L_j}}$ for all $ k \in \left[ 1 .. n \right] $.
					By the induction hypothesis, then $\safe{\channelk{s}{r}{k} \cdot \Delta }$ for all $ k \in \left[ 1 .. n \right] $.
					By Definition~\ref{def:transitionsLocalContextsWithActiveChannel} for transitions for typing contexts with an active channel, for each transition $\Lambda_1 \mapsto^{*}_{\prob_k} \Lambda_1'$ there exists a $k \in \left[ 1 .. n \right]$ such that $\channelk{s}{r}{k} \cdot \Delta \mapsto^{*}_{\prob_k} \Lambda_1'$.
					Since $\safe{\channelk{s}{r}{k} \cdot \Delta }$ for all $ k \in \left[ 1 .. n \right] $, then $\safe{\Lambda_1}$.
				\item[Case {[S-$ \Sigma $-2]}:]
					We have $\Lambda_1 = \channelk{s}{r}{1} \cdot \left( \Delta, \type{\channelk{s}{r}{1}}{\mc{i}{I'}\inpT_i + \mc{j}{J'}\outT_j} \right)$ and $\Lambda_2 = \type{\channelk{s}{r}{2}}{\mc{i}{I}L_i + \mc{j}{J}L_j}$, where $\channelk{s}{r}{1} \cdot \left( \Delta, \type{\channelk{s}{r}{1}}{\mc{i}{I'}\inpT_i} \right) \leq_{\prob} \type{\channelk{s}{r}{2}}{\mc{i}{I}L_i}$, and $\channelk{s}{r}{1} \cdot \left( \Delta, \type{\channelk{s}{r}{1}}{\mc{j}{J'}\outT_j} \right) \leq_{\prob} \type{\channelk{s}{r}{2}}{\mc{j}{J}L_j}$.
					Since $\safe{\Lambda_2}$, then $\safe{\type{\channelk{s}{r}{2}}{\mc{i}{I}L_i}}$ and $\safe{\type{\channelk{s}{r}{2}}{\mc{j}{J}L_j}}$.
					By the induction hypothesis, $\safe{\channelk{s}{r}{1} \cdot \left( \Delta, \type{\channelk{s}{r}{1}}{\mc{i}{I'}\inpT_i} \right)}$ and $\safe{\channelk{s}{r}{1} \cdot \left( \Delta, \type{\channelk{s}{r}{1}}{\mc{j}{J'}\outT_j} \right)}$.
					Then $\safe{\Lambda_1}$ as required.
				\item[Case {[S-$ \Sigma $-$\inpT$]}:]
					In this case $\Lambda_1 = \channelk{s}{r}{1} \cdot \left( \Delta, \type{\channelk{s}{r}{1}}{\mc{i}{I' \cup J'}\inpT'_i} \right)$. By Definition~\ref{def:transitionsLocalContextsWithActiveChannel} of transitions for contexts with an active channel, $ \Lambda_1 $ has no transitions and is thus safe.
				\item[Case {[S-$ \Sigma $-$\outT$]}:]
					In this case we have $\Lambda_1 = \channelk{s}{r}{1} \cdot \left( \Delta, \type{\channelk{s}{r}{1}}{\mc{i}{I'}\outT'_i} \right)$ and $\Lambda_2 = \type{\channelk{s}{r}{2}}{\mc{i}{I'}\left( \mc{k}{I_i}L_k \right) + \mc{j}{J}\outT_j}$, where for all $ i \in I' $ it holds that $\channelk{s}{r}{1} \cdot \left( \Delta, \type{\channelk{s}{r}{1}}{\outT'_i} \right) \leq_{\prob} \type{\channelk{s}{r}{2}}{\mc{k}{I_i}L_k}$.
					Since $\safe{\Lambda_2}$, also $\safe{\type{\channelk{s}{r}{2}}{\mc{k}{I_i}L_k}}$ for all $ i \in I' $.
					By the induction hypothesis, then $\safe{\channelk{s}{r}{1} \cdot \left( \Delta, \type{\channelk{s}{r}{1}}{\outT'_i} \right)}$ for all $ i \in I' $.
					Thus $\safe{\Lambda_1}$ as required.
				\item[Case {[S-$ \oplus $]}:]
					This case is similar to the previous case (with $ \bigoplus $ instead of $ \sum $).
				\item[Case {[S-$ \inpT $]}:]
					We have $ \Lambda_1 = \channelk{s}{r}{1} \cdot \left( \Delta, \type{\channelk{s}{r}{1}}{\inp{q}{p}{l}{U}{T'}} \right) $. Similar to [S-$\Sigma$-$\inpT$], $\Lambda_1$ can not perform any transitions and is thus safe.
				\item[Case {[S-$ \outT $]}:]
					In this case $ \Lambda_1 = \channelk{s}{r}{1} \cdot \left( \Delta, \type{\channelk{s}{r}{1}}{\out{p}{q}{l}{U}{T'}} \right) $ and $\participant{q} \notin \Delta$. By Definition~\ref{def:transitionsLocalContextsWithActiveChannel} for transitions for contexts with an active channel, the only transition that $\Lambda_1$ might perform is a communication using the available type $\type{\channelk{s}{r}{1}}{\out{p}{q}{l}{U}{T'}}$ of a sending action.
					Because of $\participant{q} \notin \Delta$, the channel $\participant{q}$ containing the matching receiving type is not in $\Delta$.
					Then $\Lambda_1$ has no transitions and is thus safe.
				\item[Case {[S-Link]}:]
					In this case we have contexts $\Lambda_1 = \channelk{s}{r}{1}\cdot \bigl( \Delta, \type{\channelk{s}{r}{1}}{\out{p}{q}{l}{U}{T'_1}},\linebreak \type{\channelk{s}{r}{2}}{\inp{q}{p}{l}{U}{T'_2} \kern-1pt+\kern-2pt \mc{i}{I} L_i}\kern-1pt\bigr)$ and $\Lambda_2 \kern-2pt=\kern-2pt \type{\channel{s}{r}}{T}$, with $\Delta, \type{\channelk{s}{r}{1}}{T'_1}, \type{\channelk{s}{r}{2}}{T'_2} \kern-1pt\leq_{\prob} \Lambda_2$.
					By $\safe{\Lambda_2}$ and the induction hypothesis, then $\safe{\Delta, \type{\channelk{s}{r}{1}}{T'_1}, \type{\channelk{s}{r}{2}}{T'_2}}$.
					By Def.~\ref{def:transitionsLocalContextsWithActiveChannel}, $\channelk{s}{r}{1}\cdot \bigl( \Delta, \type{\channelk{s}{r}{1}}{\out{p}{q}{l}{U}{T'_1}}, \type{\channelk{s}{r}{2}}{\inp{q}{p}{l}{U}{T'_2} +\linebreak \mc{i}{I} L_i}\bigr)$ can only perform transition reducing the output on channel $\channelk{s}{r}{1}$, resulting in a safe transition.
					Since we already have $\safe{\Delta, \type{\channelk{s}{r}{1}}{T'_1}, \type{\channelk{s}{r}{2}}{T'_2}}$, then $\safe{\Lambda_1}$ as required.
				\item[Case {[S-$\nonsend$-L]}:]
					We have $\Lambda_1 = \channelk{s}{r}{1} \cdot \left( \Delta, \type{\channelk{s}{r}{1}}{\nonsend \cont T'} \right)$ and $\Lambda_2 = \type{\channelk{s}{r}{2}}{T}$, where $\Delta, \type{\channelk{s}{r}{1}}{T'} \leq_{\prob} \Lambda_2$.
					Since $\safe{\Lambda_2}$ and by the induction hypothesis, then $\safe{\Delta, \type{\channelk{s}{r}{1}}{T'}}$.
					Therefore $\safe{\Lambda_1}$.
				\item[Case {[S-$\nonsend$-R]}:]
					We have $\Lambda_2 = \type{\channel{s}{r}}{\probOut{\prob}{\nonsend} \cont T}$ and $\Lambda_1 \leq_{1} \type{\channel{s}{r}}{T}$.
					Since $\safe{\Lambda_2}$ and by Definition~\ref{def:safetyProperty} of safety, then also $\safe{\type{\channel{s}{r}}{T}}$.
					By the induction hypothesis, then $\safe{\Lambda_1}$.
				\item[Case {[S-$ \emptyset $-1]}:]
					In this case $ \Lambda_1 = \emptyset $ and thus $\safe{\Lambda_1}$.
				\item[Case {[S-$ \emptyset $]}:]
					In this case $ \Lambda_1 = c \cdot \Delta $ and $ \stuck{\left( \Lambda_1 \right)} $.
					Then $ \Lambda_1 $ is safe as pending contexts cannot perform any transitions.
				\item[Case {[S-Split]}:]
					In this case $\Lambda_1= \Delta'_1, \Delta'_2$; $\Lambda_2 = \Delta, \type{c}{T}$; $\Delta'_1 \leq_{1} \Delta$, and $\Delta'_2 \leq_{1} \type{c}{T}$.
					Because of $\safe{\Lambda_2}$, we have $\safe{\type{c}{T}}$ and $\safe{\Delta}$.
					By the induction hypothesis, then $\safe{\Delta'_1}$ and $\safe{\Delta'_2}$.
					Since both contexts are safe individually, for $\Delta'_1, \Delta'_2$ to not be safe, it must be that $\Delta'_1 \xlongrightarrow{\type{s}{\inpPrefix{p}{q}{l'}{U'}}}_{1}$ and $\Delta'_2 \xlongrightarrow{\type{s}{\outPrefix{q}{p}{l}{U}}}_{\prob}$ but not $\Delta'_1, \Delta'_2 \xlongrightarrow{\type{s}{\comPrefix{p}{q}{l}{U}}}_{\prob} \Delta''$ (or vice versa).
					Let us assume for contradiction that this is indeed the case (the other case in which the output occurs in $\Delta_1'$ and the input in $\Delta_2'$ is symmetric).
					Then there exists a channel $\type{\channelk{s}{r}{i}}{T'} \in \Delta_2'$ such that $\type{\channelk{s}{r}{i}}{T'} \xlongrightarrow{\type{s}{\outPrefix{q}{p}{l}{U}}}_{\prob}$.
					By the subtyping rules, in particular [S-$\Sigma$-1], [S-$\Sigma$-2], [S-$\Sigma$-$\outT$], and [S-$\outT$], the same output exists in $\type{c}{T}$, too.
					Now, either $\Delta \xlongrightarrow{\type{s}{\inpPrefix{p}{q}{l''}{U''}}}_{1}$ or not.
					\begin{description}
						\item[Case {$\Delta \xlongrightarrow{\type{s}{\inpPrefix{p}{q}{l''}{U''}}}_{1}$}:]
						Then since $\type{c}{T} \xlongrightarrow{\type{s}{\outPrefix{q}{p}{l}{U}}}_{\prob}$ by $\safe{\Delta, \type{c}{T}}$, it must be that also $\Delta \xlongrightarrow{\type{s}{\inpPrefix{p}{q}{l}{U}}}_{1}$.
						By the subtyping rules, in particular {[S-$\Sigma$-1]}, $\text{[S-$\Sigma$-2]}$, [S-$\Sigma$-$\inpT$], and [S-$\inpT$], this input is also in $\Delta_1'$, thus $\Delta'_1\kern-1pt, \Delta'_2 \kern-1pt \xlongrightarrow{\type{s}{\comPrefix{p}{q}{l}{U}}}_{\prob} \Delta''$, a contradiction.
						\item[Case {$\neg\Delta \xlongrightarrow{\type{s}{\inpPrefix{p}{q}{l''}{U''}}}_{1}$}:]
						By the subtyping rules, in particular rules [S-$\Sigma$-1], \mbox{[S-$\Sigma$-2]}, [S-$\Sigma$-$\inpT$], and [S-$\inpT$], there can be no input prefix in $\Delta_1'$ which was not also in $\Delta$.
						Thus $\neg\Delta'_1 \xlongrightarrow{\type{s}{\inpPrefix{p}{q}{l'}{U'}}}_{1}$, a contradiction.
					\end{description}
					Thus, $\safe{\Delta_1', \Delta_2'}$.
			\end{description}
		\item[1.(b)]
			Assume $\Lambda_1 \leq_{\prob} \Lambda_2$, $\safe{\Lambda_2}$, and $ \Lambda_1 \mapsto_{\prob_1} \Lambda_1' $.
			By above, $\safe{\Lambda_1}$.
			By the Definition~\ref{def:safetyProperty} of safety, then $\safe{\Lambda_1'}$ and $\safe{\Lambda_2'}$ for all $ \Lambda_2' $ and all $ \prob^{*} $ such that $ \Lambda_2 \mapsto_{\prob^{*}}^{*} \Lambda_2' $.
			We have to show that there are $\Lambda_2', \prob_2, \prob_3 $ such that $\Lambda_2 \mapsto^{*}_{\prob_1\prob_2} \Lambda_2'$ and $\Lambda_1' \leq_{\prob_3} \Lambda_2'$ with $\prob = \prob_2\prob_3$ and $\safe{\Lambda_2'}$.
			We proceed by structural induction on the derivation of $\Lambda_1 \leq_{\prob} \Lambda_2$.
			\begin{description}
				\item[Case {[S-$\Sigma$-1]}:] In this case $ \Lambda_1 = \Delta $; $ \Lambda_2 = \type{\channel{s}{r}}{\mc{j}{J_1} L_j + \ldots + \mc{j}{J_n} L_j} $, and $ \channelk{s}{r}{k} \cdot \left( \Delta \right) \leq_{\prob} \type{\channel{s}{r}}{\mc{j}{J_k} L_j} $ for all $ k \in \left[ 1 .. n \right] $.
					By Definition~\ref{def:transitionsLocalContextsWithActiveChannel} for the transitions of local contexts (with or without active channels), $ \Lambda_1 \mapsto_{\prob_1} \Lambda_1' $ implies that there is one channel $ \channelk{s}{r}{i} $ for which $ \channelk{s}{r}{i} \cdot \Delta \mapsto_{\prob_1} \Lambda_1' $.
					Since $ \safe{\Lambda_2} $, then also $ \safe{\type{\channel{s}{r}}{\mc{j}{J_i} L_j}} $.
					By the induction hypothesis, then there are $ \Lambda_2', \prob_2, \prob_3 $ such that $ \type{\channel{s}{r}}{\mc{j}{J_i} L_j} \mapsto^{*}_{\prob_1\prob_2} \Lambda_2' $; $ \Lambda_1' \leq_{\prob_3} \Lambda_2' $, and $ \prob = \prob_2\prob_3 $.
					Then also $ \Lambda_2 \mapsto^{*}_{\prob_1\prob_2} \Lambda_2'$.
				\item[Case {[S-$\Sigma$-2]}:]
					In this case $ \Lambda_1 = \channelk{s}{r}{1} \cdot \left( \Delta, \type{\channelk{s}{r}{1}}{\mc{i}{I'}\inpT_i + \mc{j}{J'}\outT_j} \right) $; $ \Lambda_2 = \type{\channelk{s}{r}{2}}{\mc{i}{I}L_i + \mc{j}{J}L_j} $, and $ \channelk{s}{r}{1} \cdot \left( \Delta, \type{\channelk{s}{r}{1}}{\mc{j}{J'}\outT_j} \right) \leq_{\prob} \type{\channelk{s}{r}{2}}{\mc{j}{J}L_j} $.
					By Definition~\ref{def:transitionsLocalContextsWithActiveChannel} for transitions of local contexts, $ \Lambda_1 \mapsto_{\prob_1} \Lambda_1' $ implies that $ \channelk{s}{r}{1} \cdot \left( \Delta, \type{\channelk{s}{r}{1}}{\mc{j}{J'}\outT_j} \right) \mapsto_{\prob_1} \Lambda_1' $.
					As $ \safe{\Lambda_2} $, also $ \safe{\type{\channelk{s}{r}{2}}{\mc{j}{J}L_j}} $.
					Then, by the induction hypothesis, there exist $ \Lambda_2', \prob_2$, and $\prob_3 $ such that $ \type{\channelk{s}{r}{2}}{\mc{j}{J}L_j} \mapsto^{*}_{\prob_1\prob_2} \Lambda_2' $; $ \Lambda_1' \leq_{\prob_3} \Lambda_2' $, and $ \prob = \prob_2\prob_3 $.
					Then also $ \Lambda_2 \mapsto^{*}_{\prob_1\prob_2} \Lambda_2' $.
				\item[Case {[S-$\Sigma$-$\inpT$]}:]
					In this case $ \Lambda_1 = \channelk{s}{r}{1} \cdot \left( \Delta, \type{\channelk{s}{r}{1}}{\mc{i}{I' \cup J'}\inpT'_i} \right) $.
					By Definition~\ref{def:transitionsLocalContextsWithActiveChannel} for transitions of local contexts with an active channel, $ \Lambda_1 \nrightarrow $, \ie $ \Lambda_1 $ has no transitions.
					Accordingly, this case contradicts the assumption $ \Lambda_1 \mapsto_{\prob_1} \Lambda_1' $ and, thus, holds trivially.
				\item[Case {[S-$\Sigma$-$\outT$]}:]
					This case is similar to the Case~[S-$\Sigma$-2].
					We have $ \Lambda_1 =\channelk{s}{r}{1} \cdot \left( \Delta, \type{\channelk{s}{r}{1}}{\mc{i}{I'}\outT'_i} \right) $; $ \Lambda_2 = \type{\channelk{s}{r}{2}}{\mc{i}{I'}\left( \mc{k}{I_i}L_k \right) + \mc{j}{J}\outT_j} $, and $ \channelk{s}{r}{1} \cdot \left( \Delta, \type{\channelk{s}{r}{1}}{\outT'_i} \right) \leq_{\prob} \type{\channelk{s}{r}{2}}{\mc{k}{I_i}L_k} $ for all $ i \in I' $.
					By Definition~\ref{def:transitionsLocalContextsWithActiveChannel} for transitions of local contexts, $ \Lambda_1 \mapsto_{\prob_1} \Lambda_1' $ implies that there is one $ i \in I' $ such that $ \channelk{s}{r}{1} \cdot \left( \Delta, \type{\channelk{s}{r}{1}}{\outT'_i} \right)\mapsto_{\prob_1} \Lambda_1' $.
					Since $ \safe{\Lambda_2} $, then also $ \safe{\type{\channelk{s}{r}{2}}{\mc{k}{I_i}L_k}} $.
					Then, by the induction hypothesis, there exist $ \Lambda_2', \prob_2$, and $\prob_3 $ such that $ \type{\channelk{s}{r}{2}}{\mc{k}{I_i}L_k} \mapsto^{*}_{\prob_1\prob_2} \Lambda_2' $; $ \Lambda_1' \leq_{\prob_3} \Lambda_2' $, and $ \prob = \prob_2\prob_3 $.
					Then also $ \Lambda_2 \mapsto^{*}_{\prob_1\prob_2} \Lambda_2' $.
				\item[Case {[S-$\oplus$]}:]
					In this case we have $ \Lambda_1 = \channelk{s}{r}{1} \cdot \left( \Delta, \type{\channelk{s}{r}{1}}{\send{i}{I}}\probOut{\prob'_i}{H'_i \cont T'_i} \right) $; $ \Lambda_2 = \type{\channelk{s}{r}{2}}{\send{j}{J} \probOut{\prob_j}{H_j \cont T_j}} $, and
					\begin{equation}
						\channelk{s}{r}{1} \cdot \left( \Delta, \type{\channelk{s}{r}{1}}{H'_i \cont T'_i} \right) \leq_{\prob'_i\prob} \type{\channelk{s}{r}{2}}{\send{j}{J_i} \probOut{\prob_j}{H_j \cont T_j}} \label{eq:SOplus}
					\end{equation}
					for all $ i \in I $, where $ \prob = \sum_{j\in J}\prob_j $ and $ J = \bigcup_{i \in I}J_i $.
					By Definition~\ref{def:transitionsLocalContextsWithActiveChannel} for transitions of local contexts with an active channel, then $ \Lambda_1 \mapsto_{\prob_1} \Lambda_1' $ implies that there is some $ k \in I $ such that $ \prob'_k = \prob_1 $ and $ \type{\channelk{s}{r}{1}}{T'_k} = \Lambda_1' $.
					Then also $ \channelk{s}{r}{1} \cdot \left( \Delta, \type{\channelk{s}{r}{1}}{H'_k \cont T'_k} \right) \mapsto_{1} \Lambda_1'$.
					As in the previous cases, $ \safe{\Lambda_2} $ implies $ \safe{\type{\channelk{s}{r}{2}}{\send{j}{J_k} \probOut{\prob_j}{H_j \cont T_j}}} $.
					With $ \prob'_k = \prob_1 $ the (\ref{eq:SOplus}) above for the case $ k $ becomes:
					\begin{equation*}
						\channelk{s}{r}{1} \cdot \left( \Delta, \type{\channelk{s}{r}{1}}{H'_k \cont T'_k} \right) \leq_{\prob_1\prob} \type{\channelk{s}{r}{2}}{\send{j}{J_k} \probOut{\prob_j}{H_j \cont T_j}}
					\end{equation*}
					Then, by the induction hypothesis, there exist $ \Lambda_2', \prob_2'$, and $\prob_3 $ such that $ \type{\channelk{s}{r}{2}}{\send{j}{J_k} \probOut{\prob_j}{H_j \cont T_j}} \mapsto^{*}_{\prob'_2} \Lambda_2' $; $ \Lambda_1' \leq_{\prob_3} \Lambda_2' $, and $ \prob_1\prob = \prob'_2\prob_3 $.
					Thus also $ \Lambda_2 \mapsto^{*}_{\prob'_2} \Lambda_2' $.
					By choosing $ \prob_2 = \frac{\prob}{\prob_3} $ and with $ \prob_1\prob = \prob'_2\prob_3 $, then $ \Lambda_2 \mapsto^{*}_{\prob_1\prob_2} \Lambda_2' $ as required.
					Finally, $ \prob = \frac{\prob}{\prob_3} \prob_3 = \prob_2 \prob_3 $.
				\item[Case {[S-$\inpT$]}:] In this case $\Lambda_1 = \channelk{s}{r}{1} \cdot \left( \Delta, \type{\channelk{s}{r}{1}}{\inp{q}{p}{l}{U}{T'}} \right)$ and $\prob = 1$.
					By Definition~\ref{def:transitionsLocalContextsWithActiveChannel} for transitions of local contexts with an active channel, $\Lambda_1$ has no transitions.
					Thus, this case holds trivially.
				\item[Case {[S-$\outT$]}:] In this case $\Lambda_1 = \channelk{s}{r}{1} \cdot \left( \Delta, \type{\channelk{s}{r}{1}}{\out{p}{q}{l}{U}{T'}} \right)$ and $ \participant{q} \notin \Delta $.
					By Definition~\ref{def:transitionsLocalContextsWithActiveChannel} for transitions of local contexts with an active channel and because of $ \participant{q} \notin \Delta $; $\Lambda_1$ has no transitions.
					Thus, this case holds trivially.
				\item[Case {[S-Link]}:]
					Here, we have the contexts $\Lambda_1 = \channelk{s}{r}{1}\cdot \bigl( \Delta, \type{\channelk{s}{r}{1}}{\out{p}{q}{l}{U}{T'_1}},\linebreak \type{\channelk{s}{r}{2}}{\inp{q}{p}{l}{U}{T'_2} + \mc{i}{I} L_i} \bigr)$ and $\Lambda_2 = \type{\channelk{s}{r}{2}}{T}$, for which $( \Delta, \type{\channelk{s}{r}{1}}{T'_1},\break \type{\channelk{s}{r}{2}}{T'_2} ) \leq_{\prob} \Lambda_2 $.
					By Definition~\ref{def:transitionsLocalContextsWithActiveChannel} for transitions of local contexts with an active channel, then $ \Lambda_1 \mapsto_{\prob_1} \Lambda_1' $ implies $ \prob_1 = 1 $ and $ \Lambda_1' = \Delta, \type{\channelk{s}{r}{1}}{T'_1}, \type{\channelk{s}{r}{2}}{T'_2} $.
					We have to show that there are $ \Lambda_2', \prob_2, \prob_3 $ such that $ \Lambda_2 \mapsto^{*}_{\prob_1\prob_2} \Lambda_2' $; $ \Lambda_1' \leq_{\prob_3} \Lambda_2' $, and $ \prob = \prob_2\prob_3 $.
					By reflexivity $ \Lambda_2 \mapsto^{*}_{1} \Lambda_2 $, \ie we can choose $ \Lambda_2' = \Lambda_2 $ and $ \prob_2 = 1 $ such that $ \Lambda_2 \mapsto^{*}_{\prob_1\prob_2} \Lambda_2' $.
					From $ \Lambda_1' = \Delta, \type{\channelk{s}{r}{1}}{T'_1}, \type{\channelk{s}{r}{2}}{T'_2} $ and $ \Delta, \type{\channelk{s}{r}{1}}{T'_1}, \type{\channelk{s}{r}{2}}{T'_2} \leq_{\prob} \Lambda_2 $ and by choosing $ \prob_3 = \prob $, we obtain then $ \Lambda_1' \leq_{\prob_3} \Lambda_2' $.
					Finally, $ \prob = \prob_2\prob_3 $.
				\item[Case {[S-$\nonsend$-L]}:]
					In this case $\Lambda_1 = \channelk{s}{r}{1} \cdot \left( \Delta, \type{\channelk{s}{r}{1}}{\nonsend \cont T'}\right)$ and $\Lambda_2 = \type{\channelk{s}{r}{2}}{T}$ with $\Delta, \type{\channelk{s}{r}{1}}{T'_1} \leq_{\prob} \Lambda_2$.
					By Definition~\ref{def:transitionsLocalContextsWithActiveChannel} for transitions of local contexts with an active channel, then $ \Lambda_1 \mapsto_{\prob_1} \Lambda_1' $ implies that $ \Lambda_1' =  \Delta, \type{\channelk{s}{r}{1}}{T'} $ and $ \prob_1 = 1 $.
					By reflexivity and by choosing $ \Lambda_2' = \Lambda_2 $ and $ \prob_2 = 1 $, then $ \Lambda_2 \mapsto_{\prob_1\prob_2}^{*} \Lambda_2' $.
					By choosing $ \prob_3 = \prob $, then $\Delta, \type{\channelk{s}{r}{1}}{T'_1} \leq_{\prob} \Lambda_2$ becomes $ \Lambda_1' \leq_{\prob_3} \Lambda_2' $.
					Finally, $ \prob = \prob_2\prob_3 $.
				\item[Case {[S-$\nonsend$-R]}:]
					In this case $ \Lambda_1 = \Lambda $ and $ \Lambda_2 = \type{\channelk{s}{r}{2}}{\probOut{\prob}{\nonsend} \cont T} $, with $ \Lambda_1 \leq_1 \type{\channelk{s}{r}{2}}{T} $.
					Since $ \safe{\Lambda_2} $, then also $ \safe{\type{\channelk{s}{r}{2}}{T}} $.
					By the induction hypothesis, then there are $ \Lambda_2', \prob_2', \prob_3 $ such that $ \type{\channelk{s}{r}{2}}{T} \mapsto^{*}_{\prob_1\prob_2'} \Lambda_2' $; $ \Lambda_1' \leq_{\prob_3} \Lambda_2' $, and $ 1 = \prob_2'\prob_3 $.
					By Definition~\ref{def:transitionsLocalContexts}, then $ \Lambda_2 \mapsto^{*}_{\prob\prob_1\prob_2'} \Lambda_2' $.
					By choosing $ \prob_2 = \prob\prob_2' $, then $ \Lambda_2 \mapsto^{*}_{\prob_1\prob_2} \Lambda_2' $.
					Finally, because of $ 1 = \prob_2'\prob_3 $ and $ \prob_2 = \prob\prob_2' $, then $ \prob = \prob \prob_2' \prob_3 = \prob_2 \prob_3 $.
				\item[Case {[S-$\emptyset$-1]}:]
					In this case $ \Lambda_1 = \emptyset $ has no transitions.
					Thus, this case holds trivially.
				\item[Case {[S-$\emptyset$]}:]
					In this case $ \Lambda_1 = c\cdot \Delta $ and $ \stuck(c\cdot \Delta) $.
					By Definition~\ref{def:stuck} of pending contexts, then $ \Lambda_1 $ has no transitions.
					Thus, this case holds trivially.
				\item[Case {[S-Split]}:] In this case $ \Lambda_1 = \Delta'_1, \Delta'_2 $; $ \Lambda_2 =\Delta, \type{c}{T} $, $ \prob = 1 $, $ \Delta'_1 \leq_1 \Delta $, and $ \Delta'_2 \leq_1 \type{c}{T} $.
					From $\safe{\Lambda_2}$ and 1.(a) above, then $\safe{\Delta}$, $\safe{\type{c}{T}}$, $\safe{\Delta'_1}$, and $\safe{\Delta'_2}$.
					By Definition~\ref{def:transitionsLocalContexts} for local contexts without active channel, then $ \Lambda_1 \mapsto_{\prob_1} \Lambda_1' $ was performed by $ \Delta_1' $; $ \Delta_2' $, or by both contexts together.
					\begin{description}
						\item[Case of $ \Delta_1' \mapsto_{\prob_1} \Lambda_1' $:] By the induction hypothesis then, there are $ \Lambda_2', \prob_2, \prob_3 $ such that $ \Delta \mapsto_{\prob_1\prob_2}^{*} \Lambda_2' $; $ \Lambda_1' \leq_{\prob_3} \Lambda_2' $ and $ \prob = \prob_2\prob_3 $.
							Then also $ \Lambda_2 \mapsto_{\prob_1\prob_2}^{*} \Lambda_2' $.
						\item[Case of $ \Delta_2' \mapsto_{\prob_1} \Lambda_1' $:] By the induction hypothesis then, there are $ \Lambda_2', \prob_2, \prob_3 $ such that $ \type{c}{T} \mapsto_{\prob_1\prob_2}^{*} \Lambda_2' $; $ \Lambda_1' \leq_{\prob_3} \Lambda_2' $ and $ \prob = \prob_2\prob_3 $.
							Then also $ \Lambda_2 \mapsto_{\prob_1\prob_2}^{*} \Lambda_2' $.
						\item[Case of $ \Delta_1'\hspace{-0.5pt}, \hspace{-0.5pt}\Delta_2' \mapsto_{\prob_1}\hspace{-2.1pt} \Lambda_1' $ by an interaction between $ \Delta_1' $ and $ \Delta_2' $:] By Definition~\ref{def:transitionsLocalContexts}, then $ \Delta'_1 \xlongrightarrow{\type{s}{\inpPrefix{p}{q}{l}{U}}}_{1} \Delta_1'' $ and $ \Delta'_2 \xlongrightarrow{\type{s}{\outPrefix{q}{p}{l}{U}}}_{\prob_1} \Delta_2'' $ (or vice versa).
							We only show this case, as the other in which the output occurs in $ \Delta_1' $ and the input in $ \Delta_2' $ is symmetric.
							Then there exists a channel $\type{\channelk{s}{r}{i}}{T'} \in \Delta_2'$ such that $\type{\channelk{s}{r}{i}}{T'} \xlongrightarrow{\type{s}{\outPrefix{q}{p}{l}{U}}}_{\prob_1}$.
							By the subtyping rules, in particular [S-$\Sigma$-1], [S-$\Sigma$-2], [S-$\Sigma$-$\outT$], and [S-$\outT$], the same output exists in $\type{c}{T}$, too.
							Then $ \type{c}{T} \xlongrightarrow{\type{s}{\outPrefix{q}{p}{l}{U}}}_{\prob_1} \type{c}{T_o} $.
							Since the same output is reduced in both steps, $ \Delta_2' \leq_1 \type{c}{T} $; $ \Delta'_2 \xlongrightarrow{\type{s}{\outPrefix{q}{p}{l}{U}}}_{\prob_1} \Delta_2'' $, and $ \type{c}{T} \xlongrightarrow{\type{s}{\outPrefix{q}{p}{l}{U}}}_{\prob_1} \type{c}{T_o} $ imply $ \Delta_2'' \leq_{1} \type{c}{T_o} $.
							Now, assume for contradiction that $\neg\Delta \xlongrightarrow{\type{s}{\inpPrefix{p}{q}{l'}{U'}}}_{1}$.
							By the subtyping rules, in particular [S-$\Sigma$-1], [S-$\Sigma$-2], [S-$\Sigma$-$\inpT$], and [S-$\inpT$], there can be no input prefix in $\Delta_1'$ which was not also in $\Delta$.
							But $\Delta'_1 \xlongrightarrow{\type{s}{\inpPrefix{p}{q}{l}{U}}}_{1}$, a contradiction.
							Thus, it must be that $ \Delta \xlongrightarrow{\type{s}{\inpPrefix{p}{q}{l'}{U'}}}_{1} $.
							Since $ \safe{\Lambda_s} $, then $ \Delta \xlongrightarrow{\type{s}{\inpPrefix{p}{q}{l}{U}}}_{1} \Delta_i $ and $\Lambda_2 \mapsto_{\prob_1} \Lambda_2' $, where $ \Lambda_2' = \Delta_i, \type{c}{T_o} $.
							By choosing $ \prob_2 = 1 $, then $ \Lambda_2 \mapsto_{\prob_1\prob_2} \Lambda_2' $.
							Since the same input is reduced in both steps, $ \Delta_1' \leq_1 \Delta $; $ \Delta'_1 \xlongrightarrow{\type{s}{\inpPrefix{p}{q}{l}{U}}}_{1} \Delta_1'' $, and $ \Delta \xlongrightarrow{\type{s}{\inpPrefix{p}{q}{l}{U}}}_{1} \Delta_i $ imply $ \Delta_1'' \leq_{1} \Delta_i $.
							By [S-Split], $ \Delta_2'' \leq_{1} \type{c}{T_o} $ and $ \Delta_1'' \leq_{1} \Delta_i $ imply $ \Lambda_1' \leq_{1} \Lambda_2' $.
							By choosing $ \prob_3 = 1 $, then $ \Lambda_1' \leq_{\prob_3} \Lambda_2' $.
							Finally, $ \prob = 1 = \prob_2\prob_3 $.
					\end{description}
			\end{description}
	\item[2.(a)] Assume $ \Lambda_1 \leq_{\prob} \Lambda_2 $, $ \safe{\Lambda_2} $, and $ \dfree{\Lambda_2} $.
		We show $ \dfree{\Lambda_1} $ by contradiction.
		Assume the contrary, \ie assume that $ \Lambda_1 \mapsto_{\prob_1}^{*} \Lambda_1' $ such that $ \Lambda_1' \neq \emptyset $ and $ \Lambda_1' \nrightarrow $.
		By Theorem~\ref{thm:subtypingProperties}.\ref{thm:subtypingProperties:safe}(b) above, then $ \Lambda_2 \mapsto_{\prob_1\prob_2}^{*} \Lambda_2' $; $ \Lambda_1' \leq_{\prob_3} \Lambda_2' $, and $ \prob = \prob_2 \prob_3 $.
		By Lemma~\ref{lem:typeReductionSafe}, then $ \safe{\Lambda_2'} $.
		Since $ \dfree{\Lambda_2} $, then $ \dfree{\Lambda_2'} $.
		By Definition~\ref{def:transitionsLocalContexts}, then $ \Lambda_1' $ does not contain any unguarded $ \nonsend $ and instead all guards are inputs or outputs but there is no matching input and output.
		Moreover, we can neglect channels typed as $ \tend $.
		Hence, we can assume that all channels of $ \Lambda_1' $ are typed by an unguarded sum containing at least one summand and all unguarded summands are outputs or inputs.
		Each input and output on the left hand side of subtyping is checked in the proof tree of $ \Lambda_1' \leq_{\prob_3} \Lambda_2' $ by one of the rules [S-$ \inpT $], [S-$ \outT $], [S-Link], or [S-$ \emptyset $].
		Since all these rules require an active channel on the left hand side of the subtyping relation and since [S-$ \Sigma $-1] is the only rule introducing active channels, these inputs and outputs passed the rule [S-$ \Sigma $-1] in the proof tree of $ \Lambda_1 \leq_1 \Lambda_2 $ before moving further upwards to one of the rules [S-$ \inpT $], [S-$ \outT $], [S-Link], or [S-$ \emptyset $].
		Since [S-$ \Sigma $-1] creates a branch for every channel occurring on the left, each input and each output is in one of these branches checked with its channel marked as active.
		For every branch in the proof tree of $ \Lambda_1' \leq_{\prob_3} \Lambda_2' $ consider the lowest application of [S-$ \Sigma $-1].
		For each unguarded sum in $ \Lambda_1' $ there is one such application of [S-$ \Sigma $-1], but one application of [S-$ \Sigma $-1] may cover several unguarded sums of $ \Lambda_1' $.
		Because of the side condition $ \bigcup_{k \in \left[ 1 .. n \right]} J_k \neq \emptyset $, for each application of [S-$ \Sigma $-1], the right hand side of the conclusion cannot be empty and at least one not empty sum is produced in the precondition $ \forall k \in \left[ 1 .. n \right].\; \channelk{s}{r}{k} \cdot \left( \Delta \right) \leq_{\prob_i} \type{\channel{s}{r}}{\mc{j}{J_k} L_j} $.
		Then for each lowest application of [S-$ \Sigma $-1] at least one of the $ L_j $ becomes an input or output (possibly guarded by $\nonsend$).
		All such inputs and outputs in $ L_j $'s are introduced by [S-$ \inpT $] or [S-$ \outT $].
		Then the same inputs and outputs (with the same prefix, label, and message type) are also contained in $ \Lambda_1' $.
		Since $ \Lambda_1' $ does not contain unguarded $ \nonsend $, none of these outputs and inputs in $ \Lambda_1'$ is guarded by $ \nonsend $.
		Since we considered the lowest application of [S-$ \Sigma $-1], none of these outputs and inputs is guarded by other sums that results from inputs or outputs also present in $ \Lambda_2' $.
		These outputs and inputs can also not be guarded by other inputs and outputs that appear only in $ \Lambda_1' $.
		Such other inputs and outputs in $ \Lambda_1' $ describe interactions introduced between channels that are unified to a single channel in $ \Lambda_2' $.
		They are checked by the rule [S-Link] or [S-$ \emptyset $].
		The rule [S-Link] was not used, because else a matching pair of additional input and output would be unguarded in $ \Lambda_1' $.
		Then $ \Lambda_1' $ can do a step reducing these two actions contradicting the assumption $ \Lambda' \nrightarrow $.
		Also rule [S-$ \emptyset $] cannot be used, because without [S-Link] for such additional inputs or outputs it will create empty right hand sides, \ie empty $ L_j $'s.
		Then the outputs and inputs in $ \Lambda_1' $ that lead to non-empty $ L_j $'s in the lowest application of [S-$ \Sigma $-1] are unguarded in $ \Lambda_1' $.
		The mentioned inputs and outputs in $ L_j $'s are unguarded or guarded only by $ \nonsend $ in $ \Lambda_2' $.
		By the subtyping rules and [S-$ \Sigma $-$ \inpT $] and [S-$ \Sigma $-$ \outT $] in particular, all unguarded inputs, but not necessarily all unguarded outputs, from $ \Lambda_2' $ appear in $ \Lambda_1' $.
		Since $ \dfree{\Lambda_2'} $ and since all such inputs appear on both sides, at least one of those inputs together with a matching output can be reduced in a step of $ \Lambda_2' $ possibly after some $ \nonsend $-steps.
		By the side condition on $ \pre() $ in [S-$ \outT $], not each output of $ \Lambda_2' $ appears in $ \Lambda_1' $ but for each omitted output there is another output with the same prefix (same roles) in both $ \Lambda_1' $ and $ \Lambda_2' $.
		By $ \safe{\Lambda_2'} $, then at least one pair of output and input in both $ \Lambda_1' $ and $ \Lambda_2' $ is matching and can together perform a step.
		But then $ \Lambda_1' \mapsto_{\prob'} $ contradicting $ \Lambda_1' \nrightarrow $.
		We conclude that the assumption $ \Lambda_1' \nrightarrow $ was wrong, \ie that $ \dfree{\Lambda_1} $.
	\item[2.(b)] Follows from 2.(a) and 1.(b).
	\item[3.] We inherit the definitions of safety and deadlock-freedom from \citep{scalas2019less}.
		Accordingly, also the proof of the decidability of these predicates is similar to \citep{scalasYoshida18} (the technical report of \citep{scalas2019less}).
		The main argument (here and in \citep{scalasYoshida18}) is, that the transitive closure of the transition relation on local contexts defined in Definition~\ref{def:transitionsLocalContexts} induces a finite-state transition system.
		Hence, algorithms to check safety and deadlock-freedom can be built for each local context on its respective finite-state transition system.
		One way to do that, \ie one strategy on how to build these algorithms, is presented in \citep{scalasYoshida18}.\qedhere
\end{description}
\end{proof}

\subsection{Interface Existence}
\label{subsec:interfaceExists}

Our subtyping is very flexible.
Indeed, for every safe and deadlock-free collection of channel types there exists their interface as a single channel type.
Note that this is a major difference to \citep{DBLP:conf/concur/Horne20}.

\begin{theorem}[Interface Existence]
	\label{thm:interfaceExists}
	For all $ \Delta' = \left\lbrace \type{\channelk{s}{r}{i}}{T_i'} \right\rbrace_{i \in I} $ with $ \safe{\Delta'} $ and $\dfree{\Delta', \Delta}$ for some $ \Delta $ there are $ \vecparticipant{r} $ and $ T $ such that $ \vecparticipant{r} \subseteq \bigcup_{i \in I} \vecparticipantk{r}{i} $ and $ \Delta' \leq_1 \type{\channel{s}{r}}{T} $.
\end{theorem}

\begin{proof}
	We fix $ \Delta' = \left\lbrace \type{\channelk{s}{r}{i}}{T_i'} \right\rbrace_{i \in I} $ with $ \safe{\Delta'} $ and $\dfree{\Delta', \Delta}$ for some $ \Delta $.
	If $ I = \emptyset $ or all $ T_i' = \tend $, then we can choose $ \vecparticipant{r} = \bigcup_{i \in I} \vecparticipantk{r}{i} $ and $ T = \tend $ such that $ \Delta' \leq_1 \type{\channel{s}{r}}{T} $ by [S-$ \emptyset $].
	Else and without loss of generality, assume $ I = [1..n] $ with $ n > 1 $ and $ T_i' \neq \tend $ for all $ i \in I $.
	By [S-$ \Sigma $-1], if there are $ \channelk{s}{r}{i} \cdot \Delta' \leq_1 \type{\channel{s}{r}}{\sum_{j \in J_i} L_j} $ for all $ i \in I $ then we can choose $ T = \sum_{j \in J_1} L_j + \ldots + \sum_{j \in J_n} $ and are done.
	Hence, we fix $ i $ and show $ \channelk{s}{r}{i} \cdot \Delta' \leq_1 \type{\channel{s}{r}}{\sum_{j \in J_i} L_j} $.
	If $ \Stuck{\channelk{s}{r}{i} \cdot \Delta'} $ then, by [S-$ \emptyset $], we can choose $ J_i = \emptyset $ and are done.
	Else the type $ T_i' $ of $ \channelk{s}{r}{i} $ in $ \Delta' $ is guarded by a mixed choice that may contain $ \nonsend $, inputs, or outputs.
	We perform an induction on the structure of $ T_i' $.
	\begin{description}
		\item[Case $ T_i' = \tend $:] This case violates the assumption $ T_i' \neq \tend $ for all $ i \in I $.
		\item[Case $ T_i' = \mc{j}{J}L_j' $:] Each $ L_j' $ is of one of the following cases:
			\begin{description}
				\item[Case $ L_j' = \inp{p}{q}{l}{U}{T_j'} $:] If $ \participant{q} \notin \Delta' $, then the input is removed with [S-$ \outT $].
					Then we conclude with the induction hypothesis and the respective input in front of $ T $.
					Else if $ \participant{q} \in \Delta' $, then $ \Stuck{\channelk{s}{r}{i} \cdot \Delta''} $, where $ \Delta'' $ is obtained from $ \Delta' $ by replacing $ T_i' $ with $ L_j' $.
					Then the respective part of some $ L_x $ is empty.
					By [S-$ \emptyset $], then $ \channelk{s}{r}{i} \cdot \Delta' \leq_1 \type{\channel{s}{r}}{\sum_{j \in J_i} L_j} $.
				\item[Case $ L_j' = \send{k}{K} \probOut{\prob_k}{H_k \cont T_k'} $:] Each $ H_k \cont T_k' $ is of one of the following cases:
					\begin{description}
						\item[Case $ \nonsend\cont T_k' $:] The $ \nonsend $ can be removed with [S-$ \nonsend $-L].
							Then we conclude with the induction hypothesis.
						\item[Case $ \out{p}{q}{l}{U}{T_k'} $:] If $ \participant{q} \notin \Delta' $, then the output is removed with [S-$ \outT $].
							Then we conclude with the induction hypothesis and the respective output in front of $ T $.
							Else we have $ \participant{q} = \Delta' $.
							By $ \dfree{\Delta', \Delta} $ and $ \safe{\Delta'} $, then $ \Delta', \Delta $ can perform a step.
							If no step of $ \Delta', \Delta $ involves this output then $ \Stuck{\channelk{s}{r}{i} \cdot \Delta''} $, where $ \Delta'' $ is obtained from $ \Delta' $ by replacing $ T_i' $ with $ L_j' $.
							By [S-$ \emptyset $], then $ \channelk{s}{r}{i} \cdot \Delta' \leq_1 \type{\channel{s}{r}}{\sum_{j \in J_i} L_j} $.
							Otherwise, the matching input is unguarded in $ \Delta' $.
							By [S-Link], we can then conclude by the induction hypothesis.\qedhere
					\end{description}
			\end{description}
	\end{description}
\end{proof}

We do not prove a related statement for the opposite direction, because it is much simpler.
For example by reflexivity of $\leq_1$ (Proposition~\ref{prop:subtypeRefl}) the following holds trivially.
If $ \safe{\type{\channel{s}{r}}{T}} $ then there are $\vecparticipantk{r}{1}, \ldots, \vecparticipantk{r}{n}$ with $ \vecparticipant{r} \subseteq \bigcup_{i \in \left[ 1 .. n \right]} \vecparticipantk{r}{i} $ and $T'_1, \ldots, T'_n$ such that $ \left\lbrace \type{\channelk{s}{r}{i}}{T_i'} \right\rbrace_{i \in \left[ 1 .. n \right]} \leq_1 \type{\channel{s}{r}}{T} $.
Finding a more meaningful statement for ``a refinement always exits'' needs careful crafting, but it is quite conceivable that an interesting yet general result can be shown.

\section{Properties of Typed Processes}
\label{sec:propertiesOfTypedProcesses}
Having introduced safety and deadlock-freedom for types, in this section we will see how these properties are transferable to processes justified by safe and deadlock-free types.
The main results are Theorem~\ref{thm:subjectReduction}, subject reduction, and Theorem~\ref{thm:deadlockFreedom}, deadlock-freedom.

\subsection{Subject Reduction}
\label{subsec:subjectReduction}
This subsection is dedicated to the Subject Reduction Theorem, which states that typed processes can only reduce to typed processes.
By extension, any desirable feature enforced by the typing is preserved for all interaction behaviour and holds for all reductions.
It is clear to see why this is greatly beneficial and thus subject reduction is a key property across type theory.
For its proof, a significant number of smaller lemmas are shown first.
Most notably, these include Lemma~\ref{lem:subjectCongruence}, subject congruence, and~\ref{lem:substitution}, substitution.
The eponymous Theorem~\ref{thm:subjectReduction} is stated and proven afterwards. 

\subsubsection{Preliminary Lemmas}
We will now show eight smaller lemmas needed for the proof of Theorem~\ref{thm:subjectReduction}.
We chose to give some properties of types and subtyping here, instead of the previous section, as they are only used in the proofs of this section.

The first two results are small statements used in the proof of subject congruence.
Lemma~\ref{lem:safeEmptyset} states that the empty context is safe.
\begin{lemma}
	\label{lem:safeEmptyset}
	$ \safe{\emptyset} $.
\end{lemma}

\begin{proof}
	By Definition~\ref{def:safetyProperty}, since $ \emptyset $ has no transitions.
	%\hfill $ \square $
\end{proof}

According to Lemma~\ref{lem:globalWeakening}, adding additional variable and process variable assignments to the global context $\Gamma$ does not invalidate a type judgement. 
\begin{lemma}[Global Weakening]
	\label{lem:globalWeakening}
	If $ \Gamma \vdash P \smalltriangleright \Delta $ then also $ \Gamma, \Gamma' \vdash P \smalltriangleright \Delta $.
\end{lemma}

\begin{proof}
	Straightforward by induction on the derivation of $ \Gamma \vdash P \smalltriangleright \Delta $, since all assignments added to $ \Gamma $ in the derivation tree are on bound names.
	%\hfill $ \square $
\end{proof}

We next show subject congruence, an important property ensuring that a typing judgement of a process $P$ also holds for all processes congruent to $P$.
\begin{lemma}[Subject Congruence]
	\label{lem:subjectCongruence}
	If $ \Gamma \vdash P \smalltriangleright \Delta $ and $ P \equiv Q $ then also $ \Gamma \vdash Q \smalltriangleright \Delta $.
\end{lemma}

\begin{proof}
	Assume $ \Gamma \vdash P \smalltriangleright \Delta $ and $ P \equiv Q $.
	We prove $ \Gamma \vdash Q \smalltriangleright \Delta $ by structural induction on the derivation of $ P \equiv Q $.
	Since each typing and subtyping rule by definition respects alpha conversion and the rules that make $ \equiv $ a congruence, we consider only the rules of Definition~\ref{def:structuralCongruence} explicitly.
	\begin{description}
		\item[Case $ R \inparallel \0 \equiv R $:] Applying this rule from left to right, then $ P = R \inparallel \0 $ and $ Q = R $.
			By inverting the typing rules (and in particular [T-$ \0 $] and [T-Par] under arbitrary applications of [T-Sub]), then $ \Gamma \vdash P \smalltriangleright \Delta $ implies that $ \Gamma \vdash R \smalltriangleright \Delta_R' $ with $ \Delta_R' \leq_1 \Delta_R $, $ \Gamma \vdash \0 \smalltriangleright \emptyset $ with $ \emptyset \leq_1 \Delta_{\emptyset} $, and $ \Delta_R, \Delta_{\emptyset} \leq_1 \Delta $.
			By Lemma~\ref{lem:composeSubtyping}, then $ \Delta_R' \leq_1 \Delta_R, \Delta_{\emptyset} $.
			By Proposition~\ref{prop:subtypeTrans}, then $ \Delta_R' \leq_1 \Delta $.
			By [T-Sub], then $ \Gamma \vdash R \smalltriangleright \Delta_R' $ implies $ \Gamma \vdash Q \smalltriangleright \Delta $.

			Applying this rule from right to left, then $ P = R $ and $ Q = R \inparallel \0 $.
			By [T-$ \0 $] and [T-Par], then $ \Gamma \vdash P \smalltriangleright \Delta $ implies $ \Gamma \vdash Q \smalltriangleright \Delta $.
		\item[Case $ R_1 \inparallel R_2 \equiv R_2 \inparallel R_1 $:] Applying this rule from left to right, then $ P = R_1 \inparallel R_2 $ and $ Q = R_2 \inparallel R_1 $.
			By inverting the typing rules (and in particular [T-Par] under arbitrary applications of [T-Sub]), then $ \Gamma \vdash P \smalltriangleright \Delta $ implies $ \Gamma \vdash R_1 \smalltriangleright \Delta_{R_1}' $ with $ \Delta_{R_1}' \leq_1 \Delta_{R_1} $, $ \Gamma \vdash R_2 \smalltriangleright \Delta_{R_2}' $ with $ \Delta_{R_2}' \leq_1 \Delta_{R_2} $, and $ \Delta_{R_1}, \Delta_{R_2} \leq_1 \Delta $.
			By [T-Par] and [T-Sub], then $ \Gamma \vdash Q \smalltriangleright \Delta $.

			The case of applying this rule from right to left is symmetric.
		\item[Case $ \left( R_1 \inparallel R_2 \right) \inparallel R_3 \equiv R_1 \inparallel \left( R_2 \inparallel R_3 \right) $:] Applying this rule from left to right, then $ P = \left( R_1 \inparallel R_2 \right) \inparallel R_3 $ and $ Q = R_1 \inparallel \left( R_2 \inparallel R_3 \right) $.
			By inverting the typing rules (and in particular [T-Par] under arbitrary applications of [T-Sub]), then $ \Gamma \vdash P \smalltriangleright \Delta $ implies $ \Gamma \vdash R_1 \smalltriangleright \Delta_{R_1}' $ with $ \Delta_{R_1}' \leq_1 \Delta_{R_1} $, $ \Gamma \vdash R_2 \smalltriangleright \Delta_{R_2}' $ with $ \Delta_{R_2}' \leq_1 \Delta_{R_2} $, $ \Gamma \vdash R_3 \smalltriangleright \Delta_{R_3}' $ with $ \Delta_{R_3}' \leq_1 \Delta_{R_3} $, and $ \Delta_{R_1}, \Delta_{R_2}, \Delta_{R_3} \leq_1 \Delta $.
			By [T-Par] and [T-Sub], then $ \Gamma \vdash Q \smalltriangleright \Delta $.

			The case of applying this rule from right to left is similar.
		\item[Case $ \res{s}\0 \equiv \0 $:] Applying this rule from left to right, then $ P = \res{s}\0 $ and $ Q = \0 $.
			By inverting the typing rules (and in particular [T-Res] and [T-$ \0 $] under arbitrary applications of [T-Sub]), then $ \Gamma \vdash P \smalltriangleright \Delta $ implies $ \Gamma \vdash \0 \smalltriangleright \emptyset $ with $ \emptyset \leq_1 \Delta_{\emptyset} $ and $ \Delta_{\emptyset} \leq_1 \Delta $.
			By Proposition~\ref{prop:subtypeTrans}, then $ \emptyset \leq_1 \Delta $.
			By [T-$ \0 $] and [T-Sub], then $ \Gamma \vdash Q \smalltriangleright \Delta $.

			Applying this rule from right to left, then $ P = \0 $ and $ Q = \res{s}\0 $.
			By inverting the typing rules (and in particular [T-$ \0 $] under arbitrary applications of [T-Sub]), then $ \Gamma \vdash P \smalltriangleright \Delta $ implies $ \emptyset \leq_1 \Delta $.
			By [T-Res] and Lemma~\ref{lem:safeEmptyset}, then $ \Gamma \vdash Q \smalltriangleright \emptyset $.
			By [T-Sub], then $ \Gamma \vdash Q \smalltriangleright \Delta $.
		\item[Case $ \res{s_1}\res{s_2}R \equiv \res{s_2}\res{s_1}R $:] Applying this rule from left to right, have then $ P = \res{s_1}\res{s_2}R $ and $ Q = \res{s_2}\res{s_1}R $.
			By inverting the typing rules (and in particular [T-Res] under arbitrary applications of [T-Sub]), then $ \Gamma \vdash P \smalltriangleright \Delta $ implies $ \safe{\Delta_{s_1}} $ and $ \Gamma \vdash \res{s_2}R \smalltriangleright \Delta', \Delta_{s_1} $ with $ \Delta' \leq_1 \Delta $.
			Then $ \safe{\Delta_{s_2}} $ and $ \Gamma \vdash R \smalltriangleright \Delta_R, \Delta_{s_2} $ with $ \Delta_R \leq_1 \Delta', \Delta_{s_1} $.
			By Lemma~\ref{lem:splitSubtyping}, then $ \Delta_R \leq_1 \Delta', \Delta_{s_1} $ implies $ \Delta_R = \Delta'', \Delta_{s_1}' $; $ \Delta'' \leq_1 \Delta' $, and $ \Delta_{s_1}' \leq_1 \Delta_{s_1} $.
			By [T-Sub] and Corollary~\ref{cor:subtypePreorder}, then $ \Gamma \vdash R \smalltriangleright \Delta_R, \Delta_{s_2} $; $ \Delta_R = \Delta'', \Delta_{s_1}' $, and $ \Delta_{s_1}' \leq_1 \Delta_{s_1} $ imply $ \Gamma \vdash R \smalltriangleright \Delta'', \Delta_{s_2}, \Delta_{s_1} $.
			By [T-Res], then $ \safe{\Delta_{s_1}} $ and $ \Gamma \vdash R \smalltriangleright \Delta'', \Delta_{s_2}, \Delta_{s_1} $ imply $ \Gamma \vdash \res{s_1}R \smalltriangleright \Delta'', \Delta_{s_2} $.
			By [T-Res], then $ \safe{\Delta_{s_2}} $ and $ \Gamma \vdash \res{s_1}R \smalltriangleright \Delta'', \Delta_{s_2} $ imply $ \Gamma \vdash Q \smalltriangleright \Delta'' $.
			By [T-Sub] and Proposition~\ref{prop:subtypeTrans}, then $ \Gamma \vdash Q \smalltriangleright \Delta'' $; $ \Delta'' \leq_1 \Delta' $, and $ \Delta' \leq_1 \Delta $ imply $ \Gamma \vdash Q \smalltriangleright \Delta $.

			The case of applying this rule from right to left is similar.
		\item[Case $ R_1 \inparallel \res{s}R_2 \equiv \res{s}\left( R_1 \inparallel R_2 \right) $ if $ s \notin \fs{R_1} $:] Applying this rule from left to right, then $ P = R_1 \inparallel res{s}R_2 $, $ Q = \res{s}\left( R_1 \inparallel R_2 \right) $, and $ s \notin \fs{R_1} $.
			By inverting the typing rules (and in particular [T-Par] and [T-Res] under arbitrary applications of [T-Sub]), then $ \Gamma \vdash P \smalltriangleright \Delta $ implies $ \Gamma \vdash R_1 \smalltriangleright \Delta_1' $ with $ \Delta_1' \leq_1 \Delta_1 $, $ \safe{\Delta_{s}} $, $ \Gamma \vdash R_2 \smalltriangleright \Delta_2', \Delta_s $ with $ \Delta_2' \leq_1 \Delta_2 $, and $ \Delta = \Delta_1, \Delta_2 $, where we ensure by applying alpha-conversion that $ s $ is fresh in $ \Delta_1 $ and $ \Delta_1' $.
			By Lemma~\ref{lem:composeSubtyping}, then $ \Delta_1' \leq_1 \Delta_1 $; $ \Delta_2' \leq_1 \Delta_2 $, and $ \Delta = \Delta_1, \Delta_2 $ imply $ \Delta_1', \Delta_2' \leq_1 \Delta $.
			By [T-Par], then $ \Gamma \vdash R_1 \smalltriangleright \Delta_1' $ and $ \Gamma \vdash R_2 \smalltriangleright \Delta_2', \Delta_s $ imply $ \Gamma \vdash R_1 \inparallel R_2 \smalltriangleright \Delta_1', \Delta_2', \Delta_s $.
			By [T-Res], then $ \safe{\Delta_{s}} $ and $ \Gamma \vdash R_1 \inparallel R_2 \smalltriangleright \Delta_1', \Delta_2', \Delta_s $ imply $ \Gamma \vdash Q \smalltriangleright \Delta_1', \Delta_2' $.
			By [T-Sub], then $ \Gamma \vdash Q \smalltriangleright \Delta_1', \Delta_2' $ and $ \Delta_1', \Delta_2' \leq_1 \Delta $ imply $ \Gamma \vdash Q \smalltriangleright \Delta $.

			Applying this rule from right to left is similar.
		\item[Case \hspace{-1pt}$ \procdef{D}{\res{s}R} \hspace{-1pt}\equiv\hspace{-1.5pt} \res{s}\procdef{D}{R} $ if $ s \hspace{-1pt}\notin\hspace{-1.5pt} \fs{D} $:] \hspace{-1.8pt}Then $ D \hspace{-1pt}=\hspace{-1pt} \lbrace X_i{\left( \vecnew{x_i}, c_{i, 1}, \mydots, c_{i, n_i} \right)} \hspace{-1pt}= R_i \rbrace_{i \in I} $.
			Applying this rule from left to right, then $ P = \procdef{D}{\res{s}R} $, $ Q = \res{s}\procdef{D}{R} $, and $ s \notin \fs{D} $.
			By inverting the typing rules (and in particular [T-Def] and [T-Res] under arbitrary applications of [T-Sub]), then $ \Gamma \hspace{-1pt}\vdash\hspace{-2pt} P \smalltriangleright \Delta $ implies $ \Gamma, \type{X_1}{\left\langle \vecnew{U_1}, T_{1, 1}, \mydots T_{1, n_1} \right\rangle}, \mydots, \type{X_i}{\left\langle \vecnew{U_i}, T_{i, 1}, \mydots T_{i, n_i} \right\rangle}, \type{\vecnew{x_i}}{\vecnew{U_i}} \vdash R_i \smalltriangleright \type{c_{i, 1}}{T_{i, 1}}, \mydots, \type{c_{i, n_i}}{T_{i, n_i}} $ for all $ i \in I $, $ \Gamma, \left\lbrace \type{X_i}{\left\langle \vecnew{U_i}, T_{i, 1}, \mydots, T_{i, n_i} \right\rangle} \right\rbrace_{i \in I} \vdash \res{s}{R} \smalltriangleright \Delta' $ with $ \Delta' \leq_1 \Delta $, $ \safe{\Delta_{s}} $, and $ \Gamma, \left\lbrace \type{X_i}{\left\langle \vecnew{U_i}, T_{i, 1}, \mydots, T_{i, n_i} \right\rangle} \right\rbrace_{i \in I} \vdash R \smalltriangleright \Delta_R', \Delta_s $ with $ \Delta_R' \leq_1 \Delta' $, where we ensure by applying alpha-conversion that $ s $ is fresh in $ D $.
			By [T-Def], then for all $ i \in I $\\
            $\Gamma\hspace{-2pt}, \type{X_1}{\left\langle \vecnew{U_1}, T_{1, 1}, \mydots T_{1, n_1} \right\rangle}, \mydots, \type{X_i}{\left\langle \vecnew{U_i}, T_{i, 1}, \mydots T_{i, n_i} \right\rangle}, \type{\vecnew{x_i}}{\vecnew{U_i}} \vdash R_i \smalltriangleright \type{c_{i, 1}}{T_{i, 1}}, \mydots, \type{c_{i, n_i}}{T_{i, n_i}}$, and $ \Gamma, \left\lbrace \type{X_i}{\left\langle \vecnew{U_i}, T_{i, 1}, \mydots, T_{i, n_i} \right\rangle} \right\rbrace_{i \in I} \vdash R \smalltriangleright \Delta_R', \Delta_s $ imply $ \Gamma \vdash \procdef{D}{R} \smalltriangleright \Delta_R', \Delta_s $.
			By [T-Res], then $ \safe{\Delta_{s}} $ and $ \Gamma \vdash \procdef{D}{R} \smalltriangleright \Delta_R', \Delta_s $ imply $ \Gamma \vdash Q \smalltriangleright \Delta_R' $.
			By [T-Sub] and Proposition~\ref{prop:subtypeTrans}, then $ \Gamma \vdash Q \smalltriangleright \Delta_R' $; $ \Delta_R' \leq_1 \Delta' $, and $ \Delta' \leq_1 \Delta $ imply $ \Gamma \vdash Q \smalltriangleright \Delta $.

			Applying this rule from right to left, then $ P = \res{s}\procdef{D}{R} $ and $ Q = \procdef{D}{\res{s}R} $, where $ s \notin \fs{D} $.
			By inverting the typing rules (and in particular [T-Def] and [T-Res] under arbitrary applications of [T-Sub]), then $ \Gamma \vdash P \smalltriangleright \Delta $ implies $ \safe{\Delta_{s}} $, $ \Gamma \vdash \procdef{D}{R} \smalltriangleright \Delta', \Delta_s $ with $ \Delta' \leq_1 \Delta $, and for all $ i \in I $\\
            $\Gamma\hspace{-2pt}, \type{X_1}{\left\langle \vecnew{U_1}, T_{1, 1}, \mydots T_{1, n_1} \right\rangle}, \mydots, \type{X_i}{\left\langle \vecnew{U_i}, T_{i, 1}, \mydots T_{i, n_i} \right\rangle}, \type{\vecnew{x_i}}{\vecnew{U_i}} \vdash R_i \smalltriangleright \type{c_{i, 1}}{T_{i, 1}}, \mydots, \type{c_{i, n_i}}{T_{i, n_i}}$, and $ \Gamma, \left\lbrace \type{X_i}{\left\langle \vecnew{U_i}, T_{i, 1}, \mydots, T_{i, n_i} \right\rangle} \right\rbrace_{i \in I} \vdash R \smalltriangleright \Delta_R $ with $ \Delta_R \leq_1 \Delta', \Delta_s $, where we ensure by applying alpha-conversion that $ s $ is fresh in $ D $.
			By [T-Sub], then $ \Delta_R \leq_1 \Delta', \Delta_s $ and $ \Gamma, \left\lbrace \type{X_i}{\left\langle \vecnew{U_i}, T_{i, 1}, \mydots, T_{i, n_i} \right\rangle} \right\rbrace_{i \in I} \vdash R \smalltriangleright \Delta_R $ imply $ \Gamma, \left\lbrace \type{X_i}{\left\langle \vecnew{U_i}, T_{i, 1}, \mydots, T_{i, n_i} \right\rangle} \right\rbrace_{i \in I} \vdash R \smalltriangleright \Delta', \Delta_s $.
			By [T-Res], then $ \safe{\Delta_{s}} $ and $ \Gamma\hspace{-1pt}, \left\lbrace \type{X_i}{\left\langle \vecnew{U_i}, T_{i, 1}, \mydots, T_{i, n_i} \right\rangle} \right\rbrace_{i \in I} \hspace{-5.6pt}\vdash R \smalltriangleright \Delta', \Delta_s $ imply $ \Gamma, \left\lbrace \type{X_i}{\left\langle \vecnew{U_i}, T_{i, 1}, \mydots, T_{i, n_i} \right\rangle} \right\rbrace_{i \in I} \vdash \res{s}R \smalltriangleright \Delta' $.
			By [T-Def], then
			\begin{align*}
				\Gamma, \type{X_1}{\left\langle \vecnew{U_1}, T_{1, 1}, \mydots T_{1, n_1} \right\rangle}, \mydots, \type{X_i}{\left\langle \vecnew{U_i}, T_{i, 1}, \mydots T_{i, n_i} \right\rangle}, \type{\vecnew{x_i}}{\vecnew{U_i}} \vdash R_i \smalltriangleright \type{c_{i, 1}}{T_{i, 1}}, \mydots, \type{c_{i, n_i}}{T_{i, n_i}}
			\end{align*}
			for all $ i \in I $ and $ \Gamma, \left\lbrace \type{X_i}{\left\langle \vecnew{U_i}, T_{i, 1}, \mydots, T_{i, n_i} \right\rangle} \right\rbrace_{i \in I} \vdash \res{s}R \smalltriangleright \Delta' $ imply $ \Gamma \vdash Q \smalltriangleright \Delta' $.
			By [T-Sub], then $ \Gamma \vdash Q \smalltriangleright \Delta' $ and $ \Delta' \leq_1 \Delta $ imply $ \Gamma \vdash Q \smalltriangleright \Delta $.
		\item[Case $ \left( \procdef{D}{R_1} \right) \inparallel R_2 \equiv \procdef{D}{\left( R_1 \inparallel R_2 \right)} $ if $ \dpv{D} \cap \fpv{Q} = \emptyset $:] Then:
			\begin{align*}
				D = \left\lbrace X_i{\left( \vecnew{x_i}, c_{i, 1}, \ldots, c_{i, n_i} \right)} = R_i' \right\rbrace_{i \in I}
			\end{align*}

			Applying this rule from left to right, we then get $ P = \left( \procdef{D}{R_1} \right) \inparallel R_2 $, $ Q = \procdef{D}{\left( R_1 \inparallel R_2 \right)} $, and $ \dpv{D} \cap \fpv{Q} = \emptyset $.
			By inverting the typing rules (in particular [T-Def] and [T-Par] under arbitrary appl.\! of [T-Sub]), $ \Gamma \hspace{-2pt}\vdash\hspace{-3pt} P \smalltriangleright \Delta $ implies $\Gamma, \type{X_1}{\left\langle \vecnew{U_1}, T_{1, 1}, \mydots T_{1, n_1} \right\rangle}, \mydots, \type{X_i}{\left\langle \vecnew{U_i}, T_{i, 1}, \mydots T_{i, n_i} \right\rangle}, \type{\vecnew{x_i}}{\vecnew{U_i}} \vdash R_i' \smalltriangleright \type{c_{i, 1}}{T_{i, 1}}, \mydots, \type{c_{i, n_i}}{T_{i, n_i}}$
			for all $ i \in I $, $ \Gamma, \left\lbrace \type{X_i}{\left\langle \vecnew{U_i}, T_{i, 1}, \mydots, T_{i, n_i} \right\rangle} \right\rbrace_{i \in I} \vdash R_1 \smalltriangleright \Delta_1' $ with $ \Delta_1' \leq_1 \Delta_1 $, $ \Gamma \vdash R_2 \smalltriangleright \Delta_2' $ with $ \Delta_2' \leq_1 \Delta_2 $, and $ \Delta_1, \Delta_2 \leq_1 \Delta $.
			By Lemma~\ref{lem:composeSubtyping}, then $ \Delta_1' \leq_1 \Delta_1 $; $ \Delta_2' \leq_1 \Delta_2 $, and $ \Delta = \Delta_1, \Delta_2 $ imply $ \Delta_1', \Delta_2' \leq_1 \Delta $.
			By Lemma~\ref{lem:globalWeakening}, then $ \Gamma \vdash R_2 \smalltriangleright \Delta_2' $ implies $ \Gamma, \left\lbrace \type{X_i}{\left\langle \vecnew{U_i}, T_{i, 1}, \mydots, T_{i, n_i} \right\rangle} \right\rbrace_{i \in I} \vdash R_2 \smalltriangleright \Delta_2' $.
			By rule [T-Par], we then have that $ \Gamma, \left\lbrace \type{X_i}{\left\langle \vecnew{U_i}, T_{i, 1}, \mydots, T_{i, n_i} \right\rangle} \right\rbrace_{i \in I} \vdash R_1 \smalltriangleright \Delta_1' $ and $ \Gamma, \left\lbrace \type{X_i}{\left\langle \vecnew{U_i}, T_{i, 1}, \mydots, T_{i, n_i} \right\rangle} \right\rbrace_{i \in I} \vdash R_2 \smalltriangleright \Delta_2' $ imply $ \Gamma, \left\lbrace \type{X_i}{\left\langle \vecnew{U_i}, T_{i, 1}, \mydots, T_{i, n_i} \right\rangle} \right\rbrace_{i \in I} \vdash R_1 \inparallel R_2 \smalltriangleright \Delta_1', \Delta_2' $.
			By [T-Def], then $\Gamma, \type{X_1}{\left\langle \vecnew{U_1}, T_{1, 1}, \mydots T_{1, n_1} \right\rangle}, \mydots, \type{X_i}{\left\langle \vecnew{U_i}, T_{i, 1}, \mydots T_{i, n_i} \right\rangle}, \type{\vecnew{x_i}}{\vecnew{U_i}} \vdash R_i' \smalltriangleright \type{c_{i, 1}}{T_{i, 1}}, \mydots, \type{c_{i, n_i}}{T_{i, n_i}}$
			for all $ i \in I $ and $ \Gamma, \left\lbrace \type{X_i}{\left\langle \vecnew{U_i}, T_{i, 1}, \mydots, T_{i, n_i} \right\rangle} \right\rbrace_{i \in I} \vdash R_1 \inparallel R_2 \smalltriangleright \Delta_1', \Delta_2' $ imply $ \Gamma \vdash Q \smalltriangleright \Delta_1', \Delta_2' $.
			By [T-Sub], then $ \Gamma \vdash Q \smalltriangleright \Delta_1', \Delta_2' $ and $ \Delta_1', \Delta_2' \leq_1 \Delta $ imply $ \Gamma \vdash Q \smalltriangleright \Delta $.

			Applying this rule from right to left is similar.
		\item[Case $ \procdef{D}{\left( \procdef{D'}{R} \right)} \equiv \procdef{D \cup D'}{R} $ if $ \dpv{D} \cap \dpv{D'} = \emptyset $:] Then:
			\begin{align*}
				D &= \left\lbrace X_i{\left( \vecnew{x_i}, c_{i, 1}, \mydots, c_{i, n_i} \right)} = R_i' \right\rbrace_{i \in I}\\
				D' &= \left\lbrace Y_j{\left( \vecnew{y_j}, d_{j, 1}, \mydots, d_{j, n_j} \right)} = R_j'' \right\rbrace_{j \in J}
			\end{align*}

			Applying this rule from left to right, then $ P = \procdef{D}{\left( \procdef{D'}{R} \right)} $, $ Q = \procdef{D \cup D'}{R} $, and $ \dpv{D} \cap \dpv{D'} = \emptyset $.
			By inverting the typing rules (and in particular [T-Def] under arbitrary applications of [T-Sub]), then $ \Gamma \vdash P \smalltriangleright \Delta $ implies
			\begin{align*}
				\Gamma\hspace{-2pt}, &\type{X_1}{\left\langle \vecnew{U_1}, T_{1, 1}, \mydots T_{1, n_1} \right\rangle}\hspace{-1pt}, \mydots, \type{X_i}{\left\langle \vecnew{U_i}, T_{i, 1}, \mydots T_{i, n_i} \right\rangle}, \type{\vecnew{x_i}}{\vecnew{U_i}} \vdash R_i' \smalltriangleright \type{c_{i, 1}}{T_{i, 1}}, \mydots, \type{c_{i, n_i}}{T_{i, n_i}}\\
				\Gamma\hspace{-2pt}, &\type{X_1}{\left\langle \vecnew{U_1}, T_{1, 1}, \mydots T_{1, n_1} \right\rangle}\hspace{-1pt}, \mydots, \type{X_i}{\left\langle \vecnew{U_i}, T_{i, 1}, \mydots T_{i, n_i} \right\rangle}, \type{\vecnew{x_i}}{\vecnew{U_i}},\\
				&\type{Y_1}{\left\langle \vecnew{U_1'}, T_{1, 1}', \mydots T_{1, n_1}' \right\rangle}\hspace{-1pt}, \mydots, \type{Y_j}{\left\langle \vecnew{U_j'}, T_{j, 1}', \mydots T_{j, n_j}' \right\rangle}, \type{\vecnew{y_j}}{\vecnew{U_j'}} \vdash R_j'' \smalltriangleright \type{c_{j, 1}'}{T_{j, 1}'}, \mydots, \type{c_{j, n_j}'}{T_{j, n_j}'}
			\end{align*}
			for all $ i \in I $ and all $ j \in J $, $ \Gamma, \left\lbrace \type{X_i}{\left\langle \vecnew{U_i}, T_{i, 1}, \mydots, T_{i, n_i} \right\rangle} \right\rbrace_{i \in I} \vdash \procdef{D'}{R} \smalltriangleright \Delta' $ with $ \Delta' \leq_1 \Delta $, and
			\begin{align*}
				& \Gamma, \type{X_1}{\left\langle \vecnew{U_1}, T_{1, 1}, \mydots T_{1, n_1} \right\rangle}, \mydots, \type{X_i}{\left\langle \vecnew{U_i}, T_{i, 1}, \mydots T_{i, n_i} \right\rangle},\\
				& \quad \type{Y_1}{\left\langle \vecnew{U_1'}, T_{1, 1}', \mydots T_{1, n_1}' \right\rangle}, \mydots, \type{Y_j}{\left\langle \vecnew{U_j'}, T_{j, 1}', \mydots T_{j, n_j}' \right\rangle} \vdash R \smalltriangleright \Delta''
			\end{align*}
			with $ \Delta'' \leq_1 \Delta' $.
			By [T-Def], then $ \Gamma \vdash Q \smalltriangleright \Delta_2' $.
			By [T-Sub] and Proposition~\ref{prop:subtypeTrans}, then $ \Gamma \vdash Q \smalltriangleright \Delta'' $; $ \Delta'' \leq_1 \Delta' $, and $ \Delta' \leq_1 \Delta $ imply $ \Gamma \vdash Q \smalltriangleright \Delta $.

			Applying this rule from right to left is similar.\qedhere
	\end{description}
\end{proof}

Lemmas~\ref{lem:typeReductionSplit}, and~\ref{lem:typeReductionCompose} give a certain notion of splitting and composing the transitions of local contexts.
Lemma~\ref{lem:subtypeTypeReduction} states a useful consequence of the subtyping properties of Theorem~\ref{thm:subtypingProperties} on supertypes being able to ``follow'' the transitions of their subtypes.
\begin{lemma}
	\label{lem:typeReductionSplit}
	If $ \Delta_1, \Delta_2 \mapsto^{*}_{\prob} \Delta $ and $ \s{\Delta_1} \cap \s{\Delta_2} = \emptyset $ then there are $ \Delta_1', \Delta_2' $ such that $ \Delta_1 \mapsto^{*}_{\prob_1} \Delta_1' $; $ \Delta_2 \mapsto^{*}_{\prob_2} \Delta_2' $, $ \prob = \prob_1 \prob_2 $, and $ \Delta = \Delta_1', \Delta_2' $.
\end{lemma}

\begin{proof}
	By Definition~\ref{def:transitionsLocalContexts}, local contexts of different sessions cannot interact.
	Then every sequence of steps $ \Delta_1, \Delta_2 \mapsto^{*}_{\prob} \Delta $ can be split such that $ \Delta_1 \mapsto^{*}_{\prob_1} \Delta_1' $; $ \Delta_2 \mapsto^{*}_{\prob_2} \Delta_2' $, $ \prob = \prob_1 \prob_2 $, and $ \Delta = \Delta_1', \Delta_2' $
	%\hfill $ \square $
\end{proof}

\begin{lemma}
	\label{lem:subtypeTypeReduction}
	If $ \Delta_1 \leq_1 \Delta_2 $, $ \safe{\Delta_2} $, and $ \Delta_1 \mapsto^{*}_{\prob} \Delta_1' $ then there is some $ \Delta_2' $ such that $ \Delta_2 \mapsto^{*}_{\prob} \Delta_2' $ and $ \Delta_1' \leq_1 \Delta_2' $.
\end{lemma}

\begin{proof}
	Assume $ \Delta_1 \leq_1 \Delta_2 $, $ \safe{\Delta_2} $, and $ \Delta_1 \mapsto^{*}_{\prob} \Delta_1' $.
	By Theorem~\ref{thm:subtypingProperties}.\ref{thm:subtypingProperties:safe}(b), then there are $ \Delta_2', \prob_2, \prob_3 $ such that $ \Delta_2 \mapsto_{\prob\prob_2}^{*} \Delta_2' $; $ \Delta_1' \leq_{\prob_3} \Delta_2' $, and $ 1 = \prob_2\prob_3 $.
	Because of $ 1 = \prob_2\prob_3 $, then $ \prob_2 = 1 $ and $ \prob_3 = 1 $.
	Then $ \Delta_2 \mapsto_{\prob}^{*} \Delta_2' $ and $ \Delta_1' \leq_1 \Delta_2' $.
	%\hfill $ \square $
\end{proof}

\begin{lemma}
	\label{lem:typeReductionCompose}
	If $ \Delta_1 \mapsto^{*}_{\prob_1} \Delta_1' $; $ \Delta_2 \mapsto^{*}_{\prob_2} \Delta_2' $, $\safe{\Delta} $, and $ \Delta_1, \Delta_2 \leq_1 \Delta $ then there is some $ \Delta' $ such that $ \Delta \mapsto^{*}_{\prob_1\prob_2} \Delta' $ and $ \Delta_1', \Delta_2' \leq_1 \Delta' $.
\end{lemma}

\begin{proof}
	Assume $ \Delta_1 \mapsto^{*}_{\prob_1} \Delta_1' $; $ \Delta_2 \mapsto^{*}_{\prob_2} \Delta_2' $, $\safe{\Delta} $, and $ \Delta_1, \Delta_2 \leq_1 \Delta $.
	Then $ \Delta_1, \Delta_2 \mapsto^{*}_{\prob_1\prob_2} \Delta_1', \Delta_2' $.
	By Lemma~\ref{lem:subtypeTypeReduction}, then there is some $ \Delta' $ such that $ \Delta \mapsto^{*}_{\prob_1\prob_2} \Delta' $ and $ \Delta_1', \Delta_2' \leq_1 \Delta' $.
	%\hfill $ \square $
\end{proof}

The final lemma needed for showing subject reduction establishes the standard substitution property.
\begin{lemma}[Substitution]
	\label{lem:substitution}
	If $ \Gamma \vdash \type{v}{U} $ and $ \Gamma, \type{x}{U} \vdash P \smalltriangleright \Delta $, then $ \Gamma \vdash P\subst{x}{v} \smalltriangleright \Delta $.
\end{lemma}

\begin{proof}
	Assume $ \Gamma \vdash \type{v}{U} $ and $ \Gamma, \type{x}{U} \vdash P \smalltriangleright \Delta $.
	We prove $ \Gamma \vdash P\subst{x}{v} \smalltriangleright \Delta $ by structural induction on the derivation of $ \Gamma, \type{x}{U} \vdash P \smalltriangleright \Delta $.
	\begin{description}
		\item[Case of {[T-$ \0 $]}:] In this case $ P = \0 $ and $ \Delta = \emptyset $.
			Then $ P\subst{x}{v} = \0 $.
			By [T-$ \0 $], then $ \Gamma \vdash P\subst{x}{v} \smalltriangleright \Delta $.
		\item[Case of {[T-Res]}:] In this case $ P = \res{s}Q $ and $ \Gamma, \type{x}{U} \vdash Q \smalltriangleright \Delta, \left\lbrace \type{\channelk{s}{r}{i}}{T_i} \right\rbrace_{i \in I} $ with $ \safe{\left\lbrace \type{\channelk{s}{r}{i}}{T_i} \right\rbrace_{i \in I}} $, where we use alpha-conversion to ensure that $ s \neq x $.
			Then $ P\subst{x}{v} = \res{s}{\left( Q\subst{x}{v} \right)} $.
			By the induction hypothesis, then $ \Gamma \vdash \type{v}{U} $ and $ \Gamma, \type{x}{U} \vdash Q \smalltriangleright \Delta, \left\lbrace \type{\channelk{s}{r}{i}}{T_i} \right\rbrace_{i \in I} $ imply $ \Gamma \vdash Q\subst{x}{v} \smalltriangleright \Delta, \left\lbrace \type{\channelk{s}{r}{i}}{T_i} \right\rbrace_{i \in I} $.
			By [T-Res], then $ \safe{\left\lbrace \type{\channelk{s}{r}{i}}{T_i} \right\rbrace_{i \in I}} $ and $ \Gamma\hspace{-2pt} \vdash \hspace{-1pt}Q\subst{x}{v} \smalltriangleright \Delta, \left\lbrace \type{\channelk{s}{r}{i}}{T_i} \right\rbrace_{i \in I} $ imply $ \Gamma\hspace{-2pt} \vdash \hspace{-2pt} P\subst{x}{v} \smalltriangleright \Delta $.
		\item[Case of {[T-If]}:] In this case $ P = \cond{v_b}{Q}{R} $, where $ \Gamma, \type{x}{U} \vdash \type{v_b}{\boolT} $, and $ \Gamma, \type{x}{U} \vdash Q \smalltriangleright \Delta $ and $ \Gamma, \type{x}{U} \vdash R \smalltriangleright \Delta $.
			Then we have the substitution $ P\subst{x}{v} = \cond{{\left( v_b\subst{x}{v} \right)}}{{\left( Q\subst{x}{v} \right)}}{{\left( R\subst{x}{v} \right)}} $.
			By Definition~\ref{def:typingRules}, $ \Gamma, \type{x}{U} \vdash \type{v_b}{\boolT} $ implies one of the following three cases:
			\begin{description}
				\item[$ v_b = y $, $ x \neq y $, and $ \type{y}{\boolT} \in \Gamma, \type{x}{U} $:] Then $ \type{y}{\boolT} \in \Gamma $.
					By [T-Base], then $ \Gamma \vdash \type{y}{\boolT} $.
					Moreover, $ v_b\subst{x}{v} = v_b $.
					Hence, $ \Gamma \vdash \type{v_b\subst{x}{v}}{\boolT} $.
				\item[$ v_b = x $:] Then $ \Gamma, \type{x}{U} \vdash \type{v_b}{\boolT} $ implies $ U = \boolT $.
					Moreover, $ v_b\subst{x}{v} = v $.
					By $ \Gamma \vdash \type{v}{U} $, then $ \Gamma \vdash \type{v_b\subst{x}{v}}{\boolT} $.
				\item[$ v_b \in \left\lbrace \true, \false \right\rbrace $:] Then $ v_b\subst{x}{v} = v_b $.
					By [Bool], then $ \Gamma \vdash \type{v_b\subst{x}{v}}{\boolT} $.
			\end{description}
			In all three cases we have $ \Gamma \vdash \type{v_b\subst{x}{v}}{\boolT} $.
			By the induction hypothesis, $ \Gamma \vdash \type{v}{U} $ and $ \Gamma, \type{x}{U} \vdash Q \smalltriangleright \Delta $ imply $ \Gamma \vdash Q\subst{x}{v} \smalltriangleright \Delta $, and $ \Gamma \vdash \type{v}{U} $ and $ \Gamma, \type{x}{U} \vdash R \smalltriangleright \Delta $ imply $ \Gamma \vdash R\subst{x}{v} \smalltriangleright \Delta $.
			By [T-If], then $ \Gamma \vdash \type{v_b\subst{x}{v}}{\boolT} $, $ \Gamma \vdash Q\subst{x}{v} \smalltriangleright \Delta $, and $ \Gamma \vdash R\subst{x}{v} \smalltriangleright \Delta $ imply $ \Gamma \vdash P\subst{x}{v} \smalltriangleright \Delta $.
		\item[Case of {[T-Par]}:] In this case $ P = Q \inparallel R $, $ \Delta = \Delta_1, \Delta_2 $, $ \Gamma, \type{x}{U} \vdash Q \smalltriangleright \Delta_1 $, and $ \Gamma, \type{x}{U} \vdash R \smalltriangleright \Delta_2 $.
			Then $ P\subst{x}{v} = \left( Q\subst{x}{v} \right) \inparallel \left( R\subst{x}{v} \right) $.
			By the induction hypothesis, then $ \Gamma \vdash \type{v}{U} $ and $ \Gamma, \type{x}{U} \vdash Q \smalltriangleright \Delta_1 $ imply $ \Gamma \vdash Q\subst{x}{v} \smalltriangleright \Delta_1 $, and $ \Gamma \vdash \type{v}{U} $ and $ \Gamma, \type{x}{U} \vdash R \smalltriangleright \Delta_2 $ imply $ \Gamma \vdash R\subst{x}{v} \smalltriangleright \Delta_2 $.
			By [T-Par], then $ \Gamma \vdash Q\subst{x}{v} \smalltriangleright \Delta_1 $ and $ \Gamma \vdash R\subst{x}{v} \smalltriangleright \Delta_2 $ imply $ \Gamma \vdash P\subst{x}{v} \smalltriangleright \Delta $.
		\item[Case of {[T-Def]}:] %
            In this case we have $ P = \procdef{\procdecl{X}{\vecnew{y}}{c_1, \mydots, c_n} \defeq Q}{R} $ with $\Gamma, \type{X}{\left\langle \vecnew{U'}, T_1, \mydots, T_n \right\rangle}, \type{\vecnew{y}}{\vecnew{U'}}, \type{x}{U} \vdash Q \smalltriangleright \type{c_1}{T_1}, \mydots, \type{c_n}{T_n}$, and $ \Bigl( \Gamma, \type{X}{\left\langle \vecnew{U'}, T_1, \mydots, T_n \right\rangle},\break \type{x}{U} \Bigr) \vdash R \smalltriangleright \Delta $, where we use alpha-conversion to ensure that $ x \notin \vecnew{y} $.
			Then $ P\subst{x}{v} = \procdef{\procdecl{X}{\vecnew{y}}{c_1, \mydots, c_n} \defeq {\left( Q\subst{x}{v} \right)}}{{\left( R \subst{x}{v} \right)}} $.
			By the induction hypothesis, then $ \Gamma \vdash \type{v}{U} $ and $\Gamma, \type{X}{\left\langle \vecnew{U'}, T_1, \mydots, T_n \right\rangle}, \type{\vecnew{y}}{\vecnew{U'}}, \type{x}{U} \vdash Q \smalltriangleright \type{c_1}{T_1}, \mydots, \type{c_n}{T_n}$ imply $ \Gamma, \type{X}{\left\langle \vecnew{U'}, T_1, \mydots, T_n \right\rangle}, \type{\vecnew{y}}{\vecnew{U'}} \vdash Q\subst{x}{v} \smalltriangleright \type{c_1}{T_1}, \mydots, \type{c_n}{T_n} $.
            Moreover $ \Gamma \vdash \type{v}{U} $ and $ \Gamma, \type{X}{\left\langle \vecnew{U'}, T_1, \mydots, T_n \right\rangle}, \type{x}{U} \vdash R \smalltriangleright \Delta $ imply $ \Gamma\hspace{-1.2pt}, \type{X}{\left\langle \vecnew{U'}, T_1, \mydots, T_n \right\rangle} \vdash R\subst{x}{v} \smalltriangleright \Delta $.
			By [T-Def], then have $ \Gamma \vdash P\subst{x}{v} \smalltriangleright \Delta $.
		\item[Case of {[T-Var]}:] In this case $ P = \processcall{X}{\vecnew{v'}}{c_1, \mydots, c_n} $, $ \type{X}{\left\langle \vecnew{U'}, T_1, \mydots, T_n \right\rangle} \in \Gamma, \type{x}{U} $, $ \Delta = \type{c_1}{T_1}, \mydots, \type{c_n}{T_n} $, and $ \Gamma, \type{x}{U} \vdash \type{\vecnew{v'}}{\vecnew{U'}} $.
			Then $ P\subst{x}{v} = \processcall{X}{{\left( \vecnew{v'}\subst{x}{v} \right)}}{c_1, \mydots, c_n} $.
			By Definition~\ref{def:typingRules}, $ \Gamma, \type{x}{U} \vdash \type{\vecnew{v'}}{\vecnew{U'}} $, \ie $ \Gamma, \type{x}{U} \vdash \type{v'_i}{U'_i} $ for all $ v'_i \in \vecnew{v'} $, implies for every $ v'_i $ one of the following four cases:
			\begin{description}
				\item[Case $ v'_i = y $, $ x \neq y $, and $ \type{y}{U'_i} \in \Gamma, \type{x}{U} $:] Then $ \type{y}{U'_i} \in \Gamma $.
					By [T-Base], then $ \Gamma \vdash \type{y}{U'_i} $.
					Then, $ v'_i\subst{x}{v} = v'_i $.
					Hence, $ \Gamma \vdash \type{v'_i\subst{x}{v}}{U'_i} $.
				\item[Case $ v'_i = x $:] Then $ \Gamma, \type{x}{U} \vdash \type{v'_i}{\boolT} $ implies $ U = U'_i $.
					Moreover, $ v'_i\subst{x}{v} = v $.
					By $ \Gamma \vdash \type{v}{U} $, then $ \Gamma \vdash \type{v'_i\subst{x}{v}}{U'_i} $.
				\item[Case $ v'_i \in \mathbb{N}^{+} $:] Then $ U'_i \hspace{-1pt}=\hspace{-1pt} \natT $ and $ v'_i\subst{x}{v} \hspace{-1pt}=\hspace{-1pt} v'_i $.
					By [Nat], then $ \Gamma \vdash \type{v'_i\subst{x}{v}}{U'_i} $.
				\item[Case $ v'_i \in \left\lbrace \true, \false \right\rbrace $:] Then $ U'_i = \boolT $ and $ v'_i\subst{x}{v} = v'_i $.
					By [Bool], then $ \Gamma \hspace{-1pt}\vdash\hspace{-1pt} \type{v'_i\subst{x}{v}}{U'_i} $.
			\end{description}
			In all four cases we have $ \Gamma \vdash \type{v'_i\subst{x}{v}}{U'_i} $ and, thus, $ \Gamma \vdash \type{{\left( \vecnew{v'}\subst{x}{v} \right)}}{\vecnew{U'}} $.
			By [T-Var], then $ \Gamma \vdash P\subst{x}{v} \smalltriangleright \Delta $.
		\item[Case of {[T-Sum]}:] In this case $ P \hspace{-1pt} = c\mc{i}{I} M_i $, $ \Delta = \Delta_0, \type{c}{\mc{i}{I} L_i} $, and for all $ i \in I $ we have $ \Gamma, \type{x}{U} \vdash \insess{c}M_i \smalltriangleright \Delta_0, \type{c}{L_i} $.
			Then $ P\subst{x}{v} = c\mc{i}{I} {\left( M_i\subst{x}{v} \right)} $.
			By the induction hypothesis, $ \Gamma \vdash \type{v}{U} $ and $ \Gamma, \type{x}{U} \vdash \insess{c}M_i \smalltriangleright \Delta_0, \type{c}{L_i} $ imply $ \Gamma \vdash \insess{c}{\left( M_i\subst{x}{v} \right)} \smalltriangleright \Delta_0, \type{c}{L_i} $ for all $ i \in I $.
			BY [T-Sum], then $ \Gamma \vdash P\subst{x}{v} \smalltriangleright \Delta $.
		\item[Case of {[T-Sub]}:] In this case $ \Gamma, \type{x}{U} \vdash P \smalltriangleright \Delta' $ and $ \Delta' \leq_1 \Delta $.
			By the induction hypothesis, then $ \Gamma \vdash \type{v}{U} $ and $ \Gamma, \type{x}{U} \vdash P \smalltriangleright \Delta' $ imply $ \Gamma \vdash P\subst{x}{v} \smalltriangleright \Delta' $.
			By [T-Sub], then $ \Gamma \vdash P\subst{x}{v} \smalltriangleright \Delta' $ and $ \Delta' \leq_1 \Delta $ imply $ \Gamma \vdash P\subst{x}{v} \smalltriangleright \Delta $.
		\item[Case of {[T-$ \nonsend $]}:] In this case $ P = \insess{c}{\probOut{\prob}{\nonsend \cont Q}} $ with $ \Delta = \Delta_0, \type{c}{\probOut{\prob}{\nonsend \cont T}} $, where $ \Gamma, \type{x}{U} \vdash Q \smalltriangleright \Delta_0, \type{c}{T} $.
			Then we have $ P\subst{x}{v} = \insess{c}{\probOut{\prob}{\nonsend \cont {\left( Q\subst{x}{v} \right)}}} $.
			By the induction hypothesis, then the judgements $ \Gamma \vdash \type{v}{U} $ and $ \Gamma, \type{x}{U} \vdash Q \smalltriangleright \Delta_0, \type{c}{T} $ imply that $ \Gamma \vdash Q\subst{x}{v} \smalltriangleright \Delta_0, \type{c}{T} $.
			By [T-$ \nonsend $], then $ \Gamma \vdash P\subst{x}{v} \smalltriangleright \Delta $.
		\item[Case of {[T-Inp]}:] In this case $ P = \insess{c}\inp{p}{q}{l}{y}{Q} $, $ \Delta = \Delta_0, \type{c}{\inp{p}{q}{l}{U_y}{T}} $, $ \participant{p} \in c $, and $ \Gamma, \type{y}{U_y}, \type{x}{U} \vdash Q \smalltriangleright \Delta_0, \type{c}{T} $, where we use alpha-conversion to ensure that $ x \neq y $.
			Then, since $ x \neq y $, $ P\subst{x}{v} = \insess{c}\inp{p}{q}{l}{y}{{\left( Q\subst{x}{v} \right)}} $.
			By the induction hypothesis, $ \Gamma \vdash \type{v}{U} $ and $ \Gamma, \type{y}{U_y}, \type{x}{U} \vdash Q \smalltriangleright \Delta_0, \type{c}{T} $ imply $ \Gamma, \type{y}{U_y} \vdash Q\subst{x}{v} \smalltriangleright \Delta_0, \type{c}{T} $.
			By [T-Inp], then $ \participant{p} \in c $ and $ \Gamma, \type{y}{U_y} \vdash Q\subst{x}{v} \smalltriangleright \Delta_0, \type{c}{T} $ imply $ \Gamma \vdash P\subst{x}{v} \smalltriangleright \Delta $.
		\item[Case of {[T-Out]}:] In this case $ P = \insess{c}\probOut{\prob}{\out{p}{q}{l}{v'}{Q}} $, $ \Delta = \Delta_0, \type{c}{\probOut{\prob}{\out{p}{q}{l}{U'}{T}}} $, $ \participant{p} \in c $, $ \Gamma, \type{x}{U} \vdash \type{v'}{U'} $, and $ \Gamma, \type{x}{U} \vdash Q \smalltriangleright \Delta, \type{c}{T} $.
			Then have the substitution $ P\subst{x}{v} = \insess{c}\probOut{\prob}{\out{p}{q}{l}{{\left( v'\subst{x}{v} \right)}}{{\left( Q\subst{x}{v} \right)}}} $.
			By Definition~\ref{def:typingRules}, $ \Gamma, \type{x}{U} \vdash \type{v'}{U'} $ implies one of the following four cases:
			\begin{description}
				\item[Case $ v' = y $, $ x \neq y $, and $ \type{y}{U'} \in \Gamma, \type{x}{U} $:] Then $ \type{y}{U'} \in \Gamma $.
					By [T-Base], then $ \Gamma \vdash \type{y}{U'} $.
					Then, $ v'\subst{x}{v} = v' $.
					Hence, $ \Gamma \vdash \type{v'\subst{x}{v}}{U'} $.
				\item[Case $ v' = x $:] Then $ \Gamma, \type{x}{U} \vdash \type{v'}{\boolT} $ implies $ U = U' $.
					Moreover, $ v'\subst{x}{v} = v $.
					By $ \Gamma \vdash \type{v}{U} $, then $ \Gamma \vdash \type{v'\subst{x}{v}}{U'} $.
				\item[Case $ v' \in \mathbb{N}^{+} $:] Then $ U' \hspace{-1pt}=\hspace{-1pt} \natT $ and $ v'\subst{x}{v} \hspace{-1pt}=\hspace{-1pt} v' $.
					By [Nat], then $ \Gamma \vdash \type{v'\subst{x}{v}}{U'} $.
				\item[Case $ v' \in \left\lbrace \true, \false \right\rbrace $:] Then $ U' = \boolT $ and $ v'\subst{x}{v} = v' $.
					By [Bool], then $ \Gamma \vdash \type{v'\subst{x}{v}}{U'} $.
			\end{description}
			In all four cases we have $ \Gamma \vdash \type{v'\subst{x}{v}}{U'} $.
			By the induction hypothesis, then $ \Gamma \vdash \type{v}{U} $ and $ \Gamma, \type{x}{U} \vdash Q \smalltriangleright \Delta_0, \type{c}{T} $ imply $ \Gamma \vdash Q\subst{x}{v} \smalltriangleright \Delta_0, \type{c}{T} $.
			By [T-Out], then $ \participant{p} \in c $, $ \Gamma \vdash \type{v'\subst{x}{v}}{U'} $, and $ \Gamma \vdash Q\subst{x}{v} \smalltriangleright \Delta_0, \type{c}{T} $ imply $ \Gamma \vdash P\subst{x}{v} \smalltriangleright \Delta $.
		\item[Case of {[T-Prob]}:] In this case $ P = \insess{c}\send{i}{I} \probOut{\prob_i}{N_i{\cont}P_i} $, $ \Delta = \Delta_0, \type{c}{\send{i}{I}\probOut{\prob_i}{H_i \cont T_i}} $, and for all $ i \in I $ we have $ \Gamma, \type{x}{U} \vdash \insess{c}\probOut{\prob_i}{N_i{\cont}P_i} \smalltriangleright \Delta_0, \type{c}{\probOut{\prob_i}{H_i \cont T_i}} $.
			Then $ P\subst{x}{v} = \insess{c}\send{i}{I} \probOut{\prob_i}{{\left( {\left( N_i{\cont}P_i \right)} \subst{x}{v} \right)}} $.
			By the induction hypothesis, then judgements $ \Gamma \vdash \type{v}{U} $ and $ \Gamma, \type{x}{U} \vdash \insess{c}N_i{\cont}P_i \smalltriangleright \Delta_0, \type{c}{\probOut{\prob_i}{H_i \cont T_i}} $ imply $ \Gamma \vdash \insess{c}{\probOut{\prob_i}{\left( {\left( N_i{\cont}P_i \right)}\subst{x}{v} \right)}} \smalltriangleright\break \Delta_0, \type{c}{\probOut{\prob_i}{H_i \cont T_i}} $ for all $ i \in I $.
			By [T-Prob], then $ \Gamma \vdash P\subst{x}{v} \smalltriangleright \Delta $.\qedhere
	\end{description}
\end{proof}

\subsubsection{Theorem Statement and Proof}

With the properties we have shown above, the proof of subject reduction is a straightforward induction on the derivation of $ P \longrightarrow_{\prob} P' $.
In each case we construct the typing for $ P' $ from the information gained by $ \Gamma \vdash P \smalltriangleright \Delta $.

\begin{theorem}[Subject Reduction]
	\label{thm:subjectReduction}
	If $ \Gamma \vdash P \smalltriangleright \Delta $, $ \safe{\Delta} $, and $ P \longrightarrow_{\prob} P' $ then there is some $ \Delta' $ such that $ \Delta \mapsto^{*}_{\prob} \Delta' $ and $ \Gamma \vdash P' \smalltriangleright \Delta' $.
\end{theorem}

\begin{proof}
	Assume $ \Gamma \vdash P \smalltriangleright \Delta $, $ \safe{\Delta} $, and $ P \longrightarrow_{\prob} P' $.
	We prove $ \Delta \mapsto^{*} \Delta' $ and $ \Gamma \vdash P' \smalltriangleright \Delta' $ by structural induction on the derivation of $ P \longrightarrow_{\prob} P' $.
	\begin{description}
		\item[Case of {[R-Cond-$ \true $]}:] In this case $ P = \cond{\true}{Q}{R} $ and $ P' = Q $.
			By inverting the typing rules (and in particular [T-If] under arbitrary applications of [T-Sub]), then $ \Gamma \vdash P \smalltriangleright \Delta $ implies $ \Gamma \vdash P' \smalltriangleright \Delta_Q $ with $ \Delta_Q \leq_1 \Delta $.
			By [T-Sub], then $ \Gamma \vdash P' \smalltriangleright \Delta_Q $ and $ \Delta_Q \leq_1 \Delta $ imply $ \Gamma \vdash P' \smalltriangleright \Delta $ and $ \Delta \mapsto^{*} \Delta $.
		\item[Case of {[R-Cond-$ \false $]}:] This case is similar to the case above.
		\item[Case of {[R-Def-$ \0 $]}:] In this case we have $ P = \procdef{D}{\0} $ and $ P' = \0 $.
			Let $ D = \left\lbrace X_i{\left( \vecnew{x_i}, c_{i, 1}, \ldots, c_{i, n_i} \right)} = R_i \right\rbrace_{i \in I} $.
			By inverting the typing rules (in particular [T-Def] and [T-$ \0 $] under arbitrary appl.\! of [T-Sub]), then $\Gamma, \left\lbrace \type{X_i}{\left\langle \vecnew{U_i}, T_{i, 1}, \ldots, T_{i, n_i} \right\rangle} \right\rbrace_{i \in I}\hspace{-5pt} \vdash \0 \smalltriangleright \emptyset$
			with $ \emptyset \leq_1 \Delta $.
			By [T-$ \0 $] and [T-Sub], then $ \Gamma \vdash P' \smalltriangleright \Delta $ and $ \Delta \mapsto^{*} \Delta $.
		\item[Case of {[R-$ \nonsend $]}:] In this case $ P = \insess{c}\left( \probOut{\prob}{\nonsend{\cont}Q} \oplus N \right) + M $ and $ P' = Q $.
			By inverting the typing rules (and in particular [T-Sum], [T-Prob] and [T-$ \nonsend $] under arbitrary applications of [T-Sub]), then $ \Gamma \vdash P \smalltriangleright \Delta $ implies $ \Gamma \vdash P' \smalltriangleright \Delta' $ with $ \Delta' = \Delta_{Q}, \type{c}{T} $ and $ \Delta_{Q}, \type{c}{\left( \probOut{\prob}{\nonsend\cont{T}} \oplus N \right) + M} \leq_1 \Delta $.
			By [TR-$ \nonsend $], then $ \Delta \mapsto^{*} \Delta' $.
		\item[Case of {[R-Com]}:] In this case we have processes $ P = \insess{\channelk{s}{r}{1}}\inp{p}{q}{l}{x}{Q} + \choice_Q \break \inparallel \insess{\channelk{s}{r}{2}}\left( \probOut{\prob}{\out{q}{p}{l}{v}{R}} \oplus \choiceProb \right) + \choice_R $ and $ P' = Q\subst{x}{v} \inparallel R $.
			By inverting the typing rules (and in particular [T-Par], [T-Sum], [T-Inp], [T-Prob], and [T-Out] under arbitrary applications of [T-Sub]), then $ \Gamma \vdash P \smalltriangleright \Delta $ implies
			\begin{align*}
				& \Gamma \vdash \insess{\channelk{s}{r}{1}}\inp{p}{q}{l}{x}{Q} + \choice_Q \smalltriangleright \Delta_1'\\
				& \Gamma \vdash \insess{\channelk{s}{r}{2}}\left( \probOut{\prob}{\out{q}{p}{l}{v}{R}} \oplus \choiceProb \right) + \choice_R \smalltriangleright \Delta_2'
			\end{align*}
			with $ \Delta_1' \leq_1 \Delta_1 $; $ \Delta_2' \leq_1 \Delta_2 $, and $ \Delta_1, \Delta_2 \leq_1 \Delta $,
			\begin{align*}
				& \Gamma \vdash \insess{\channelk{s}{r}{1}}\inp{p}{q}{l}{x}{Q} \smalltriangleright \Delta^{T}_{1}, \type{\channelk{s}{r}{1}}{\inp{p}{q}{l}{U_1}{T_1}}\\
				& \Gamma, \type{x}{U_1} \vdash Q \smalltriangleright \Delta^{T}_{1}, \type{\channelk{s}{r}{1}}{T_1}
			\end{align*}
			with $ \Delta^{T}_{1}, \type{\channelk{s}{r}{1}}{\inp{p}{q}{l}{U_1}{T_1} + \choice_Q'} \leq_1 \Delta_1' $, $ \participant{p} \in \vecnew{\participant{r}_{1}} $,
			\begin{align*}
				& \Gamma \vdash \insess{\channelk{s}{r}{2}}\left( \probOut{\prob}{\out{q}{p}{l}{v}{R}} \oplus \choiceProb \right) \smalltriangleright  \Delta^{T}_{2}, \type{\channelk{s}{r}{2}}{\probOut{\prob}{\out{q}{p}{l}{U_2}{T_2}} \oplus \choiceProb'}
			\end{align*}
			with $ \Delta^{T}_{2}, \type{\channelk{s}{r}{2}}{\left( \probOut{\prob}{\out{q}{p}{l}{U_2}{T_2}} \oplus \choiceProb' \right) + \choice_R'} \leq_1 \Delta_2' $,
			\begin{align*}
				& \Gamma \vdash \insess{\channelk{s}{r}{2}}\probOut{\prob}{\out{q}{p}{l}{v}{R}} \smalltriangleright \Delta^{T}_{2}, \type{\channelk{s}{r}{2}}{\probOut{\prob}{\out{q}{p}{l}{U_2}{T_2}}}\\
				& \Gamma \vdash R \smalltriangleright \Delta^{T}_{2}, \type{\channelk{s}{r}{2}}{T_2}
			\end{align*}
			with $ \participant{q} \in \vecnew{\participant{r}_{2}} $ and $ \Gamma \vdash \type{v}{U_2} $.
			By Definition~\ref{def:safetyProperty} and Theorem~\ref{thm:subtypingProperties}.\ref{thm:subtypingProperties:safe}, $ \safe{\Delta} $ and the above subtyping relations imply $ U_1 = U_2 $.
			By Lemma~\ref{lem:substitution}, then $ \Gamma \vdash \type{v}{U_2} $ and $ \Gamma, \type{x}{U_1} \vdash Q \smalltriangleright \Delta^{T}_{1}, \type{\channelk{s}{r}{1}}{T_1} $ imply $ \Gamma \vdash Q\subst{x}{v} \smalltriangleright \Delta^{T}_{1}, \type{\channelk{s}{r}{1}}{T_1} $.
			By Proposition~\ref{prop:subtypeTrans} and Lemma~\ref{lem:composeSubtyping}, then:
			\begin{align*}
				\Delta^{T}_{1}, \type{\channelk{s}{r}{1}}{\inp{p}{q}{l}{U_1}{T_1} + \choice_Q'}, \Delta^{T}_{2}, \type{\channelk{s}{r}{2}}{\left( \probOut{\prob}{\out{q}{p}{l}{U_2}{T_2}} \oplus \choiceProb' \right) + \choice_R'} \leq_1 \Delta
			\end{align*}
			By Definition~\ref{def:transitionsLocalContexts}, then:
			\begin{align*}
				& \Delta^{T}_{1}, \type{\channelk{s}{r}{1}}{\inp{p}{q}{l}{U_1}{T_1} + \choice_Q'}, \Delta^{T}_{2}, \type{\channelk{s}{r}{2}}{\left( \probOut{\prob}{\out{q}{p}{l}{U_2}{T_2}} \oplus \choiceProb' \right) + \choice_R'} \mapsto\\
				& \Delta^{T}_{1}, \type{\channelk{s}{r}{1}}{T_1}, \Delta^{T}_{2}, \type{\channelk{s}{r}{2}}{T_2}
			\end{align*}
			By Lemma~\ref{lem:subtypeTypeReduction}, then there is some $ \Delta' $ such that $ \Delta \mapsto^{*} \Delta' $ and $\bigl( \Delta^{T}_{1}, \type{\channelk{s}{r}{1}}{T_1},\break \Delta^{T}_{2}, \type{\channelk{s}{r}{2}}{T_2} \bigr) \leq_1 \Delta' $.
			By [T-Par] and [T-Sub], then $ \Gamma \vdash Q\subst{x}{v} \smalltriangleright \Delta^{T}_{1}, \type{\channelk{s}{r}{1}}{T_1} $ and $ \Gamma \vdash R \smalltriangleright \Delta^{T}_{2}, \type{\channelk{s}{r}{2}}{T_2} $ imply $ \Gamma \vdash P' \smalltriangleright \Delta' $.
		\item[Case of {[R-Def]}:] Have $ P \hspace{-1pt}=\hspace{-1pt} \procdef{D}{\left( \processcall{X}{\vecnew{v}}{\vecnew{c}} \inparallel Q \right)} $ and $ P' \hspace{-1.5pt}=\hspace{-1pt} \procdef{D}{\left( R\subst{\vecnew{x}}{\vecnew{v}} \inparallel Q \right)} $, where  $ \procdecl{X}{\vecnew{x}}{\vecnew{c}} \defeq R \in D $ with $ \vecnew{c} = c_1, \mydots, c_n $.
			By applying structural congruence and by using the argumentation of case [R-Struct] in this proof, we can ensure that $ \procdecl{X}{\vecnew{x}}{\vecnew{c}} $ is the last/innermost declaration in P.
			By inverting the typing rules (and in particular [T-Def], [T-Var], and [T-Par] under arbitrary applications of [T-Sub]), then $ \Gamma \vdash P \smalltriangleright \Delta $ implies $ \Gamma, \Gamma', \type{X}{\left\langle \vecnew{U}, T_1, \mydots, T_n \right\rangle}, \type{\vecnew{x}}{\vecnew{U}} \vdash R \smalltriangleright \type{c_1}{T_1}, \mydots, \type{c_n}{T_n} $, $ \Gamma, \Gamma' \vdash \type{\vecnew{v}}{\vecnew{U}} $, $ \Gamma, \Gamma', \type{X}{\left\langle \vecnew{U}, T_1, \mydots, T_n \right\rangle} \vdash \processcall{X}{\vecnew{v}}{\vecnew{c}} \smalltriangleright \type{c_1}{T_1}, \mydots, \type{c_n}{T_n} $, and $ \Gamma, \Gamma' \vdash Q \smalltriangleright \Delta_Q $ with $ \type{c_1}{T_1}, \mydots, \type{c_n}{T_n}, \Delta_Q \leq_1 \Delta $, where $ \Gamma' $ lists exactly one assignment for each process variable contained in $ D $ besides $ X $ and nothing else and for each such process variable we have the corresponding type judgement for the declared process similar to $ \Gamma, \Gamma', \type{X}{\left\langle \vecnew{U}, T_1, \mydots, T_n \right\rangle}, \type{\vecnew{x}}{\vecnew{U}} \vdash R \smalltriangleright \type{c_1}{T_1}, \mydots, \type{c_n}{T_n} $.
			By Lemma~\ref{lem:substitution}, then $ \Gamma, \Gamma' \vdash \type{\vecnew{v}}{\vecnew{U}} $ and $ \Gamma, \Gamma', \type{X}{\left\langle \vecnew{U}, T_1, \mydots, T_n \right\rangle}, \type{\vecnew{x}}{\vecnew{U}} \vdash R \smalltriangleright \type{c_1}{T_1}, \mydots, \type{c_n}{T_n} $ imply $ \Gamma, \Gamma', \type{X}{\left\langle \vecnew{U}, T_1, \mydots, T_n \right\rangle} \vdash R\subst{\vecnew{x}}{\vecnew{v}} \smalltriangleright \type{c_1}{T_1}, \mydots, \type{c_n}{T_n} $.
			Then, by [T-Par], the judgements $ \Gamma, \Gamma' \vdash Q \smalltriangleright \Delta_Q $ and $ \Gamma, \Gamma', \type{X}{\left\langle \vecnew{U}, T_1, \mydots, T_n \right\rangle} \vdash R\subst{\vecnew{x}}{\vecnew{v}} \smalltriangleright \type{c_1}{T_1}, \mydots, \type{c_n}{T_n} $ imply $ \Gamma, \Gamma', \type{X}{\left\langle \vecnew{U}, T_1, \mydots, T_n \right\rangle} \vdash R\subst{\vecnew{x}}{\vecnew{v}} \inparallel Q \smalltriangleright \type{c_1}{T_1}, \mydots, \type{c_n}{T_n}, \Delta_{Q} $.
			Then, by several applications of [T-Def] using the type judgements for the process variables besides $ X $, obtain $ \Gamma \vdash P' \smalltriangleright \type{c_1}{T_1}, \mydots, \type{c_n}{T_n}, \Delta_{Q} $.
			By [T-Sub], then $ \type{c_1}{T_1}, \mydots, \type{c_n}{T_n}, \Delta_Q \leq_1 \Delta $ and $ \Gamma \vdash P' \smalltriangleright \type{c_1}{T_1}, \mydots, \type{c_n}{T_n}, \Delta_{Q} $ imply $ \Gamma \vdash P' \smalltriangleright \Delta $ and $ \Delta \mapsto^{*} \Delta $.
		\item[Case of {[R-Par]}:] In this case $ P = Q \inparallel R $, $ P' = Q' \inparallel R $, and $ Q \longrightarrow_{\prob} Q' $.
			By inverting the typing rules (and in particular [T-Par] under arbitrary applications of [T-Sub]), then $ \Gamma \vdash P \smalltriangleright \Delta $ implies $ \Gamma \vdash Q \smalltriangleright \Delta_Q'' $ with $ \Delta_Q'' \leq_1 \Delta_Q $, $ \Gamma \vdash R \smalltriangleright \Delta_R' $ with $ \Delta_R' \leq_1 \Delta_R $, and $ \Delta_Q, \Delta_R \leq_1 \Delta $.
			By [T-Sub], then $ \Gamma \vdash Q \smalltriangleright \Delta_Q'' $ and $ \Delta_Q'' \leq_1 \Delta_Q $ imply $ \Gamma \vdash Q \smalltriangleright \Delta_Q $.
			By [T-Sub], then $ \Gamma \vdash R \smalltriangleright \Delta_R' $ and $ \Delta_R' \leq_1 \Delta_R $ imply $ \Gamma \vdash R \smalltriangleright \Delta_R $.
			By the induction hypothesis, $ \Gamma \vdash Q \smalltriangleright \Delta_Q $ and $ Q \longrightarrow_{\prob} Q' $ imply $ \Delta_Q \mapsto^{*} \Delta_Q' $ and $ \Gamma \vdash Q' \smalltriangleright \Delta_Q' $.
			By Lemma~\ref{lem:typeReductionCompose}, then $ \Delta_Q \mapsto^{*} \Delta_Q' $; $ \Delta_R \mapsto^{*} \Delta_R $, and $ \Delta_Q, \Delta_R \leq_1 \Delta $ imply $ \Delta \mapsto^{*} \Delta' $ and $ \Delta_Q', \Delta_R \leq_1 \Delta' $.
			By [T-Par], then $ \Gamma \vdash Q' \smalltriangleright \Delta_Q' $ and $ \Gamma \vdash R \smalltriangleright \Delta_R $ imply $ \Gamma \vdash P' \smalltriangleright \Delta_Q', \Delta_R $.
			By [T-Sub], then $ \Gamma \vdash P' \smalltriangleright \Delta_Q', \Delta_R $ and $ \Delta_Q', \Delta_R \leq_1 \Delta' $ imply $ \Gamma \vdash P' \smalltriangleright \Delta' $.
		\item[Case of {[R-Struct]}:] In this case $ P \equiv Q $, $ Q \longrightarrow_{\prob} Q' $, and $ Q' \equiv P' $.
			By Lemma~\ref{lem:subjectCongruence}, $ \Gamma \vdash P \smalltriangleright \Delta $ and $ P \equiv Q $ imply $ \Gamma \vdash Q \smalltriangleright \Delta $.
			By the induction hypothesis, then $ \Gamma \vdash Q \smalltriangleright \Delta $ and $ Q \longrightarrow_{\prob} Q' $ imply $ \Delta \mapsto^{*} \Delta' $ and $ \Gamma \vdash Q' \smalltriangleright \Delta' $.
		\item[Case of {[R-Res]}:] In this case $ P = \res{s}Q $, $ P' = \res{s}Q' $, and $ Q \longrightarrow_{\prob} Q' $.
			By inverting the typing rules (and in particular [T-Res] under arbitrary applications of [T-Sub]), then $ \Gamma \vdash P \smalltriangleright \Delta $ implies $ \safe{\Delta_s} $ and $ \Gamma \vdash Q \smalltriangleright \Delta_Q $ with $ \Delta_Q \leq_1 \Delta, \Delta_s $ and $ \s{\Delta} \cap \s{\Delta_s} = \emptyset $.
			By the induction hypothesis, then $ \Gamma \vdash Q \smalltriangleright \Delta_Q $ and $ Q \longrightarrow_{\prob} Q' $ imply $ \Delta_Q \mapsto^{*} \Delta_Q' $ and $ \Gamma \vdash Q' \smalltriangleright \Delta_Q' $.
			By Lemma~\ref{lem:subtypeTypeReduction}, then $ \Delta_Q \leq_1 \Delta, \Delta_s $ and $ \Delta_Q \mapsto^{*} \Delta_Q' $ imply $ \Delta, \Delta_s \mapsto^{*} \Delta'' $ and $ \Delta_Q' \leq_1 \Delta'' $.
			By Lemma~\ref{lem:typeReductionSplit}, then $ \Delta, \Delta_s \mapsto^{*} \Delta'' $ and $ \s{\Delta} \cap \s{\Delta_s} = \emptyset $ imply $ \Delta \mapsto^{*} \Delta' $; $ \Delta_s \mapsto^{*} \Delta_s' $, and $ \Delta'' = \Delta', \Delta_s' $.
			By Lemma~\ref{lem:typeReductionSafe}, then $ \Delta_s \mapsto^{*} \Delta_s' $ and $ \safe{\Delta_s} $ imply $ \safe{\Delta_s'} $.
			By [T-Sub], then $ \Gamma \vdash Q' \smalltriangleright \Delta_Q' $; $ \Delta_Q' \leq_1 \Delta'' $, and $ \Delta'' = \Delta', \Delta_s' $ imply $ \Gamma \vdash Q' \smalltriangleright \Delta', \Delta_s' $.
			By [T-Res], then $ \safe{\Delta_s'} $ and $ \Gamma \vdash Q' \smalltriangleright \Delta', \Delta_s' $ imply $ \Gamma \vdash P' \smalltriangleright \Delta' $.
		\item[Case of {[R-Def-In]}:] In this case $ P = \procdef{D}{Q} $, $ P' = \procdef{D}{Q'} $, and $ Q' \longrightarrow_{\prob} Q' $.
			Let:
			\begin{align*}
				D = \left\lbrace X_i{\left( \vecnew{x_i}, c_{i, 1}, \mydots, c_{i, n_i} \right)} = R_i \right\rbrace_{i \in I}
			\end{align*}
			By inverting the typing rules (and in particular [T-Def] under arbitrary applications of [T-Sub]), then $ \Gamma \vdash P \smalltriangleright \Delta $ implies
			\begin{align*}
				\Gamma, \type{X_1}{\left\langle \vecnew{U_1}, T_{1, 1}, \mydots T_{1, n_1} \right\rangle}, \mydots, \type{X_i}{\left\langle \vecnew{U_i}, T_{i, 1}, \mydots T_{i, n_i} \right\rangle}, \type{\vecnew{x_i}}{\vecnew{U_i}} \vdash R_i' \smalltriangleright \type{c_{i, 1}}{T_{i, 1}}, \mydots, \type{c_{i, n_i}}{T_{i, n_i}}
			\end{align*}
			for all $ i \in I $ and $ \Gamma, \left\lbrace \type{X_i}{\left\langle \vecnew{U_i}, T_{i, 1}, \mydots, T_{i, n_i} \right\rangle} \right\rbrace_{i \in I} \vdash Q \smalltriangleright \Delta_Q $ with $ \Delta_Q \leq_1 \Delta $.
			By the induction hypothesis, then $ \Gamma, \left\lbrace \type{X_i}{\left\langle \vecnew{U_i}, T_{i, 1}, \mydots, T_{i, n_i} \right\rangle} \right\rbrace_{i \in I} \vdash Q \smalltriangleright \Delta_Q $ and $ Q \longrightarrow_{\prob} Q' $ imply $ \Gamma, \left\lbrace \type{X_i}{\left\langle \vecnew{U_i}, T_{i, 1}, \mydots, T_{i, n_i} \right\rangle} \right\rbrace_{i \in I} \vdash Q' \smalltriangleright \Delta_Q' $ and $ \Delta_Q \mapsto^{*} \Delta_Q' $.
			By [T-Def], then
			\begin{align*}
				\Gamma, \type{X_1}{\left\langle \vecnew{U_1}, T_{1, 1}, \mydots T_{1, n_1} \right\rangle}, \mydots, \type{X_i}{\left\langle \vecnew{U_i}, T_{i, 1}, \mydots T_{i, n_i} \right\rangle}, \type{\vecnew{x_i}}{\vecnew{U_i}} \vdash R_i' \smalltriangleright \type{c_{i, 1}}{T_{i, 1}}, \mydots, \type{c_{i, n_i}}{T_{i, n_i}}
			\end{align*}
			for all $ i \in I $ and $ \Gamma, \left\lbrace \type{X_i}{\left\langle \vecnew{U_i}, T_{i, 1}, \mydots, T_{i, n_i} \right\rangle} \right\rbrace_{i \in I} \vdash Q' \smalltriangleright \Delta_Q' $ imply $ \Gamma \vdash P' \smalltriangleright \Delta_Q' $.
			By Lemma~\ref{lem:subtypeTypeReduction}, then $ \Delta_Q \leq_1 \Delta $ and $ \Delta_Q \mapsto^{*} \Delta_Q' $ imply $ \Delta \mapsto^{*} \Delta' $ and $ \Delta_Q' \leq_1 \Delta' $.
			By [T-Sub], then $ \Gamma \vdash P' \smalltriangleright \Delta_Q' $ and $ \Delta_Q' \leq_1 \Delta' $ imply $ \Gamma \vdash P' \smalltriangleright \Delta' $.\qedhere
	\end{description}
\end{proof}

\subsection{Error- and Deadlock-Freedom}
\label{subsec:errorAndDeadlockFreedom}

From the Subject Reduction Theorem, two crucial results follow rather naturally, error-freedom and deadlock-freedom.
Errors are essentially the opposite of safety and therefore by having a safe typing context, we seek to infer processes to be error-free when typed by it.
The definition of error processes is similar to the definition of session errors in \citep{peters2024separation}.
In systems with mixed choice, instead of considering all pairs of parallel processes (for non-MC systems, see \citep{scalas2019less}), the error definition must consider all available input-output pairs.
Thus, we define an error process to be a process in which any available output with an equally available dual input does not have an input for which the labels match.

\begin{definition}[Error Process]
	\label{def:errorProcess}
	A process $P$ has a communication error if
	\begin{gather*}
		P \equiv \insess{\channelk{s}{r}{1}}\left( \outPrefix{p}{q}{l}{U}\cont Q \oplus \choiceProb \right) + \choice \inparallel \insess{\channelk{s}{r}{2}}{\inp{q}{p}{l_q}{U_q}{Q_q}+ \choice^q} \inparallel P'
	\end{gather*}
	Where $l \neq l_q$ and for all $M^q_i = \inp{q}{p}{l_i}{U_i}{Q_i} \in \choice^q$, the labels do not match, \ie $l \neq l_i$.
	A process $P$ has a value error if
	\begin{gather*}
		P \equiv \cond{v}{P}{Q} \inparallel P' \qquad v \notin \{\true, \false\}
	\end{gather*}
	$P$ is an error process if it has either a communication error or value error.
\end{definition}

From Theorem~\ref{thm:subjectReduction} and the fact that error processes are untypable by a safe context it follows that:

\begin{corollary}[Error-Freedom]
	\label{cor:errorFreedom}
	If $ \Gamma \vdash P \smalltriangleright \Delta$ with $\safe{\Delta}$ and $ P \Longrightarrow_{\prob} P' $, then $ P' $ is not an error process.
\end{corollary}

We now show the Deadlock-Freedom Theorem, stating that a process typed by a deadlock-free and safe local context, is deadlock-free.
Similar to \citep{scalas2019less}, we require two additional side conditions on the shape of these processes.
The first prohibits restriction within the process and the second enforces each participant in a session to be typed by exactly one corresponding type.
While our system does not inherently impose these conditions on all processes, some other formalism, like \citep{peters2024separation}, do.
Thus, we consider these conditions quite natural and reasonable (for results on interleaved, interfering sessions, see \citep{DBLP:journals/mscs/CoppoDYP16}).
Indeed, without these conditions, deadlock-freedom for a local type does not imply deadlock-freedom for processes typed by it.
Let us illustrate why this is with the following example.

\begin{example}[Deadlock Freedom]
\addxcontentsline{loe}{example}[\theexample]{Deadlock Freedom}
    Consider the following deadlock-free local context $\Delta$, which we will use to type two different processes, each exemplifying why one of the conditions is needed.
    \begin{gather*}
        \Delta = \type{\channelsingle{s}{a}}{\inpPrefix{a}{b}{l}{}}, \type{\channelsingle{s}{b}}{\outPrefix{b}{a}{l}{}}
    \end{gather*}
    Firstly, take process $P_1$ which can be justified by it, \ie $ \vdash P_1 \smalltriangleright \Delta $.
    \begin{gather*}
        P_1 = \res{s'} \left( \insess{\channelsingle{s}{a}}{\outPrefix{a}{b}{l}{}} \;\inparallel\; \insess{\channelsingle{s}{b}}{\outPrefix{b}{a}{l}{}} \;\inparallel\; \insess{\channelsingle{s'}{x}}{\outPrefix{x}{y}{l}{}} \right)
    \end{gather*}
    $P_1$ will deadlock after one step, despite $\Delta$ being deadlock-free, as restricting a session within a process ``removes'' the types used for that session from the context.
    For the second condition, take the process $P_2$ for which also holds $ \vdash P_2 \smalltriangleright \Delta $.
    \begin{gather*}
        P_2 = \insess{\channelsingle{s}{a}}{\inpPrefix{a}{b}{l}{}}\cont \insess{\channelsingle{s}{b}}{\outPrefix{b}{a}{l}{}}
    \end{gather*}
    Clearly, $P_2$ is deadlocked, too.
    \exampleDone
\end{example}

We now formalize these conditions. 

\begin{definition}[Canonical]
    Assume $\vdash P \smalltriangleright \Delta$, then $P$ is \emph{canonical for $\Delta$} iff
    \begin{compactenum}
        \item there is no occurrence of restriction, $\res{s'} P'$, in $P$, and
        \item $P \equiv \big|_{k \in [1..n]} P_k$ and $\Delta = \type{c_1}{T_1}, \dots, \type{c_n}{T_n}$ where for all $k \in [1..n]$ have $\vdash P_k \smalltriangleright \type{c_k}{T_k}$ and for all subterms of the form $c \mc{i}{I} M_i$ occurring in $P_k$, have $c = c_k$.
    \end{compactenum}
\end{definition}

To show deadlock-freedom, we first prove an intermediary lemma stating that if a canonical process has no reductions, then its corresponding local context has no reductions either.
We will use the shorthand $\xlongrightarrow{\type{s}{\nonsend}}^{*}_{1}$ to mean a reflexive transitive closure of the labelled transition $\xlongrightarrow{\type{s}{\nonsend}}_{\prob}$, \ie a reduction sequence of $\probOut{1}{\nonsend}$ actions.

\begin{lemma}
	\label{lem:deadlock}
	   If $ P \nrightarrow $, $ \vdash P \smalltriangleright \Delta $, and $P$ canonical for $\Delta$ then $\Delta \xlongrightarrow{\type{s}{\nonsend}}^{*}_{1} \Delta' $ with $ \vdash P \smalltriangleright \Delta' $, and $\Delta' \nrightarrow$.
\end{lemma}

\begin{proof}
    We will first show that given $ P \nrightarrow $, $ \vdash P \smalltriangleright \Delta $, and $P$ canonical for $\Delta$, the transition sequence $\Delta \xlongrightarrow{\type{s}{\nonsend}}^{*}_{1} \Delta' $ implies $ \vdash P \smalltriangleright \Delta' $.
    Assume that there is at least one step in said transition sequence, as otherwise the statement holds trivially by reflexivity.
    As $P$ canonical for $\Delta$ we have that $P \equiv \big|_{k \in [1..n]} P_k$ and $\Delta = \type{c_1}{T_1}, \dots, \type{c_n}{T_n}$ where for all $k \in [1..n]$ have $\vdash P_k \smalltriangleright \type{c_k}{T_k}$ and for all subterms of the form $c \mc{i}{I} M_i$ occurring in $P_k$, have $c = c_k$.
    W.l.o.g. let us now fix a typed channel $\type{c_i}{T_i}$ from $\Delta$ which can perform said sequence of at least one internal action, \ie $\type{c_i}{T_i} \xlongrightarrow{\type{s}{\nonsend}}^{*}_{1}$ and $\type{c_i}{T_i} \xlongrightarrow{\type{s}{\nonsend}}_{1}$.
    As $ P \nrightarrow $, in particular also $P_i \nrightarrow$.
    In other words $P_i$ has no unguarded internal action $\probOut{1}{\nonsend}$.
    Therefore, in the typing derivation of $\vdash P_i \smalltriangleright \type{c_i}{T_i}$, there occurs an application of [T-Sub] such that $\vdash P_i \smalltriangleright \type{c_i}{T_i'}$ with $\type{c_i}{T_i'} \leq_1 \type{c_i}{T_i}$ and $T_i'$ has \emph{no} unguarded internal action $\probOut{1}{\nonsend}$.
    By Corollary~\ref{cor:subtypePreorder} and the subtyping rules in Definition~\ref{def:subtypingCompositional}, $\type{c_i}{T_i'} \leq_1 \type{c_i}{T_i}$ where $\type{c_i}{T_i} \xlongrightarrow{\type{s}{\nonsend}}_{1}$ and $\type{c_i}{T_i'} \nrightarrow$ implies that [S-$\nonsend$-R] is the only applicable rule.
    Thus, $\type{c_i}{T_i} \xlongrightarrow{\type{s}{\nonsend}}^{*}_{1} \type{c_i}{T_i'}$.
    Let $\Delta'$ be obtained from $\Delta$ by replacing $ \type{c_i}{T_i} $ with $ \type{c_i}{T_i'}$.
    Then indeed $\Delta \xlongrightarrow{\type{s}{\nonsend}}^{*}_{1} \Delta'$.
    As $P$ canonical for $\Delta$ with $\vdash P_i \smalltriangleright \type{c_i}{T_i}$ and $\vdash P_i \smalltriangleright \type{c_i}{T_i'}$, we have $ \vdash P \smalltriangleright \Delta' $ as required.
    
    Given this, we may restrict our attention to contexts which have no internal transitions of probability one. To show the main statement given \emph{not} $\Delta \xlongrightarrow{\type{s}{\nonsend}}_{1}$, assume the contrary, \ie assume $ P \nrightarrow $, $ \vdash P \smalltriangleright \Delta $, and $ P $ is canonical for $\Delta$ but not $\Delta \nrightarrow$.
    Then $ \Delta \xlongrightarrow{\type{s}{\comPrefix{p}{q}{l}{U}}}_{\prob_1}$ or $ \Delta \xlongrightarrow{\type{s}{\nonsend}}_{\prob_2} $ with $\prob_2 \neq 1$.
	In the former case, by [TR-Com], [TR-Inp], and [TR-Out], then $ \Delta \xlongrightarrow{\type{s}{\comPrefix{p}{q}{l}{U}}}_{\prob} \Delta' $ and $P$ canonical for $\Delta$ imply that $ P $ contains an unguarded output and a matching unguarded input.
	Then $ P $ can perform a step using [R-Com] contradicting $ P \nrightarrow $.
	In the latter case, by [TR-$ \nonsend $], then $ \Delta \xlongrightarrow{\type{s}{\nonsend}}_{\prob} \Delta' $ with $P$ canonical for $\Delta$ imply that $ P $ contains an unguarded subterm of the form $ \nonsend \cont Q $.
	Then $ P $ can perform a step using [R-$ \nonsend $] again contradicting $ P \nrightarrow $.
	In both cases we have a contradiction.
	Hence, $ \Delta \nrightarrow $.
\end{proof}

\begin{theorem}[Deadlock-Freedom]
	\label{thm:deadlockFreedom}
	If $ {} \vdash P \smalltriangleright \Delta $, $ \safe{\Delta} $, $ \dfree{\Delta} $, and $P$ canonical for $\Delta$, then $ P $ is deadlock-free.
\end{theorem}

The proof of Deadlock-Freedom is similar to the corresponding proof in \citep{peters2024separation}.

\begin{proof}
	Assume $ P \Longrightarrow_{\prob} P' \nrightarrow $, $ {} \vdash P \smalltriangleright \Delta $, $ \safe{\Delta} $, $ \dfree{\Delta} $, and $P$ canonical for $\Delta$.
	By Theorem~\ref{thm:subjectReduction}, then there is some $ \Delta' $ such that $ \Delta \mapsto_{\prob}^{*} \Delta' $ and $ {} \vdash P' \smalltriangleright \Delta' $.
	By Lemma~\ref{lem:deadlock}, then $\Delta' \xlongrightarrow{\type{s}{\nonsend}}^{*}_{1} \Delta''$ with $\vdash P' \smalltriangleright \Delta''$ and $ \Delta'' \nrightarrow $.
	By Definition~\ref{def:deadlockFreedom}, then $ \Delta'' = \emptyset $.
	Since the global environment is empty and $ {} \vdash P' \smalltriangleright \Delta'' $, for all conditionals in $ P' $ the boolean condition is already a value and, since $ P' \nrightarrow $, then $ P' $ cannot contain unguarded conditionals.
	Since $ P' \nrightarrow $ also cannot contain (modulo structural congruence), unguarded subterms of the form $ \procdef{D}{\0} $, by inverting the typing rules, then $ {} \vdash P' \smalltriangleright \Delta'' $ and $ \Delta'' = \emptyset $ imply $ P' \equiv \0 $.
	%\hfill $ \square $
\end{proof}

In this chapter we have presented numerous important results of the \emph{\calculusname{} $\pi$-calculus}, its types, and refinement-based multi-channel subtyping.
With Corollaries~\ref{cor:subtypePreorder} ($\leq_1$ is a preorder) and \ref{cor:errorFreedom} (error-freedom), as well as Theorems~\ref{thm:subjectReduction} (subject reduction) and \ref{thm:deadlockFreedom} (deadlock-freedom), we have established that our system fulfils all expected and necessary properties of MPST systems.
Additionally, the Interface Existence Theorem~\ref{thm:interfaceExists} guarantees great flexibility of the multi-channel subtyping by ensuring that a single-channel interface can always be assembled from almost any local context, which distinguishes our work from \citep{DBLP:conf/concur/Horne20}.
Having finished all technical contributions, we will now summarize the presented work, highlighting context and significance, and discuss interesting avenues for future research.

\chapter{Conclusion and Future Work}

This thesis set out to create a previously unexplored approach to compositional refinement in multiparty session types using subtyping.
Towards this, we first presented a powerful calculus framework, the \emph{\calculusname{} $\pi$-calculus}, whose combination of the expressive mixed choice setting from \citep{peters2024separation} with probabilistic choices (see \citep{aman2019probabilities, inverso_et_al:LIPIcs.CONCUR.2020.14}) is, to the best of our knowledge, entirely new.
We then extended said calculus with the refinement-based multi-channel subtyping (Definition~\ref{def:subtypingCompositional}).
This unique subtyping allows a typed channel (the interface) to be safely substituted by a collection of several typed channels (the refinement), if their collective behaviour models that of the interface.
This framework enables robust stepwise refinement.
A single channel within a protocol may be taken as interface and safely replaced by several channels, that in turn can each be interfaces for another refinement.
Not only is stepwise refinement useful for systematic, collaborative programming, but, as we have already hinted at with the ``conflict of interest'' in our courthouse example, having the option to perform refinement in this way can enable designers to easily patch security concerns.
For example, a compromised system in which an actor has too many access rights can be repaired by replacing this actor with several and distributing their power (\cf \cite{DBLP:conf/concur/Horne20}).
Moreover, by considering a refinement as the starting point, we can combine its channels into a single channel for whom the interactions within the refinement are concealed.
Cleverly doing so will greatly reduce the complexity and size of the overall system.
Crucially, our subtyping relation is maximally flexible, allowing to unify any selection of well-typed, safe, and deadlock-free channels (the refinement) into an interface consisting of a single channel (Theorem~\ref{thm:interfaceExists}).
Hence, our framework facilitates the strong compositional verifiability of protocols that we sought to create.

\medbreak
Currently the system only supports natural numbers and booleans as base types, and subtyping does thus not include the types of payloads.
Adding more complex types would be a straightforward addition.
Similarly, we currently do not allow session delegation, \ie sending and receiving channels as payload.
While session delegation is not uncommon, it would add significant complexity, especially in our system.

So far, we have been requiring deadlock-freedom to be preserved in $100\%$ of cases when constructing a refinement.
As the system is probabilistic, we believe it would be very interesting to explore \emph{imperfect refinement}, by allowing the refinement to deadlock with a bounded probability.
Much of the system is already designed to accommodate such extensions: We imagine expanding on the semantics of the probability $\prob$ at the subtyping relation $\leq_{\prob}$ to be especially fruitful in this regard.
Furthermore, bounded-probability safety may also be worth exploring.
For example one might allow protocols which are still under development to be deployed if their error probability is below a small threshold.

We are also interested in analysing behavioural properties other than deadlock-freedom and safety.
\citep{scalas2019less} uses the safety predicate $\varphi$ to statically verify typing contexts to be, for example, terminating and live (see \citep{DBLP:journals/toplas/KobayashiS10, DBLP:conf/coordination/PadovaniVV14}) in addition to safe.
As before, typed processes would then be enforced to fulfil these run-time properties.
The notion of \emph{imperfect refinement} could then be taken even further by allowing bounded-probability termination and liveness, too.

\bibliography{msc-thesis}
\bibliographystyle{plainnat}

\end{document}